\documentclass[twocolumn,twocolappendix]{aastex631}

\usepackage{newtxtext,newtxmath}
\usepackage[T1]{fontenc}

\DeclareRobustCommand{\VAN}[3]{#2}
\let\VANthebibliography\thebibliography
\def\thebibliography{\DeclareRobustCommand{\VAN}[3]{##3}\VANthebibliography}

\usepackage{graphicx}	
\usepackage{amsmath}	

\makeatletter
\newcommand{\rmnum}[1]{\romannumeral #1}
\newcommand{\Rmnum}[1]{\expandafter\@slowromancap\romannumeral #1@}

\begin{document}

\title{Disruption of Dark Matter Minihalos in the Milky Way environment: Implications for Axion Miniclusters and Early Matter Domination}

\author[0000-0002-6196-823X]{Xuejian Shen}
\affiliation{TAPIR, California Institute of Technology, Pasadena, CA 91125, USA}

\author[0000-0003-2485-5700]{Huangyu Xiao}
\affiliation{Department of Physics, University of Washington,  Seattle, WA 98195, USA}
\affiliation{Astrophysics Theory Department, Theory Division, Fermilab, Batavia, IL 60510, USA}

\author[0000-0003-3729-1684]{Philip F. Hopkins}
\affiliation{TAPIR, California Institute of Technology, Pasadena, CA 91125, USA}

\author[0000-0002-2629-337X]{Kathryn M. Zurek}
\affiliation{Walter Burke Institute for Theoretical Physics, California Institute of Technology, Pasadena, CA 91125, USA}



\begin{abstract}
Many theories of dark matter beyond the Weakly Interacting Massive Particles (WIMP) paradigm feature an enhanced matter power spectrum on sub-parsec scales, leading to the formation of dense dark matter minihalos. Future local observations are promising to search for and constrain such substructures. The survival probability of these dense minihalos in the Milky Way environment is crucial for interpreting local observations. In this work, we investigate two environmental effects: stellar disruption and (smooth) tidal disruption. These two mechanisms are studied using semi-analytic models and idealized N-body simulations. For stellar disruption, we perform a series of N-body simulations of isolated minihalo-star encounters to test and calibrate analytic models of stellar encounters before applying the model to the realistic Milky Way disk environment. For tidal disruption, we perform N-body simulations to confirm the effectiveness of the analytic treatment. Finally, we propose a framework to combine the hierarchical assembly and infall of minihalos to the Milky Way with the late-time disruption mechanisms. We make predictions for the mass functions of minihalos in the Milky Way. The mass survival fraction (at $M_{\rm mh}\geq 10^{-12}\,{\rm M}_{\odot}$) of dense dark matter minihalos, {\em e.g.} for axion miniclusters and minihalos from Early Matter Domination, is $\sim 60\%$ with the relatively low-mass, compact population surviving. The survival fraction is insensitive to the detailed model parameters. We discuss various implications of the framework and future direct detection prospects.
\end{abstract}



\section{Introduction}

The gravitational clustering of dark matter has been well measured on galactic scales and super-galactic scales and is consistent with a nearly scale-invariant spectrum of primordial fluctuations \citep[{\em e.g.}\,][]{Planck:2018vyg}. However, the matter power spectrum on extremely small scales ($k\gtrsim\rm pc^{-1}$), which corresponds to sub-planetary-mass structures, is still weakly constrained and is sensitive to both the nature of dark matter and the thermal history of the early Universe. There have been proposals to detect small-scale structures in the mass range $\sim 10^{-13} \operatorname{-} 10^{2}\,{\rm M}_{\odot}$ in the future with Pulsar Timing Arrays \citep[PTAs; {\em e.g.},][]{Siegel:2007fz,Baghram:2011is,Dror:2019twh,Ramani:2020hdo,Lee:2020wfn,Lee:2021zqw} and lensing effects \citep[{\em e.g.},][]{Kolb:1995bu,Metcalf:2001ap,DiazRivero:2017xkd,Fairbairn:2017sil,Katz2018,VanTilburg:2018ykj,Dai_2020}.  

Many well-motivated dark matter theories can leave unique fingerprints on the primordial perturbations at small scales~\footnote{We note that the ``small scale'' here is fundamentally different from the small-scale problem of CDM (at kpc scale) discussed in astrophysical studies \citep[see review by][]{Bullock2017}.} ($k\gtrsim\rm pc^{-1}$), such as the quantum chromodynamics (QCD) axion/axion-like particles (ALPs) with the Peccei-Quinn (PQ) symmetry \citep{Peccei1977} broken after inflation \citep[{\em e.g.},][]{Hogan:1988mp,Kolb:1993zz,Kolb1994,Zurek:2006sy}, Early Matter Domination \citep[EMD; {\em e.g.},][]{Erickcek:2011us,Fan:2014zua} and vector dark matter produced during inflation \citep[{\em e.g.},][]{Nelson:2011sf,Graham:2015rva}. Therefore, small dark matter substructures provide unique insights into the microphysics of dark matter. The model space of interest here differs from that in more common WIMP-like collisionless cold dark matter (CDM) models. In those models, adiabatic fluctuations produced at the end of inflation can also seed small CDM subhalos down to the kinetic decoupling and free-streaming limit ($k\sim \mathcal{O}({\rm pc}^{-1})$), roughly corresponding to the Earth mass~\citep[$\sim 10^{-6}\,{\rm M}_{\odot}$; {\em e.g.},][]{Hofmann2001, Berezinsky2003, Green2005, Loeb2005}. The evolution of these subhalos in the Milky Way environment has been studied in the past~\citep[{\em e.g.},][]{Angus2007,Goerdt2007,Green2007,Zhao2007,Schneider2010, Berezinsky2014,Delos:2019tsl,Facchinetti2022}. However, WIMP-like CDM formed its first non-linear structures at relatively late times with correspondingly low density ({\em e.g.} $z\sim 60 \ll z_{\rm eq}$ as shown in \citealt{Green2005}), and the minihalos are thus subject to significant disruption due to tidal stripping and disk shocking after falling onto their host halos \citep[{\em e.g.},][]{Ostriker1972,Gnedin1999,Goerdt2007,Zhao2007,Schneider2010}. The typical minihalos of WIMP-like CDM are out of reach for PTAs and other observations we discuss here~\citep[{\em e.g.},][]{Lee:2020wfn}. 

On the other hand, dark matter minihalos in the theories we consider here formed in the early Universe. Therefore, they are much denser and less likely to be disrupted by tidal forces than WIMP-like CDM subhalos. For instance, a (pseudo)scalar field ({\em e.g.} the QCD axion) with the PQ symmetry broken after inflation \citep[{\em e.g.},][hereafter called the ``post-inflationary axion'']{Hogan:1988mp,Zurek:2006sy} can induce order-unity isocurvature fluctuations on the horizon scale during the symmetry breaking. Regions with order-unity overdensities tend to collapse gravitationally very early, even before matter-radiation equality ($z_{\rm eq} \sim 3000$), into small axion miniclusters (AMCs). The miniclusters underwent subsequent hierarchical clustering until the large-scale adiabatic perturbations intervened. As another example, EMD models can introduce a non-standard thermal history that is not constrained by any current data, but adiabatic fluctuations within the horizon can grow during this early period of matter domination. The characteristic mass of dark matter minihalos formed in early matter domination models is determined by the reheating temperature of this period. In vector dark matter models, the longitudinal modes of the vector DM produced at the end of inflation give rise to a peak in the matter power spectrum on small scales, with the scale directly determined by the dark matter particle mass~\citep{Graham:2015rva,Lee:2020wfn}. Those models have interesting dynamics in the early Universe, which are difficult to probe directly. However, the remnants of those early Universe dynamics, dark matter minihalos, may be detectable in local observations~\citep[{\em e.g.},][]{Dror:2019twh,Ramani:2020hdo,Lee:2020wfn}. 

While the evolution of (some versions of) these dark matter minihalos in the non-linear regime has been studied both semi-analytically with the Press-Schechter model \citep{Zurek:2006sy,Fairbairn:2017sil,Enander:2017ogx,Blinov:2019jqc,Lee:2020wfn,Blinov:2021axd} and numerically with N-body simulations \citep{Zurek:2006sy,Buschmann:2019icd,Eggemeier:2019khm,Xiao:2021nkb}, the gravitational interactions between dark matter minihalos and the large-scale dark matter structures or baryonic structures are not well studied, due to the large dynamic range and non-linear behaviors involved. Dark matter minihalos formed from non-standard early universe dynamics can be as light as $\sim 10^{-12}\,{\rm M}_{\odot}$, while the Milky Way has a halo mass of $\sim 10^{12}\,{\rm M}_{\odot}$. Therefore it is challenging to resolve these small structures while simultaneously simulating the dynamics of the largest structures. However, it is critical to study the survival probability of minihalos in the Milky Way and to determine the prospects of detecting such structures in the local environment. This aspect is actively studied in the literature. For example, \citet{kavanagh2020stellar} made an analytic estimate of the effects of stellar disruption on AMCs as well as the survival fraction of miniclusters. \citet{Dandoy2022} proposed a quantum mechanical description of AMCs and studied the impact of stellar encounters using standard perturbation theories. However, these studies neglect some important non-linear effects, such as the realistic minihalo concentration distribution, multiple mechanisms of disruption, and their interplay. In this paper, we improve these estimates with detailed numerical simulations and semi-analytic treatment to combine all the expected dominant disruption terms. To deal with the enormous dynamic range involved, our strategy is to numerically simulate the dynamics of the dark matter minihalos in individual encounters (varying {\em e.g.} halo parameters and impact parameters), using these to build detailed semi-analytic models which can be used to treat the large-scale behavior. This allows us to capture the key non-linear physics on small scales while making realistic predictions for the overall behavior of dark matter minihalos in the Milky Way galaxy (in ways that can be generalized, in principle, to a broad class of minihalo-like models). Broadly speaking, the disruption of dark matter minihalos in the Milky Way can be divided into two parts: stellar disruption and tidal disruption. The stellar disruption term is sensitive to close encounters with individual stars, while the tidal term depends on the gradient of the collective gravitational potential on large scales. We study these separately using N-body simulations and then combine them using semi-analytic models.

This paper is organized as follows. In Section~\ref{sec:analytic_model}, we discuss the minihalo mass function, mass-concentration relation, and the analytic models for stellar disruptions due to individual stellar encounters and tidal disruptions. In Section~\ref{sec:stellar_sim}, we present our numerical results from a series of idealized N-body simulations for isolated minihalos encountering a star and simulations of tidal stripping. In Section~\ref{sec:MW_disruption}, we apply the semi-analytic model calibrated using idealized simulations to a realistic Milky Way environment and combine both stellar disruption and tidal disruption. In Section~\ref{sec:dm_mf_change}, we study the survival fraction of dark matter minihalos by applying the framework to different physical models. 

We assume a $\Lambda$CDM cosmology with parameters given as $h=0.697$, $\Omega_{\rm m}=0.2814$, and $\Omega_{\Lambda}=0.7186$, and adopt scalar spectral index $n_s= 0.9667$. These are consistent with the recent Planck results \citep{Planck:2018vyg}, and our conclusions are relatively insensitive to variations in these parameters (compared to the much larger uncertainties in {\em e.g.}\ minihalo properties from different physical models).

\section{Analytic model}\label{sec:analytic_model}

\begin{figure}
    \centering
    \includegraphics[width=0.49\textwidth]{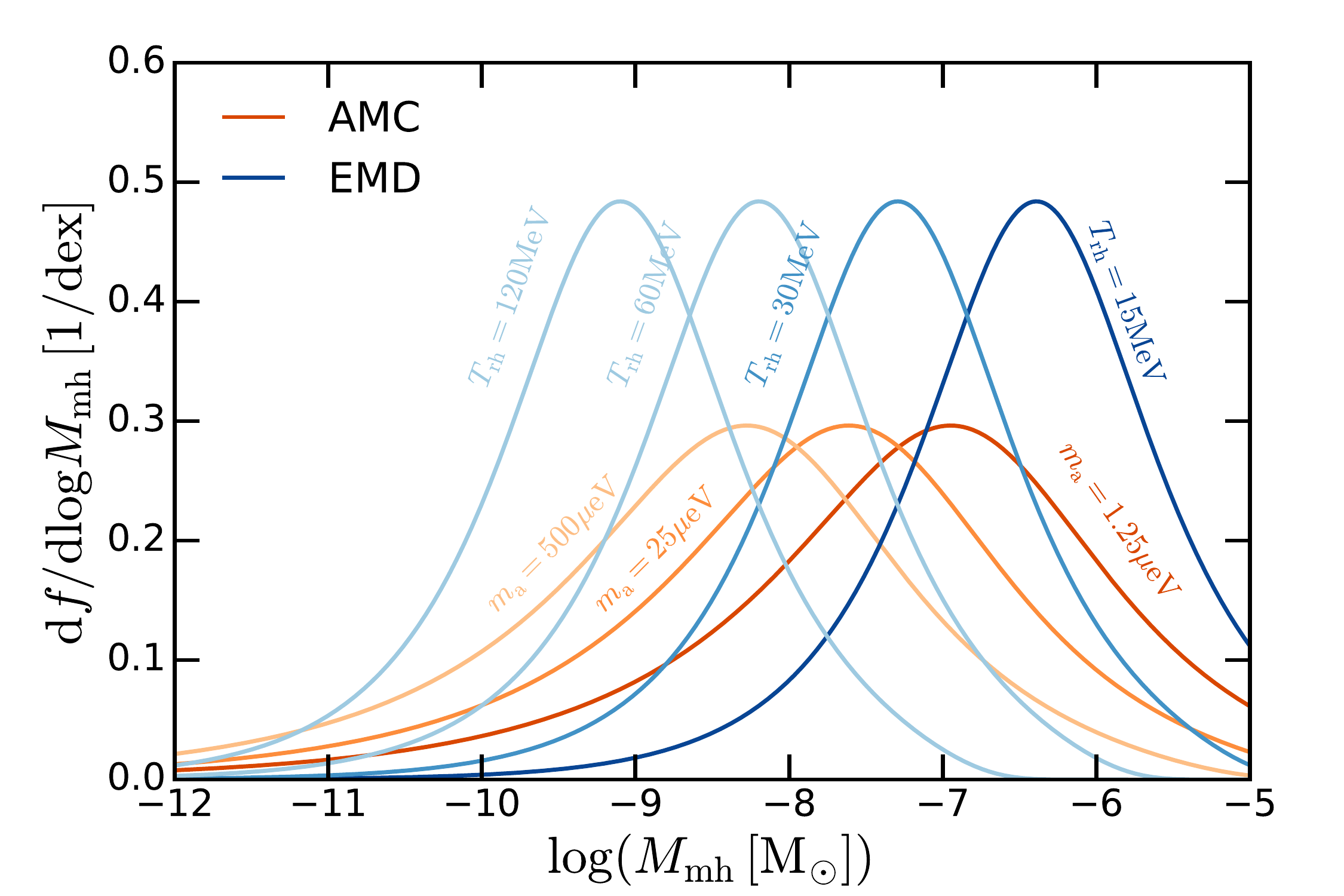}
    \includegraphics[width=0.49\textwidth]{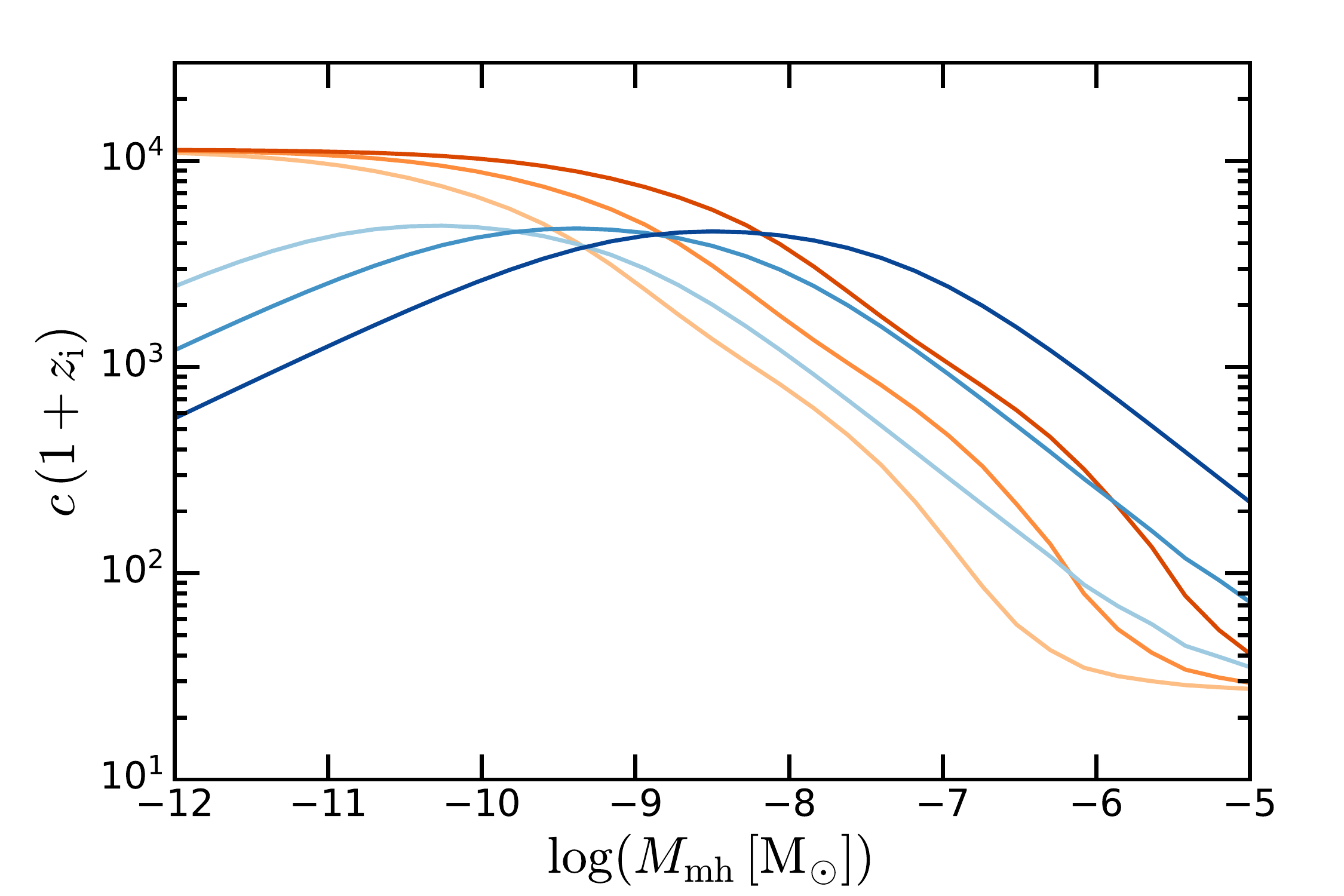}
    \caption{{\it Top}: Initial (pre-disruption) mass function of minihalos in different physics models ($f$ is the mass fraction with respect to the total dark matter mass in the Universe). Here we show the mass functions for the AMC models with axion mass $m_{\rm a} = 1.25, 25, 500\, \mu{\rm eV}$ and the EMD models with reheating temperature $T_{\rm rh}=15,30,60,120\,{\rm MeV}$. The model parameter variations manifest as constant horizontal shifts of the mass function (see Appendix~\ref{app:mass_function} for details). The EMD models exhibit sharper peaks than the AMC models. {\it Bottom}: Mass-concentration relation of minihalos in different models shown in the top panel. The mass-concentration relation in the EMD model is peaked due to its peaked matter power spectrum. In general, the halo concentration hits a floor at the massive end as the adiabatic CDM power spectrum takes over. At the low-mass end, since one would not expect minihalos to form before matter-radiation equality, a cap to the concentration appears at around $c(1+z_{\rm i})=10^{4}$. }
    \label{fig:mass_function_init}
\end{figure}

\subsection{Initial mass functions and concentrations of small-scale structures}
\label{sec:models}

The two models discussed in the introduction that lead to the formation of dense minihalos in the early Universe are: \rmnum{1}. (pseudo)scalars ({\em e.g.} the QCD axion) with symmetry broken after inflation~\citep{Hogan:1988mp,Zurek:2006sy}, and \rmnum{2}. the EMD model~\citep{Erickcek:2011us}. These two models are physically well-motivated and are also representatives of models enhancing the matter power spectrum at small scales. Although the cosmological perturbations in those models have different origins, they share common features. They are generated at extremely small scales before matter-radiation equality, forming dark matter substructures as light as $10^{-12}\,{\rm M}_{\odot}$ ubiquitously, decoupled from the usual adiabatic fluctuations. The model parameters are the axion mass $m_{\rm a}$ for the AMC model and the reheating temperature $T_{\rm rh}$ for the EMD model. We vary the axion mass from $1.25\,\mu{\rm eV}$ to $500\,\mu{\rm eV}$, which is roughly the mass window that can produce the correct dark matter relic abundance, given uncertainties introduced by the axion emission from strings as shown in \citet{Buschmann:2021sdq,Gorghetto:2020qws}. On the contrary, the reheating temperature is loosely constrained although $T_{\rm rh}$ cannot be below a few MeV. Otherwise, Big Bang Nucleosynthesis will be spoiled (see {\em e.g.} \citealt{Kawasaki:1999na}). We study the value from $15\,{\rm MeV}$ to $120\,{\rm MeV}$. We choose the fiducial values $m_{\rm a}=25\mu {\rm eV}$ and $T_{\rm rh}=30\,{\rm MeV}$, such that the minihalo mass function in both models peaks at $\sim 10^{-8}\operatorname{-}10^{-7}\,{\rm M}_{\odot}$.

The formation of dense substructures from these small-scale perturbations occurs at early times (see \citealt{Lee:2020wfn} for an analytic study and \citealt{Xiao:2021nkb} for simulations of axion miniclusters). The formation and hierarchical mergers of minihalos are described by the redshift-dependent mass function ${\rm d}n_0/{\rm d}M(z)$, calibrated by simulations in \citet{Xiao:2021nkb}. Meanwhile, the minihalos can fall onto CDM substructures formed from adiabatic perturbations. The redshift of infall, $z_{\rm i}$, which occurs after the minihalo formation, defines the time when they stop merging or accretion, owing to the high virial velocities in the normal CDM halos, and their properties (mass, concentration) remain unchanged afterward.

Therefore, the final mass function, including minihalos falling onto the CDM structures, can be expressed as
\begin{equation}
\frac{{\rm d}n_{\rm f}}{{\rm d}M}(z) =\int_{z_{\rm eq}}^{z} {\rm d}z_{\rm i} \frac{{\rm d}f_{\rm col}^{\rm CDM}(z_{\rm i})}{{\rm d}z_{\rm i}}\frac{{\rm d}n_{0 }}{{\rm d}M}(z_{\rm i})
\label{eq:current_mf}
\end{equation}
where $f_{\rm col}^{\rm CDM}(z)$ is the collapse fraction of normal CDM halos formed from adiabatic fluctuations. The probability of infall at $z=z_{\rm i}$ is proportional to $\frac{{\rm d}f_{\rm col}^{\rm CDM}(z_{\rm i})}{{\rm d}z_{\rm i}}$, where we have assumed there is no dynamic decoupling between collapsed minihalos and the overall dark matter content. 

We can use the Press–Schechter model \citep{1974ApJ...187..425P} to compute the collapse fraction
\begin{equation}
    f_{\rm col}^{\rm CDM}(z)={\rm erfc}\left(\frac{\delta_{\rm c}}{\sqrt{2}\,\sigma_{\rm CDM}(M_{\rm min})\,D(z)}\right),
    \label{eqn:fcol}
\end{equation}
where $\delta_{\rm c} = 1.686$ is the critical overdensity for spherical collapse, $D(z)$ is the growth function, $\sigma^{2}_{\rm CDM}(M)$ is the variance of the adiabatic fluctuations calculated using the Code for Anisotropies in the Microwave Background~\footnote{\href{https://camb.info/}{{\sc Camb} documentation}} \citep[{\sc Camb};][]{Lewis:1999bs,Howlett:2012mh,camb_notes}, and $M_{\rm min}$ is the smallest CDM structure we consider formed from adiabatic fluctuations. We take $M_{\rm min}$ to be $10^{-2}\,{\rm M}_{\odot}$, corresponding to a scale where the CDM power spectrum dominates while the enhanced small-scale perturbations become subdominant. The result is insensitive to $M_{\rm min}$ as $\sigma(M)$ depends logarithmically on $M$ at small scales. There could be even smaller CDM halos that can form, which will increase the collapse fraction $f_{\rm col}^{\rm CDM}(z)$, especially at high redshifts. However, below a certain mass, the normal CDM halos may have comparable masses to the enhanced substructures, and our assumptions will no longer hold. A detailed study of this infall process will ultimately be required for more detailed predictions and we leave it for future study. 

The method described above allows us to compute the collapse fraction at various redshifts analytically and obtain the final mass function for the enhanced substructures. This will also give us the distribution of the infall redshift $z_{\rm i}$, $\mathcal{P}(z_{\rm i})$~\footnote{Integrated over the entire mass range, $\mathcal{P}(z_{\rm i}) \propto {\rm d}f_{\rm col}^{\rm CDM}(z)/{\rm d}z$ is purely determined by CDM cosmology. However, if looking at minihalos at $z=0$ in a certain mass range, the conditional probability distribution of $z_{\rm i}$ will be mass dependent. In the hierarchical formation of minihalos, more massive ones form at later times. Therefore, more massive minihalos found at $z=0$ tend to have lower $z_{\rm i}$.}, which determines the time when the structural changes of minihalos halt and thus the average density of minihalos. Including minihalos falling onto CDM halos, we obtain the initial (pre-disruption) minihalo mass function at $z=0$ from different models, as shown in Figure~\ref{fig:mass_function_init}.

Regarding the concentration of minihalos, we adopt the model in \citet{Lee:2020wfn} to calculate the concentration from the power spectra of different models. For each mass, it evaluates the redshift when the corresponding primordial fluctuation mode collapse ($z_{\rm c}$) and assumes $c(M) \propto (1+z_{\rm c}(M))$, which is similar to the CDM results \citep[{\em e.g.}\,][]{Bullock2001}. The only difference here is that we assume the merger or smooth accretion of minihalos stops at $z=z_{\rm i}$ instead of $z=0$. Therefore, effectively, we have $c(M) \propto (1+z_{\rm c}(M))/(1+z_{\rm i})$. In Figure~\ref{fig:mass_function_init}, we show the initial (pre-disruption) mass functions and the mass-concentration relations of minihalos in different physics models. At the massive end, the adiabatic CDM power spectrum will dominate and one should reproduce the mass-concentration relation of normal CDM halos. There is a cap for concentration at the low-mass end since one will not expect minihalo formation before matter-radiation equality. The axion mass for the AMC models or the reheating temperature of the EMD models only creates constant mass shifts in the mass function and the mass-concentration relation. For comparison, the typical concentration of WIMP-like CDM halos studied in {\em e.g.} \citet{Green2007} and \citet{Delos:2019tsl} are $c \sim 1\operatorname{-}20$ at the redshift when minihalos are initialized (usually $z\sim 60-100$ in these studies). In terms of the central density, they are comparable to the relatively massive minihalos studied here.

\subsection{Disruption in late-time evolution}

Owing to the ultra-compact structure of the minihalos, they are largely immune to external perturbations through their evolutionary history after decoupling from the Hubble flow. However, after falling into a massive host system like the Milky Way, non-linear gravitational interactions with the host halo and the dense baryonic structures in the host halo could lead to significant disruption of minihalos. The two leading disruption mechanisms are tidal disruption from the host halo (and the baryonic disk) and close encounters with stars (referred to as stellar disruption). The relevant spatial and time scales on which these two mechanisms operate are drastically different. In this section, we will review the analytic model developed in the literature. We also note that the term ``disruption'' used in this paper describes the mass loss of minihalo from external perturbations at various levels and is not restricted to the case where a minihalo is completely ``destroyed'' as in some literature.

\subsubsection{Stellar disruption}
\label{sec:stellar_analytic}

First, we consider the consequence of the encounter between a minihalo and a star. The virial radius of minihalos with a mass of interest ($\sim 10^{-10}\,{\rm M}_{\odot}$) is of the order of $0.01\,{\rm pc} \sim 2000\,{\rm AU}$, which is still much larger than the radius of main sequence stars even though we are considering minihalos. Therefore, for simplicity, stars can be treated as point-like objects during encounters. In addition, after a stellar encounter, the structure of the minihalo cannot immediately react to the energy imparted during the encounter. It takes roughly a dynamical time for the minihalos to relax to the final state after disruption, which is given by
\begin{equation}
    \label{eq:tdyn}
    t_{\rm dyn} = \sqrt{\dfrac{3\pi}{16 G \bar{\rho}_{\rm mh}}} 
    \approx 0.3\,{\rm Gyr} \,\left(\dfrac{1+z_{\rm i}}{1+5}\right)^{-3/2},
\end{equation}
where $\bar{\rho}_{\rm mh}$ is the average density of the minihalo (assumed to be $\Delta_{\rm c}$ times the critical density of the Universe at $z_{\rm i}$, see the discussion after Equation~\ref{eq:single_encounter}). The dynamical time $t_{\rm dyn}$ is comparable to the Hubble time scale for $z_{\rm i}=0$, but is much shorter than that if the minihalos fall into CDM structures at a high redshift. The duration of the star-minihalo encounter, however, is several orders of magnitude shorter than $t_{\rm dyn}$. Therefore, the impulse approximation holds and the encounter can be treated as an instantaneous interaction \citep{Spitzer1958}. In the distant-tide approximation (when the impact parameter is much larger than minihalo size), the imparted energy from a single star encounter can be expressed as \citep{Spitzer1958}
\begin{equation}
    \label{eq:deltaE}
        \Delta E \approx \dfrac{4}{3} \dfrac{G^2 m^{2}_{\ast} M_\text{mh}\langle r^2\rangle}{v^{2}_{\ast} b^4} = \dfrac{4 \alpha^{2}}{3} \dfrac{G^2 m^{2}_{\ast} M_\text{mh} R_{\rm mh}^2}{v^{2}_{\ast} b^4},
\end{equation}
where $\Delta E$ is the increase of internal energy of the minihalo, $M_{\rm mh}$ ($R_{\rm mh}$) is the virial mass (radius) of the minihalo, $\langle r^2\rangle$ represents the mean-squared radius of particles with respect to the center of the minihalo. $m_{\ast}$, $b$, and $v_{\ast}$ are the mass, impact parameter, and relative velocity of the stellar object, respectively. The mean-squared radius can be parameterized as $\langle r^2\rangle=\alpha^2 R_{\rm mh}^2 $, where $\alpha$ is a dimensionless parameter determined by the density profile $\rho(r)$  of the dark matter minihalo
\begin{equation}\label{eq:alpha2_int}
    \alpha^2=\frac{\langle r^2\rangle}{R_{\rm mh}^2}=\frac{1}{M_{\rm mh} R_{\rm mh}^2}\int_0^{R_{\rm mh}}{\rm d}^{3}\mathbf{r}\, r^2\rho(r).
\end{equation}
Assuming that the minihalo has the NFW profile~\citep{Navarro1996,Navarro1997}, one obtains 
\begin{equation}
    \label{eq:alpha}
    \alpha^{2}(c) = \frac{c(-3-3c/2+c^2/2)+3(1+c)\,{\rm ln}(1+c)}{c^2(-c+(1+c)\,{\rm ln}(1+c))},
\end{equation}
where $c$ is the concentration number of the minihalo.

However, when the impact parameter becomes comparable to the size of the minihalo, the distant-tide approximation made by Equation~\ref{eq:deltaE} breaks down. The strong $b^{-4}$ dependence of $\Delta E$ will disappear once the star passes through the minihalo and the disruption effect is suppressed \citep[{\em e.g.},][]{Gerhard1983,Moore1993,Carr1999,Green2007}. For a single encounter, \citet{Green2007} proposed a more general treatment of the imparted energy calibrated using simulations
\begin{equation}
    \label{eq:deltaE_general}
    \Delta E = \begin{cases} 
    \dfrac{4 \alpha^{2}(c)}{3} \dfrac{G^{2} m_{\ast}^{2} M_{\rm mh} R^{2}_{\rm mh}}{v_{\ast}^2} \dfrac{1}{b^4} & \text{$(b>b_{\rm s})$} \\
    \dfrac{4 \alpha^{2}(c)}{3} \dfrac{G^{2} m_{\ast}^{2} M_{\rm mh} R^{2}_{\rm mh}}{v_{\ast}^2} \dfrac{1}{b_{\rm s}^4} & \text{$(b\leq b_{\rm s})$}
    \end{cases}
\end{equation}
where $b_{\rm s} = f_{\rm b}\,(2\alpha/3\beta)^{1/2} R_{\rm mh}$ is the transition radius, which is close to the physical size of the minihalo up to a factor determined by structure parameters, and $f_{\rm b}$ is an order-unity correction factor we introduce~\footnote{The original formula proposed in \citet{Green2007}, who were studying low concentration halos, does not have the correction term (equivalently $f_{\rm b}=1$). However, minihalos with higher concentrations are studied in this work and we find the correction term is necessary empirically to fit the simulation results.} to be determined by our simulations, which will be discussed in Section~\ref{sec:stellar_sim}. $\beta$ is another structural parameter defined as 
\begin{align}
    \beta^{2} &= \langle r^{-2} \rangle R_{\rm mh}^{2} = \dfrac{R_{\rm mh}^{2}\,\int_{r_{\rm c}}^{R_{\rm mh}} {\rm d}^{3}\mathbf{r}\,\,r^{-2}\,\rho(r)}{M_{\rm mh}} \nonumber \\
    &\simeq \dfrac{c^{2}\,\ln{(r_{\rm s}/r_{\rm c})} + c^{2}/2 - 1/2}{\ln{(1+c) - c/(1+c)}},
\end{align}
where the NFW profile is assumed and $r_{\rm c}$ is the smallest radius that the profile extends to, which we assume to be $0.01\,r_{\rm s}$. In principle, an axion star formed in the center of a minihalo can provide a natural physical scale for $r_{\rm c}$. However, the size of an axion star sensitively depends on particle physics parameters as well as the growth rate of axion stars \citep[{\em e.g.},][]{Visinelli:2017ooc,Helfer:2016ljl, Chen:2020cef}. We note that the choice of $r_{\rm c}$ has weak effects since (1) it only appears in the logarithm; (2) the uncertainties of $r_{\rm c}$ and $\beta$ have been effectively captured by the free parameter $f_{\rm b}$; (3) $\beta$ in fact only matters for rare close encounters. Compared to full analytic calculations under the impulse and distant-tide approximations, Equation~\ref{eq:deltaE_general} gives better agreement with simulations in the transitional regime. 

Roughly speaking, disruption of a minihalo is expected to occur when the increase in internal energy of the minihalo given by Equation~\ref{eq:deltaE} exceeds the binding energy of the minihalo
\begin{equation}
\label{eq:ebinding}
E_{\rm b}= \gamma G M_\text{mh}^2/R_{\rm mh}.    
\end{equation}
Here $\gamma$ is a dimensionless parameter again determined by the mass profile of the dark matter halo. For the NFW profile, it takes the form~\citep{1998MNRAS.295..319M}
\begin{equation}\label{eq:gamma}
    \gamma(c)=\frac{c}{2}\frac{1-1/(1+c)^2-2{\rm ln}(1+c)/(1+c)}{[c/(1+c)-{\rm ln}(1+c)]^2}.
\end{equation}

Utilizing Equation~\ref{eq:deltaE_general} and \ref{eq:ebinding}, we obtain the normalized energy input to the minihalo as
\begin{equation}
    \label{eq:single_encounter}
    \dfrac{\Delta E}{E_{\rm b}} = \begin{cases}
    \dfrac{4\alpha^2(c)}{3 \gamma(c)} \dfrac{G m^{2}_{\ast}}{v_{\ast}^{2} b^{4}}  \dfrac{R^{3}_{\rm mh}}{M_{\rm mh}} = 
    \dfrac{\alpha^2(c)}{\pi \gamma(c)} \dfrac{G m^{2}_{\ast}}{v_{\ast}^{2} b^{4}}  \dfrac{1}{\bar{\rho}_{\rm mh}} & \text{$(b>b_{\rm s})$} \\ \\
    
    \dfrac{4\alpha^2(c)}{3 \gamma(c)} \dfrac{G m^{2}_{\ast}}{v_{\ast}^{2} b_{\rm s}^{4}} \dfrac{R^{3}_{\rm mh}}{M_{\rm mh}} =
    \dfrac{3 \beta^2(c)}{f^{4}_{\rm b}\,\gamma(c)} \dfrac{G m^{2}_{\ast}}{v_{\ast}^{2}} \dfrac{1}{M_{\rm mh} R_{\rm mh}} & \text{$(b\leq b_{\rm s})$}
   \end{cases}
\end{equation}
where $\bar{\rho}_{\rm mh} \equiv M_{\rm mh}/(4\pi\,R^{3}_{\rm mh}/3)$ is the average density of the minihalo. Assuming the minihalos are in virial equilibrium with respect to the background density at redshift $z_{\rm i}$ (the infall redshift), we obtain $\bar{\rho}_{\rm mh}=\Delta_{\rm c}\,\rho_{\rm crit}(z_{\rm i})$, where $\rho_{\rm crit}(z_{\rm i})$ is the critical density of the Universe at $z_{\rm i}$ and $\Delta_{\rm c}=200$ is the critical overdensity of collapsed objects (neglecting corrections from {\em e.g.}\ $\Omega_{\Lambda}$, $\rho_{\rm crit}(z_{\rm i}) \sim (1+z_{\rm i})^{3}$). Since $\bar{\rho}_{\rm mh}$ is independent of minihalo mass, the energy input in the large-$b$ case in Equation~\ref{eq:single_encounter} will be independent of minihalo mass while having a strong dependence on the impact parameter. The only free parameter that is left to be determined by simulations in Equation~\ref{eq:single_encounter} is the correction factor $f_{\rm b}$.

From Equation~\ref{eq:single_encounter} (assuming the $b>b_{\rm s}$ case), we can compute a characteristic impact parameter when $\Delta E / E_{\rm b} = 1$
\begin{align}
    \label{eq:bmin}
    & b_\text{min} = \sqrt{\frac{m_{\ast}}{v_{\ast}}}\left(\frac{\alpha^2(c) G}{\pi \gamma(c) \bar{\rho}_{\rm mh}}\right)^{1/4} \nonumber \\
    & \approx 0.07\,{\rm pc}\,\left(\dfrac{\alpha^2(c)}{\gamma(c)}\right)^{1/4}\,\left(\dfrac{m_{\ast}}{1\,{\rm M}_{\odot}}\right)^{1/2} \left(\dfrac{v_{\ast}}{200\,{\rm km/s}}\right)^{-1/2} \left(\dfrac{1+z_{\rm i}}{1+5}\right)^{-3/4},
\end{align}
which gives a crude estimate of the condition for the destruction of minihalos. It is worth noting that the average density of a virialized halo is much larger when formed at higher redshifts, given as $\bar{\rho}_{\rm mh} \propto (1+z_{\rm i})^3$. Minihalos that collapsed and fell into the host halo earlier in cosmic time should be less vulnerable to stellar disruptions due to higher central densities. Although $b_{\rm min}$ serves as an indicator for ``significant'' disruption of the minihalo, the actual mass loss of the minihalo (as a function of $\Delta E/E_{\rm b}$) after a single encounter should be calibrated by numerical simulations presented in the following section.

\begin{figure*}
    \centering
    \includegraphics[width=0.33\textwidth]{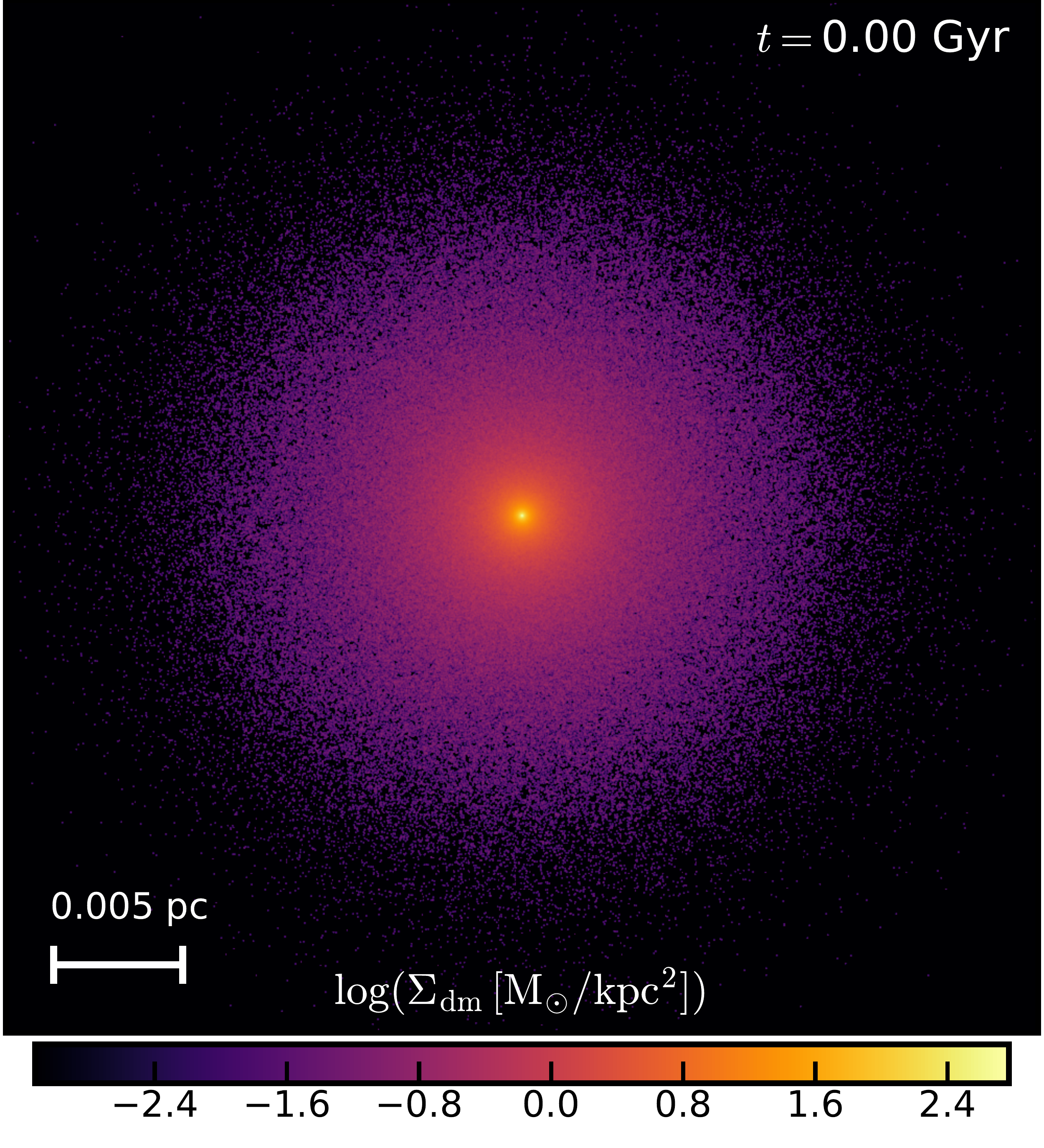}
    \includegraphics[width=0.33\textwidth]{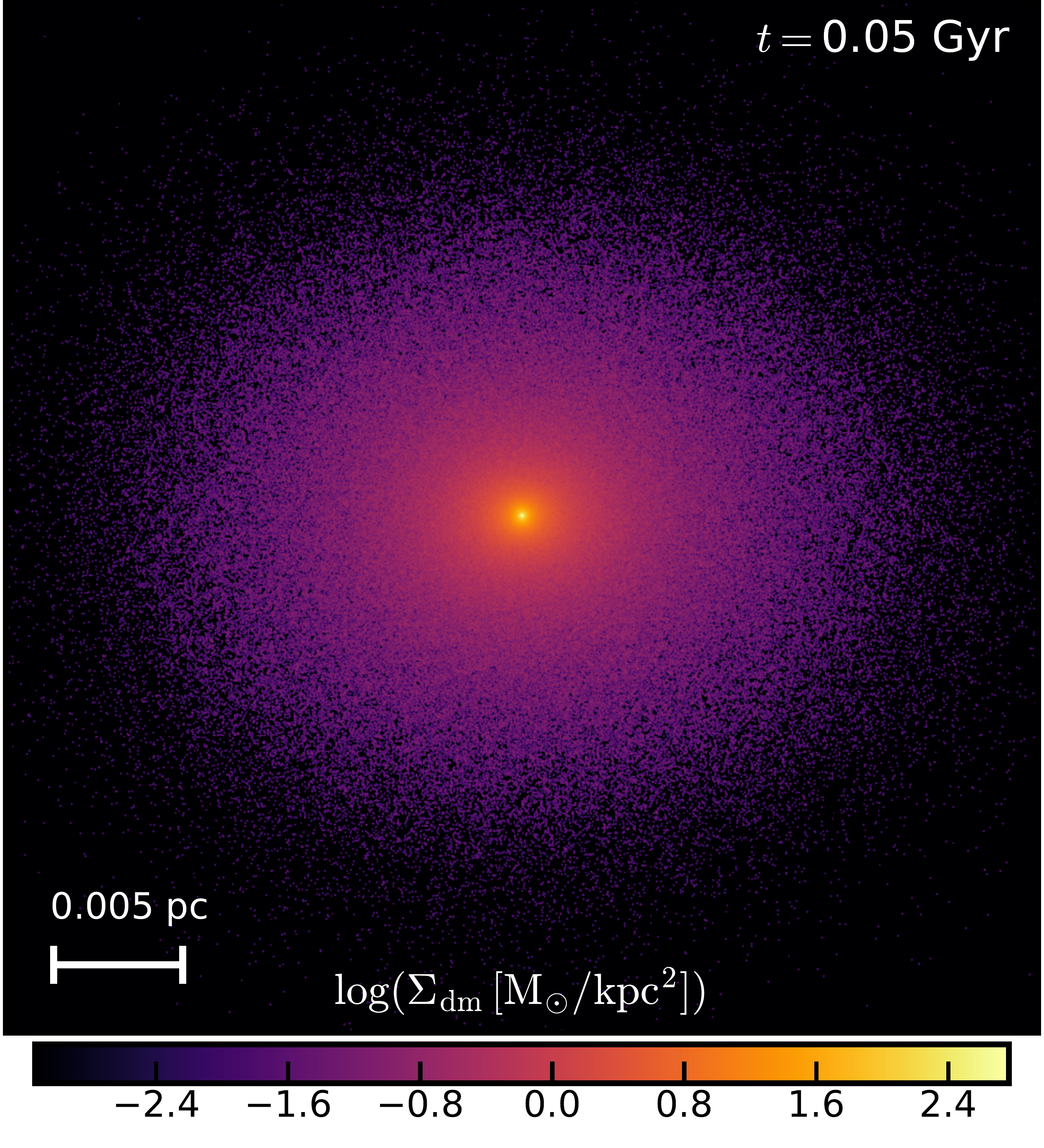}
    \includegraphics[width=0.33\textwidth]{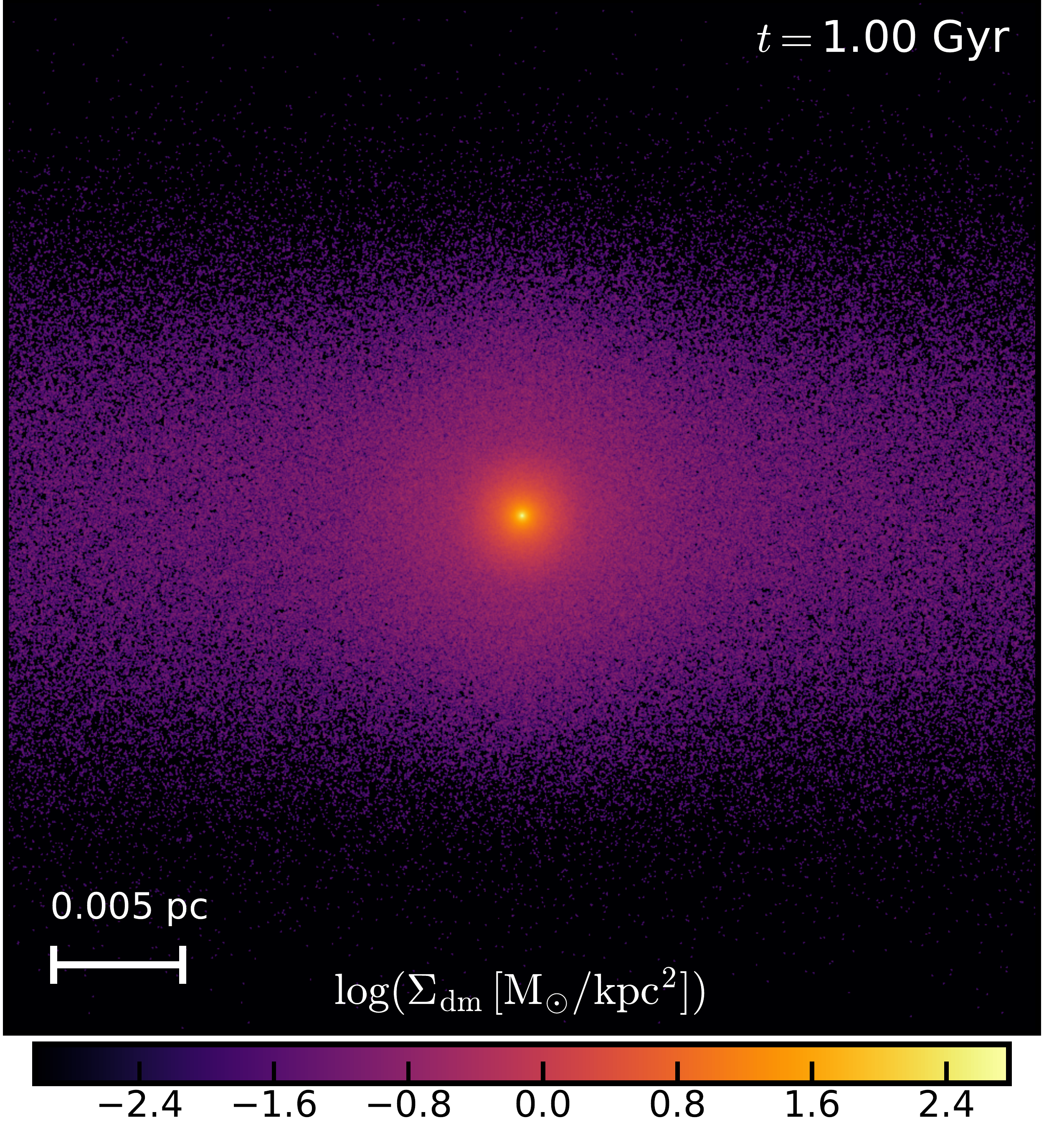}
    \caption{The evolution of a minihalo during stellar disruption is visualized. This is a simulation of an encounter between a minihalo of mass $M_{\rm mh}=10^{-10}\,{\rm M}_{\odot}$ and a star of mass $m_{\ast}= 1\,{\rm M}_{\odot}$, with the impact parameter $b=10^{-4}\,{\rm kpc}$. The star passed by in the y-direction of the image. We show the dark matter surface density distribution of the minihalo. The left panel shows the initial dark matter distribution. Right after the encounter, a large fraction of dark matter has been heated up by the gravitational interaction with the passing star and become unbounded. Although this energy is transferred impulsively, it requires roughly the minihalo dynamical time ($t_{\rm dyn} \gtrsim 1\,{\rm Gyr}$) for the disruption to be reflected in the minihalo density distribution, as shown in the middle and right panel.}
\label{fig:halo_relaxation}
\end{figure*}

\subsubsection{Tidal disruption}
\label{sec:tidal-disruption}

Another important disruption mechanism for minihalos is tidal disruption. After the minihalos fall into the Milky Way halo, they experience tidal forces from the Milky Way dark matter halo and the Galactic disk. In contrast to stellar disruption, which sensitively depends on close encounters with stars, the tidal forces are determined by gravity at galactic scales, dominated by the collective effects of the smooth gravitational potential rather than the fluctuating component from individual objects. Moreover, rather than an impulsive event, tidal disruption is secular and usually treated differently from stellar disruptions. 

In a sufficiently strong and smooth external tidal field, the outskirts of the minihalo will be stripped away where the tidal force exceeds the self-gravity of the minihalo. This tidal radius is given by \citep[{\em e.g.},][]{King1962,TaylorBabul2001,ZentnerBullock2003}
\begin{equation}
    r_{ t}=R \left[\frac{M_{\rm mh}(r<r_{ t})/M_{\rm MW}(r<R)}{2-\frac{{\rm d}{\rm ln}M_{\rm MW}}{{\rm d}{\rm ln}R}\big\vert_{R} + \dfrac{V_{\rm t}(\bf R)}{V_{\rm c}(R)}  }\right]^{1/3},
    \label{eq:tidal_radius}
\end{equation}
where $R$ is the galactocentric distance of the minihalo, $M_{\rm MW}(r<R)$ is the mass of the Milky Way (including both dark and baryonic matter) enclosed within radius $R$ and $M_{\rm mh}(r<r_{\rm t})$ is the minihalo mass enclosed within $r_{ t}$. $V_{\rm t}(\bf R)$ is the instantaneous tangential speed of the minihalo, which equals the circular velocity $V_{\rm c}(R)$ when the minihalo is on a circular orbit. The minihalo mass outside the tidal radius will be stripped away roughly over a dynamical time scale of the host, and the instantaneous mass-loss rate is often expressed as \citep[{\em e.g.},][]{Taffoni2003, Zentner2003, Oguri2004, Pullen2014}
\begin{equation}
    \dot{M}_{\rm mh} = - \dfrac{M_{\rm mh}(r>r_{\rm t})}{t_{\rm ts}(R)},
    \label{eq:mass_tidal_loss}
\end{equation}
where $t_{\rm ts}$ is the characteristic time scale of tidal stripping, which is proportional to the dynamical time scale of the host halo (not the minihalo)
\begin{align}
    \label{eq:tidal_time}
    t_{\rm ts}(R) & = A\,t^{\rm host}_{\rm dyn}(R) \nonumber \\
    & \approx 90\,{\rm Myr}\, \left(\dfrac{A}{1}\right)\,\left(\dfrac{\bar{\rho}_{\rm host}(R)}{1.6\times 10^{7} {\rm M}_{\odot}\,{\rm kpc}^{-3}}\right)^{-1/2}
\end{align}
where $\bar{\rho}_{\rm host}(R)$ is the averaged density of the host halo within radius $R$ and $A$ is a constant fudge factor of order unity that is found to be $\sim 0.5 \operatorname{-} 3$ in several previous studies \citep[{\em e.g.},][]{Zentner2003, Zentner2005, Pullen2014, vandenBosch2018, Green2021,Errani2023}. If the tidal stripping occurs in the Solar neighborhood, the time scale will be much shorter than the lifetime of minihalo in the Milky Way environment. As a simple estimation, the characteristic formation time (defined as when ${\rm d}\log{M}/{\rm d}\log{a}$ falls below a threshold as proposed in \citealt{Wechsler2002}) of a Milky Way-mass halo is about $a=0.3$ (corresponding to $z \sim 2.5$ and lookback time $\gtrsim 10\,{\rm Gyr}$). If the minihalos are dynamically coupled to the smooth dark matter content accreted by the Milky Way halo, the typical lifetime of minihalos in the Milky Way environment will be of the same order. Meanwhile, since $t_{\rm ts}$ is approximately the orbital period of the minihalo at the pericenter, a reasonable assumption is that the minihalo outskirts will be tidally disrupted ``immediately'' in the first pericenter passage and before any form of stellar disruptions takes place.

\section{Idealized simulations for stellar encounters and tidal stripping}
\label{sec:stellar_sim}

In this section, we use N-body simulations to systematically study the stellar disruption and tidal stripping effects for isolated minihalos initialized with the NFW profile. The goal is to test the analytic models described in Section~\ref{sec:analytic_model} and calibrate them against various minihalo parameters.

\subsection{Stellar encounters}

We perform a suite of N-body simulations for minihalos initialized with the NFW profile by varying minihalo concentrations, masses, and impact parameters of the encountering stars. The simulations adopt the code {\sc Gizmo}~\footnote{\href{http://www.tapir.caltech.edu/~phopkins/Site/GIZMO_files/gizmo_documentation.html}{{\sc Gizmo} documentation}}~\citep{Hopkins2015}, which has been widely used in cosmological N-body or hydrodynamical simulations~\citep[{\em e.g.},][]{Hopkins2014,Hopkins2018,Feldmann2022}. The simulations aim to test and calibrate analytic predictions of the minihalo mass fraction disrupted in stellar encounters. The relaxation process of minihalos and the evolution of density profiles are also studied, which provides important insights into modeling the disruption effects under multiple encounters. 

\begin{figure*}
\includegraphics[width=0.98\textwidth]{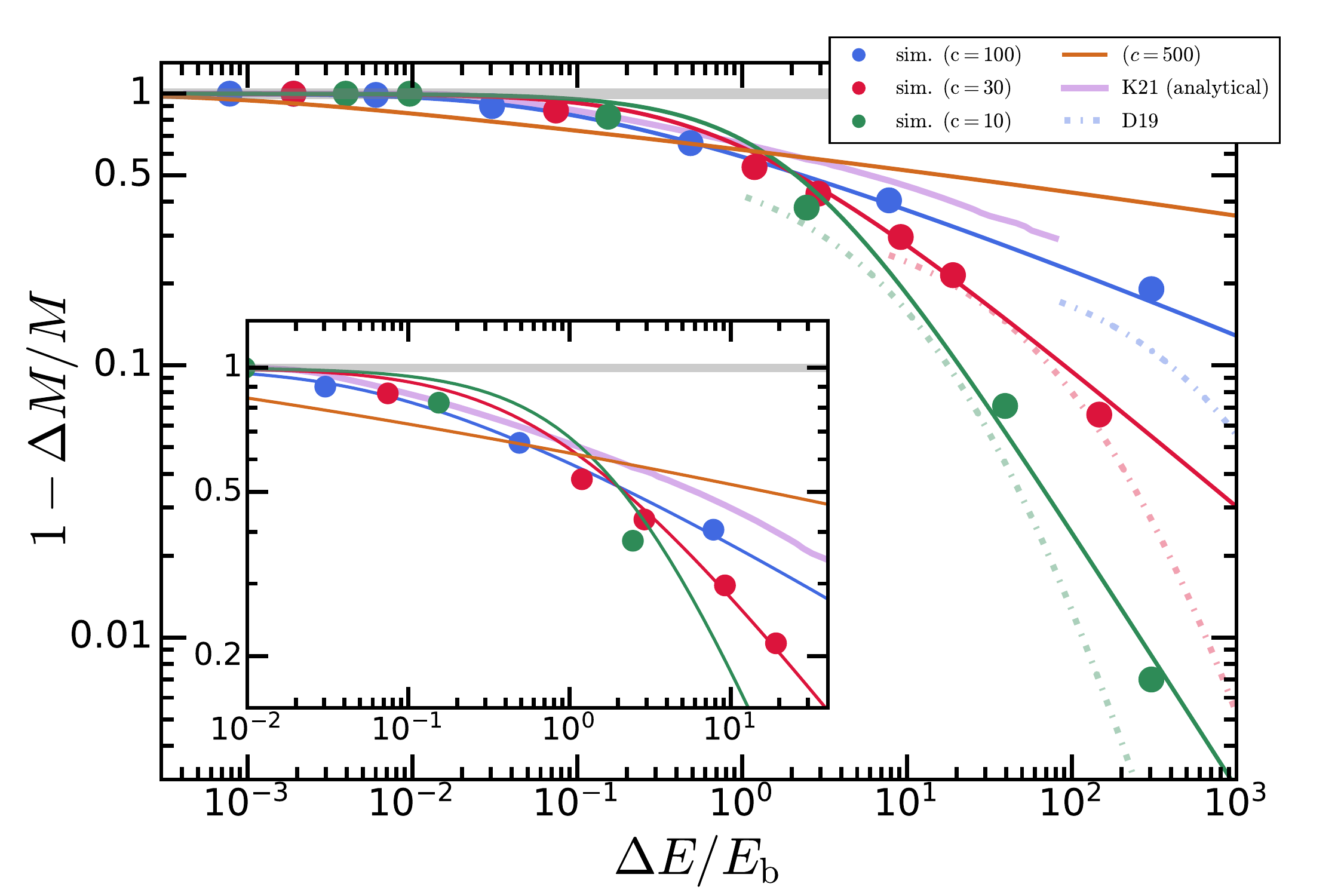}
\caption{ Fractional mass loss as a function of the normalized imparted energy from stellar encounters with minihalos. Results are shown for minihalo concentrations $c=10,30,100,500$. The simulation results and the best-fit model are shown in solid points and lines. For comparison, the purple curve is the analytic prediction from \citet{kavanagh2020stellar} for $c=100$ minihalos. The dot-dashed lines show the semi-analytical model developed in \citet{Delos:2019tsl}. The zoom-in subplot in the bottom left corner shows the transitional regime around $\Delta E/E_{\rm b} \sim 1$. The asymptotic behavior of the response curve at large $\Delta E/E_{\rm b}$ has a significant concentration dependence. The response curves calibrated from our simulations are in reasonably good agreement with results in \citet{kavanagh2020stellar} and \citet{Delos:2019tsl} in the shared dynamical ranges.}
\label{fig:deltamM_dE}
\end{figure*}

In these simulations, an isolated minihalo (composed of collisionless N-body dark matter particles) is initialized at $z=0$~\footnote{The simulations are not cosmological, but the redshift value is required for initializing the NFW halo. The response of minihalos to encounters we test in this section is not sensitive to the density normalization or the redshift we set up the minihalo.} with a star (represented by a single point-mass particle) having mass $m_{\ast} =1\,{\rm M}_{\odot}$ at a large distance ($10\,{\rm pc}$) moving towards the minihalo with a relative velocity of $200\,{\rm km\,s^{-1}}$. Since the code is Lagrangian, it makes no difference whether the star or the minihalo is moving and in which frame we solve the dynamics equations. In Figure~\ref{fig:halo_relaxation}, the evolution of a minihalo during disruption is visualized. The encounter with a star will impart a certain amount of energy into the minihalo, disrupting the halo outskirts at roughly a minihalo dynamical time. Our default simulations resolve the minihalo with $10^6$ dark matter particles initialized in an equilibrium NFW halo following the method in \citet{Springel2005}. We generate the initial condition using the package {\sc pyICs} \footnote{\href{http://jakobherpich.github.io/pyICs/}{{\sc pyICs} documentation}}\citep{2017MNRAS.470.4941H}. To systematically study stellar disruption effects, we vary the impact parameter of the star, the minihalo mass, and the minihalo concentration (or equivalently scale radius). The gravitational softening length of the dark matter particles is taken to be $10^{-9}\,{\rm kpc}$, which is small enough to resolve the dense core of a minihalo with mass $10^{-10}\,{\rm M}_{\odot}$, concentration $c=100$, and scale radius $r_{\rm s}=9.6\times 10^{-8}\,{\rm kpc}$. The time-stepping in the simulation requires more careful consideration since most disruption occurs when the distance between the star and the minihalo is around its minimum. We choose the maximum size of the timestep to be $10^{-8}\,{\rm Gyr}$ even when the star is still at a large distance ($10\,{\rm pc}$) from the minihalo and study the subsequent disruption as the star moves closer towards the minihalo. It is worth noting that this timestep is about an order of magnitude smaller than the crossing time of the star, $\sim R_{\rm mh}/v_{\ast}$ so that the trajectory of the star around the minihalo can be well resolved. After the close encounter, we can relax the upper limit on the timestep to integrate the relaxation of the minihalo after the stellar disruption to arbitrarily long times (at a fairly low computational cost). A detailed discussion of numerical convergence is given in Appendix~\ref{app:conv}. Experimenting with the maximum timestep and/or the gravitational softening length shows that smaller values produce essentially identical results (with larger computational costs). However, order-of-magnitude larger values of softening length can risk allowing the simulation to ``over smooth'' gravity or take excessively large timesteps for some particles, which ``overshoot'' the very brief duration of the encounter (for detailed numerical tests, see \citealt{Hopkins:2017ycn} and \citealt{Grudi__2020}).

\subsubsection{Disruption under different impact parameters and calibration of the response function}
\label{subsec:mloss_vs_e}

We first run three sets of simulations for encounters with fixed minihalo mass and concentration but different impact parameters. In all these simulations, we set a fixed minihalo mass $M_{\rm mh}=10^{-10}\,{\rm M}_{\odot}$, star mass $m_{\ast}=1\,{\rm M}_{\odot}$ and initial star-minihalo relative velocity $v_{\ast}=200\,{\rm km/s}$. The halo concentration has been fixed to be $c=10,30,100$ for each set of simulations, respectively. The goal is to characterize the relation between the mass loss of the minihalo after a stellar encounter and the normalized energy input. The imparted energy can be related to the impact parameter with Equation~\ref{eq:single_encounter}. After the minihalo becomes fully relaxed after the stellar encounter (at $t= 1\,{\rm Gyr} > t_{\rm dyn}(z=0)$), dark matter particles, with kinetic energy (in the center-of-momentum frame) larger than the absolute value of their gravitational potential energy, are identified as unbound and disrupted. The remaining mass of the minihalo is measured as the fraction of bound dark matter particles after minihalo relaxation, and we have verified that the remaining minihalo mass has converged at the time of measurement.

In Figure~\ref{fig:deltamM_dE}, we show the minihalo mass loss as a function of the normalized imparted energy for different choices of halo concentrations. The imparted energy is calculated using Equation~\ref{eq:single_encounter}. The mass loss is negligible when $\Delta E/E_{\rm b} \ll 1$, and quickly increases in a power-law fashion with respect to $\Delta E/E_{\rm b}$. In this regime, minihalos with low concentrations are more vulnerable to stellar encounters with steeper power-law slopes. In the following, the mass loss as a function of the imparted energy and halo concentration will be referred to as the response curve, $\mathcal{F}(\Delta E/E_{\rm b}, c)$. We propose the following functional form to fit the response curve 
\begin{equation}
    \label{eq:responce}
    \dfrac{M_{\rm mh} -\Delta M_{\rm mh}}{M_{\rm mh}} \equiv \mathcal{F}(\Delta E/E_{\rm b}, c) = \dfrac{2}{1 + \left(1 + \dfrac{\Delta E/E_{\rm b}}{p(c)}\right)^{k(c)} },
\end{equation}
where $\Delta E/E_{\rm b}$ should be evaluated using Equation~\ref{eq:single_encounter} (the free order-unity parameter $f_{\rm b}$ is yet to be determined, but the calibration here is done in the $b > b_{\rm s}$ regime, so $f_{\rm b}$ does not have any real impact). After exploring the response curve of each choice of minihalo concentration, we propose the following fitting formula for the parameters $p(c)$ and $k(c)$
\begin{align}
    \log{p}(c) & = a_1\, (\log{c} - \eta) + a_2\, (\log{c} - \eta)^{2} + a_3\, (\log{c} - \eta)^{3} \nonumber \\
    \log{k}(c) & = b_0 + b_1\, (\log{c} - 2)
\end{align}
Then we perform the least-square fits jointly on the results of all sets of simulations. The best-fit parameters are $\eta = 0.987, a_1 = -0.8 ({\rm fixed}), a_2 = -0.586, a_3 = -0.034, b_0 = -0.583 , b_1 = -0.559$. The best-fit model is also shown in Figure~\ref{fig:deltamM_dE} and is in good agreement with the simulation results of minihalos with various concentrations. This best-fit model of the response curve will be the foundation we use to understand the disruption of minihalos with different masses or concentrations. For comparison, we also show the analytical model in \citet{kavanagh2020stellar} (where $c=100$ is fixed) and the semi-analytical model developed in \citet{Delos:2019tsl} (which is calibrated for large $\Delta E/E_{\rm b}$ values). Our model agrees reasonably well with their results for shared dynamical ranges and minihalo parameters.

\subsubsection{Disruption under different halo concentrations}

\begin{figure}
\includegraphics[width=0.49\textwidth]{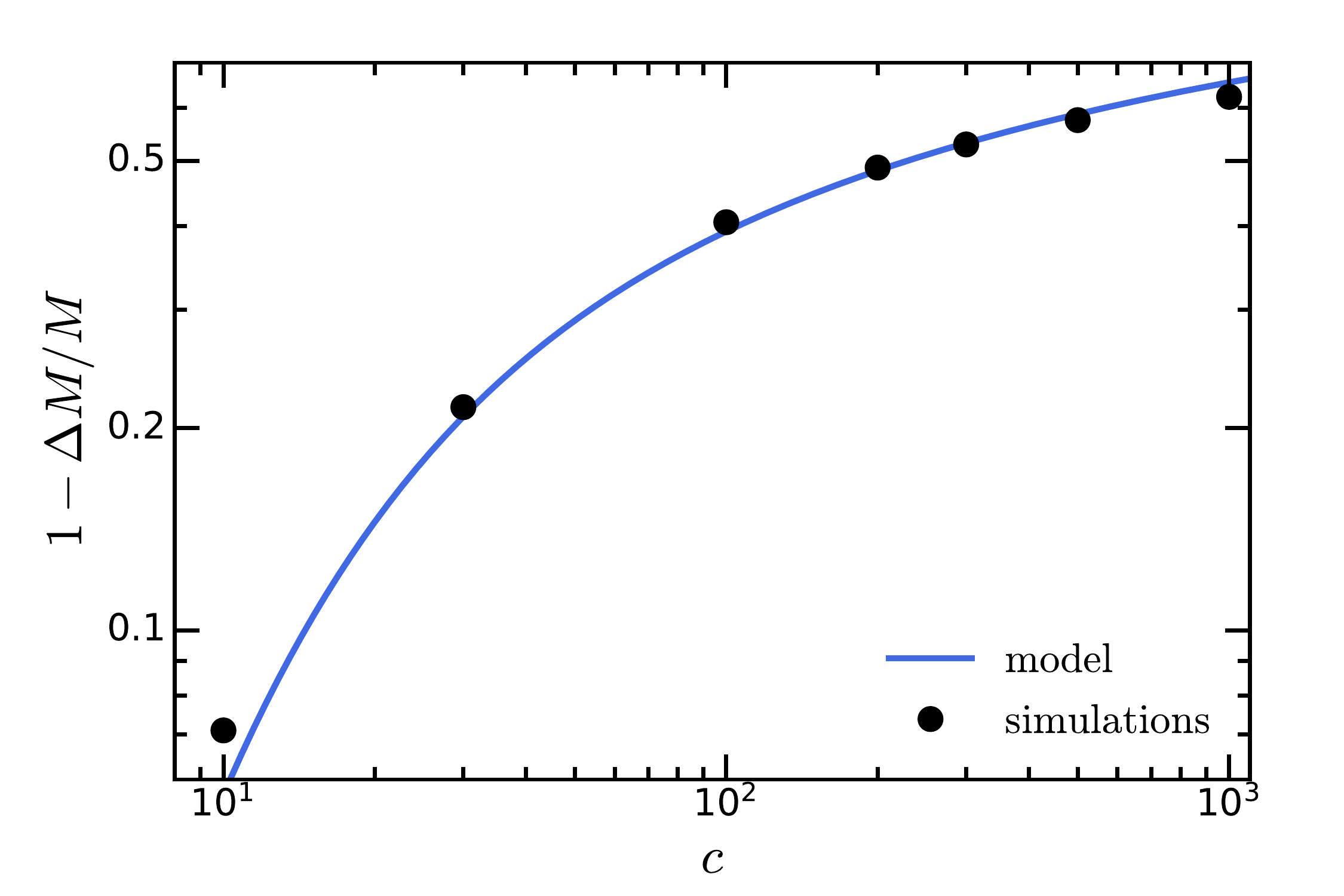}
\caption{Fractional mass loss as a function of concentration with impact parameter $b=0.05\,{\rm pc}$. The blue solid curve is the semi-analytic model prediction while the data points are obtained from simulations. 
}
\label{fig:deltam_c}
\end{figure}

Given the calibrated response function, we next run an additional set of simulations with a fixed impact parameter $b=0.05\,\rm pc$ but for minihalos with different concentrations. The goal is to further validate this semi-analytic model (the analytic calculation of imparted energy plus simulation calibrated response function) for minihalos with various compactness. The minihalo mass is still fixed to $M_{\rm mh} = 10^{-10}\,{\rm M}_{\odot}$ and the properties of the encountering star are the same as in Section~\ref{subsec:mloss_vs_e}. 

The halo concentration will affect the mass loss from stellar encounters in two ways. First, the structural parameters $\alpha$, $\beta$, and $\gamma$ all have an explicit dependence on minihalo concentration, which will propagate to the calculation of energy imparted ({\em i.e.} $\Delta E/E_{\rm b} \propto \alpha^2(c)/\gamma(c) \sim \ln{(c)}/c$ when $b \gg b_{\rm s}$ and $c\gg 1$). Secondly, the response function also has a strong dependence on concentration (see Equation~\ref{eq:responce}), especially when $\Delta E/E_{\rm b} \gg 1$. Less concentrated minihalos will become increasingly vulnerable to disruptive stellar encounters. Therefore, it is non-trivial for the semi-analytic model to correctly capture the concentration dependence of minihalo mass loss. In Figure~\ref{fig:deltam_c}, we show the mass loss of minihalo versus minihalo concentration. The best-fit semi-analytic model agrees with the simulation results over a wide range of concentrations, even extrapolating into regimes not covered in our original calibration step above. 

\subsubsection{Disruption under different minihalo masses}
\label{subsec:mloss_vs_m}

\begin{figure}
\centering
\includegraphics[width=0.49\textwidth]{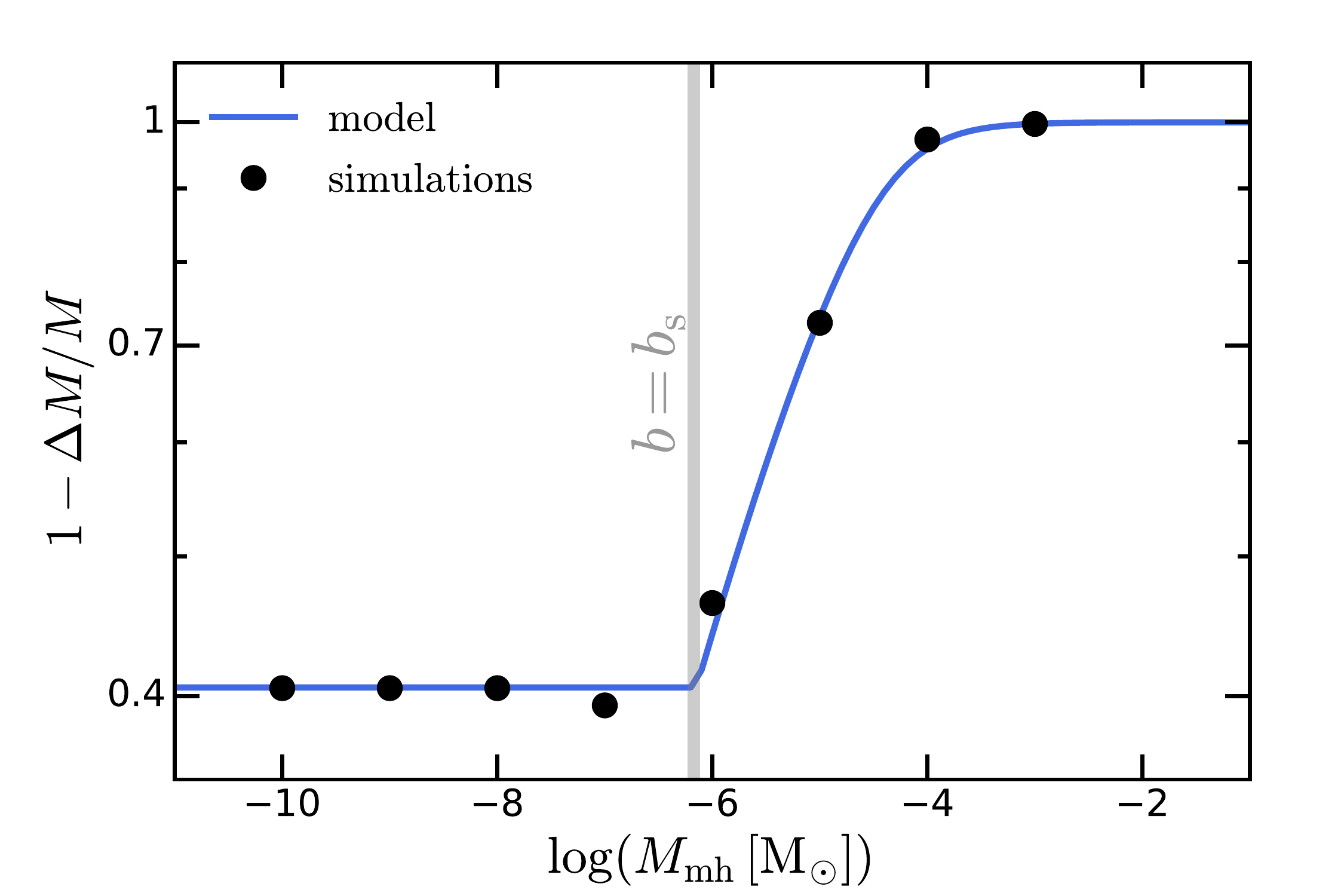}
\caption{Fractional mass loss as a function of minihalo mass with concentration $c=100$ and impact parameter $b = 0.05\, \rm pc$. The black solid points are simulation results. The blue curve is the semi-analytic prediction using the best-fit response curve. The vertical line indicates where the transition impact parameter $b_{\rm s}$ is reached. The semi-analytical model accurately predicts the location and the shape of this transition.
}
\label{fig:deltam_M}
\end{figure}

One remaining ingredient of the semi-analytic model that needs to be calibrated is the disruption behavior in the $b\lesssim b_{\rm s}$ regime. According to Equation~\ref{eq:single_encounter}, the imparted energy will stop rising as $b$ decreases when a characteristic scale $b_{\rm s}$ is reached. The free, order-unity correction factor should be calibrated by simulations. To fulfill that, we run an additional set of simulations for minihalos of different masses ranging from $10^{-10}$ to $10^{-3}\,{\rm M}_{\odot}$ but fixing the impact parameter $b=0.05\rm pc$ and minihalo concentration $c=100$. In low-mass minihalos, $b_{\rm s}$ is small enough that the encounter is in the $b>b_{\rm s}$ regime, where mass loss is independent of minihalo mass. As minihalo mass increases (and $b_{\rm s}$ increases), test cases with massive minihalos enter the $b<b_{\rm s}$ regime, where stellar disruptions are suppressed. In Figure~\ref{fig:deltam_M}, we show the mass loss as a function of minihalo mass. A transition of the response of the minihalo occurs, and we use the mass of this transition to calibrate the free parameter to be $f_{\rm b} = 6$. The predicted mass loss using the best-fit response curve for $c=100$ halos is shown with the solid line in the figure. In general, the semi-analytic model gives the correct location and shape of the transition at $b\sim b_{\rm s}$. The mass loss remains a constant at low minihalo masses as indicated by Equation~\ref{eq:single_encounter} in the $b\gg b_{\rm s}$ regime. A sharp transition occurs at $b\simeq \, b_{\rm s}$ such that the mass-loss rate reduces to almost zero at high minihalo masses. 

\begin{figure}
\centering
\includegraphics[width=0.49\textwidth]{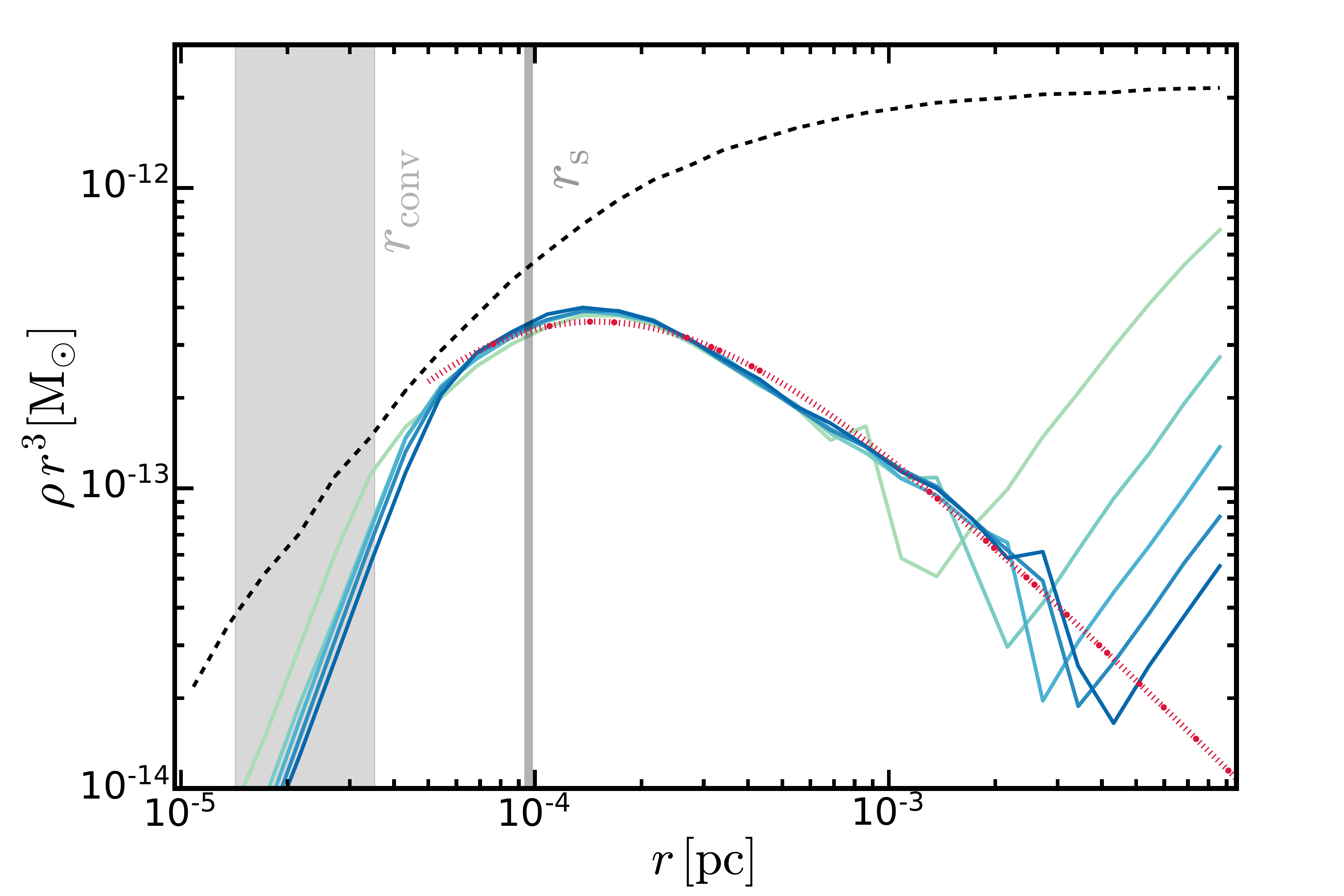}
\includegraphics[width=0.49\textwidth]{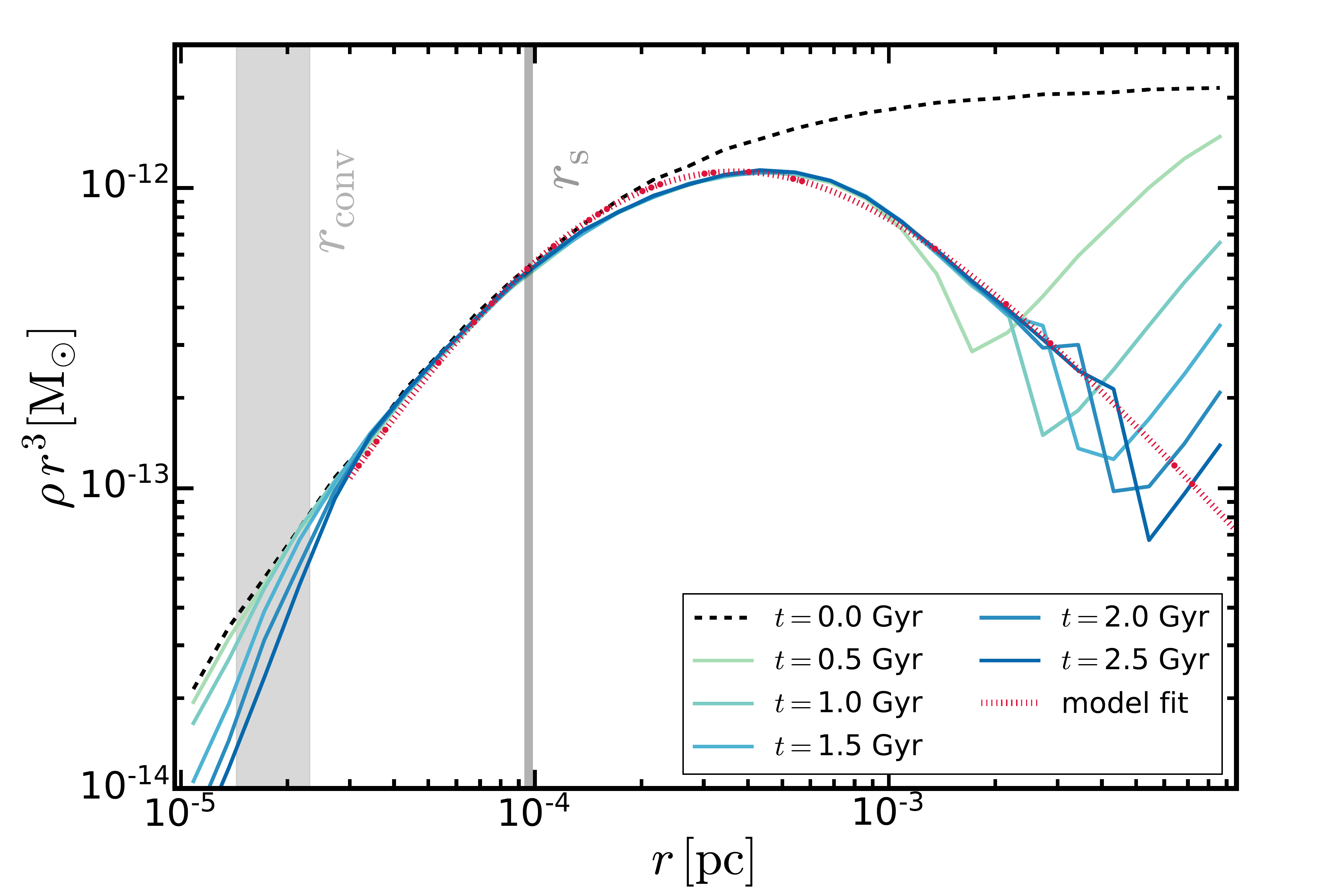}
\caption{The evolution of minihalo density profiles after a close encounter with a star. For the tests here, the minihalo mass is set as $10^{-10}\,{\rm M}_{\odot}$ initialized with the NFW profile at redshift zero (with virial radius $R_{\rm mh} \sim 0.01\,{\rm pc}$ and dynamical time scale $t_{\rm dyn} \sim 2.2\,{\rm Gyr}$). The halo concentration is $c=100$. The impact parameter is $b=2\times 10^{-5}\,\rm kpc$ and $b=5\times 10^{-5}\,\rm kpc$ in the top and bottom panels, respectively. The vertical shaded regions indicate the range of convergence radii at different times. After the close encounter with the star, the outskirts of the minihalos are dominantly disrupted in less than a dynamical time. Eventually $81\%$ ($60\%$) of the minihalo mass is disrupted for the $b=2\times 10^{-5}\,{\rm kpc}$ ($b=5\times 10^{-5}\,{\rm kpc}$) case. The density profile of minihalo after disruption can be well-fitted by a broken power-law profile with best-fit asymptotic slopes around $-4$ analogous to that of the Hernquist profile. The ``model fit'' curve provides an asymptotic limit of the density profile after infinite time. The upturn at the outskirt of the minihalo represents a propagating shell of unbound particles. As this upturn propagates towards the outer part of the minihalo, the density profile gradually converges to the asymptotic limit.
}
\label{fig:density_profile}
\end{figure}

\subsubsection{Density profiles} \label{subsec:density_profile}

The density profile will not immediately change after the close encounter with stars, but will gradually relax to the final minihalo profile within a few minihalo dynamical times \citep[{\em e.g.}\,][]{Delos:2019tsl}. This is mainly because the close encounter with a star occurs on a timescale much shorter than the minihalo dynamical time. 

In Figure~\ref{fig:density_profile}, we show the evolution of minihalo density profiles for minihalos with mass $10^{-10}\,{\rm M}_{\odot}$ and concentration $c=100$. In the top (bottom) panel, we show the case with impact parameter $b=2\times10^{-5}\,{\rm pc}$ ($5\times10^{-5}\,{\rm pc}$). The density profile is shown as $\rho\,r^{3}$, which is proportional to $\Delta M/ \Delta \log{r}$, the contribution to total mass per unit logarithmic interval of radius. The vertical shaded region shows the convergence radius for collisionless particles based on the \citet{Power2003} criterion (this is roughly the radius interior to which the numerical two-body relaxation time drops below the Hubble time, a conservative indication of where N-body integration error could be significant). In the $b=5\times10^{-5}\,{\rm pc}$ case, the outskirts of the halo are predominantly disrupted while the core of the halo remains relatively unperturbed, leaving approximately $60\%$ of the minihalo mass disrupted. Although the central density exhibits a small decrease, its contribution to the total mass loss is negligible and the scale of this decrease is close to the convergence radius of dark matter properties, which makes it hard to distinguish the decrease from a numerical artifact. At the outskirts of the minihalo, the density profile turns up, corresponding to a shell of unbound particles (heated by the encounter) propagating outward. For the $b=2\times10^{-5}\,{\rm pc}$ case, due to higher imparted energy, the disruption is more significant with over $80\%$ of the minihalo mass disrupted. However, the behavior of the final density profile is rather similar to the previous case. 

It is worth noting the remaining minihalos never relax to a new NFW profile. Instead, the final density profile (after the shell of unbound particles escapes) can be well described by a broken power law of the form
\begin{equation}
    \rho(r) = \dfrac{\rho_{0}}{\dfrac{r}{r_0}\left(1+\dfrac{r}{r_0}\right)^{k}},
\end{equation}
with the best-fit asymptotic slope $k = 3.2$ for $b=2\times 10^{-5}\,\rm kpc$ and $k = 3.3$ for $b=5\times 10^{-5}\,\rm kpc$. This slope implies that the density profile is close to a \citet{Hernquist1990} profile ($k=3$), or in a more general sense, the $\eta$-profile family \citep{Dehnen1993, Tremaine1994} with an asymptotic slope of $-4$. Unsurprisingly, given the scale-free physics involved, this is similar to well-studied simulations of impulsive high-mass ratio galaxy-galaxy encounters \citep[{\em e.g.},][]{hernquist1988,barneshernquist92,hopkins:disk.heating,hopkins:cores,boylankolchin:mergers.fp}, particularly the structure of shell galaxies \citep{hernquist.quinn.87,hernquist.spergel.92} and the closely-related slopes of the marginally-bound layer of particles at apocenter in cosmological simulations defined as the halo ``splashback'' radius \citep[{\em e.g.},][]{Diemer2014, More2015}. Generically, this slope arises from relaxation after dynamical mass ejection events, which (by definition) excite some material to a broad distribution of energies crossing the specific binding energy $\mathcal{E}=0$, as the outer slope $k+1=4$ corresponds to the only finite-mass asymptotic power-law distribution function which is continuous through $\mathcal{E}=0$ \citep{hernquist:phasespace}. 

\begin{figure}
    \centering
    \includegraphics[width=0.49\textwidth]{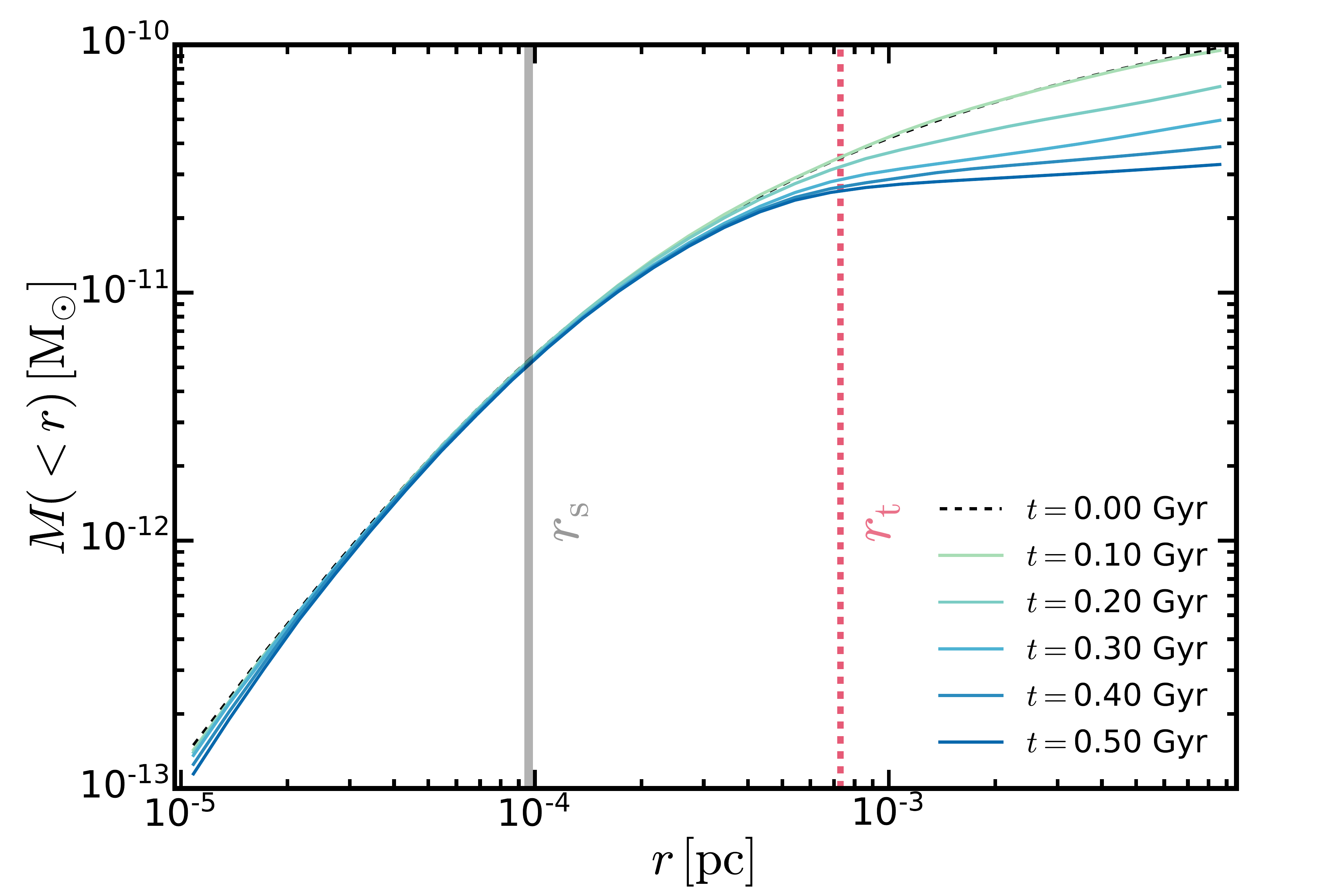}
    \includegraphics[width=0.49\textwidth]{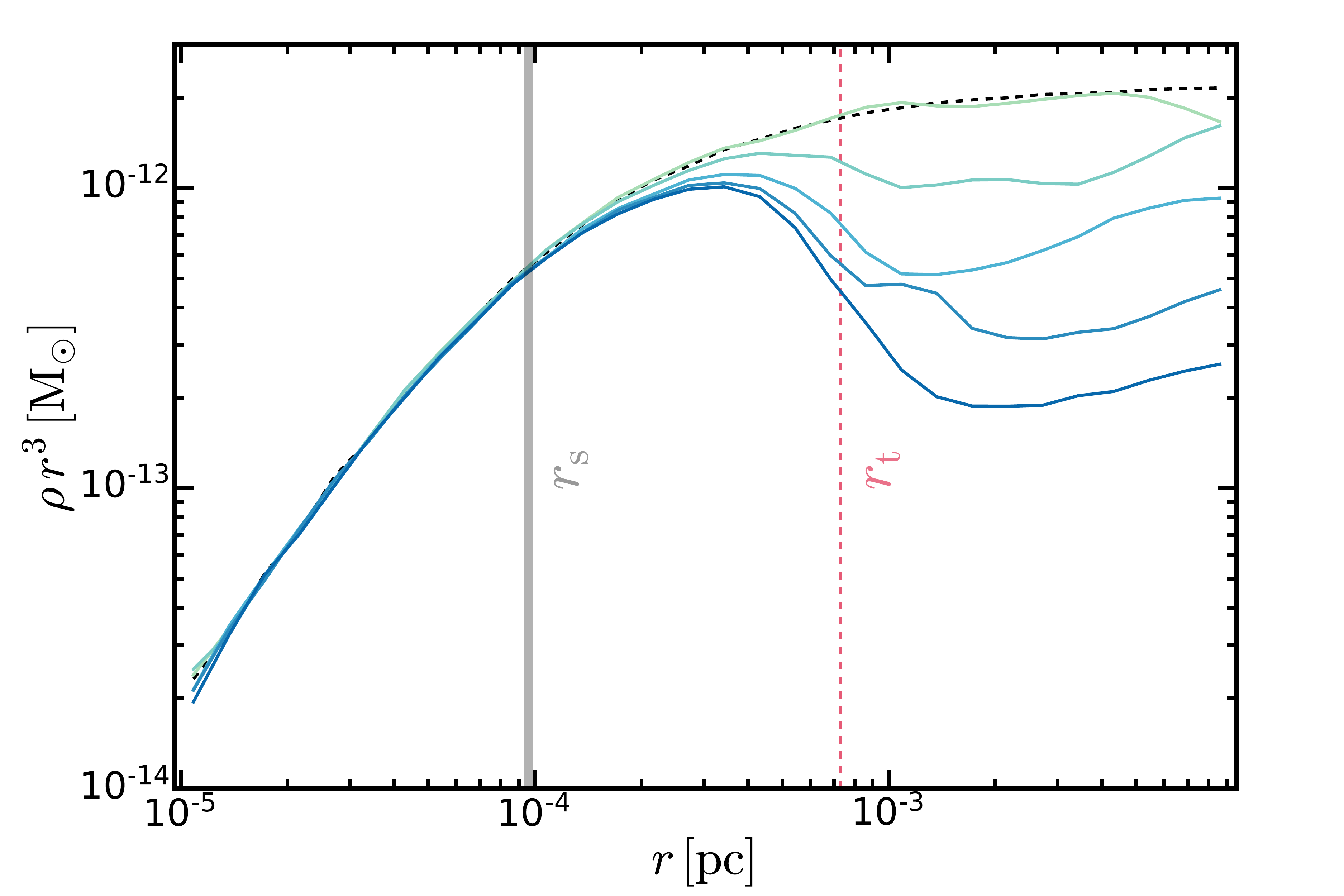}
    \caption{{\it Top}: The enclosed mass profile of a minihalo with $M=10^{-10}\,{\rm M}_{\odot}$ at $t=0-0.5\,{\rm Gyr}$. The minihalo is on a circular orbit at $R=R_{\odot} = 8\,{\rm kpc}$ in a Milky Way-mass dark matter halo. For reference, the dynamical time of the host halo within this radius is about $0.1\,{\rm Gyr}$. The gray and red vertical lines indicate the scale radius and the tidal radius of the minihalo. The mass outside the tidal radius is stripped away at roughly the dynamical time scale of the host halo while the mass inside is marginally perturbed. {\it Bottom}: The density profile of the minihalo at the same time as the top panel. The plot illustrates the behavior of the matter distribution at the outskirts of the minihalo during tidal stripping.}
    \label{fig:tidal}
\end{figure}

\begin{figure}
    \centering
    \includegraphics[width=0.49\textwidth]{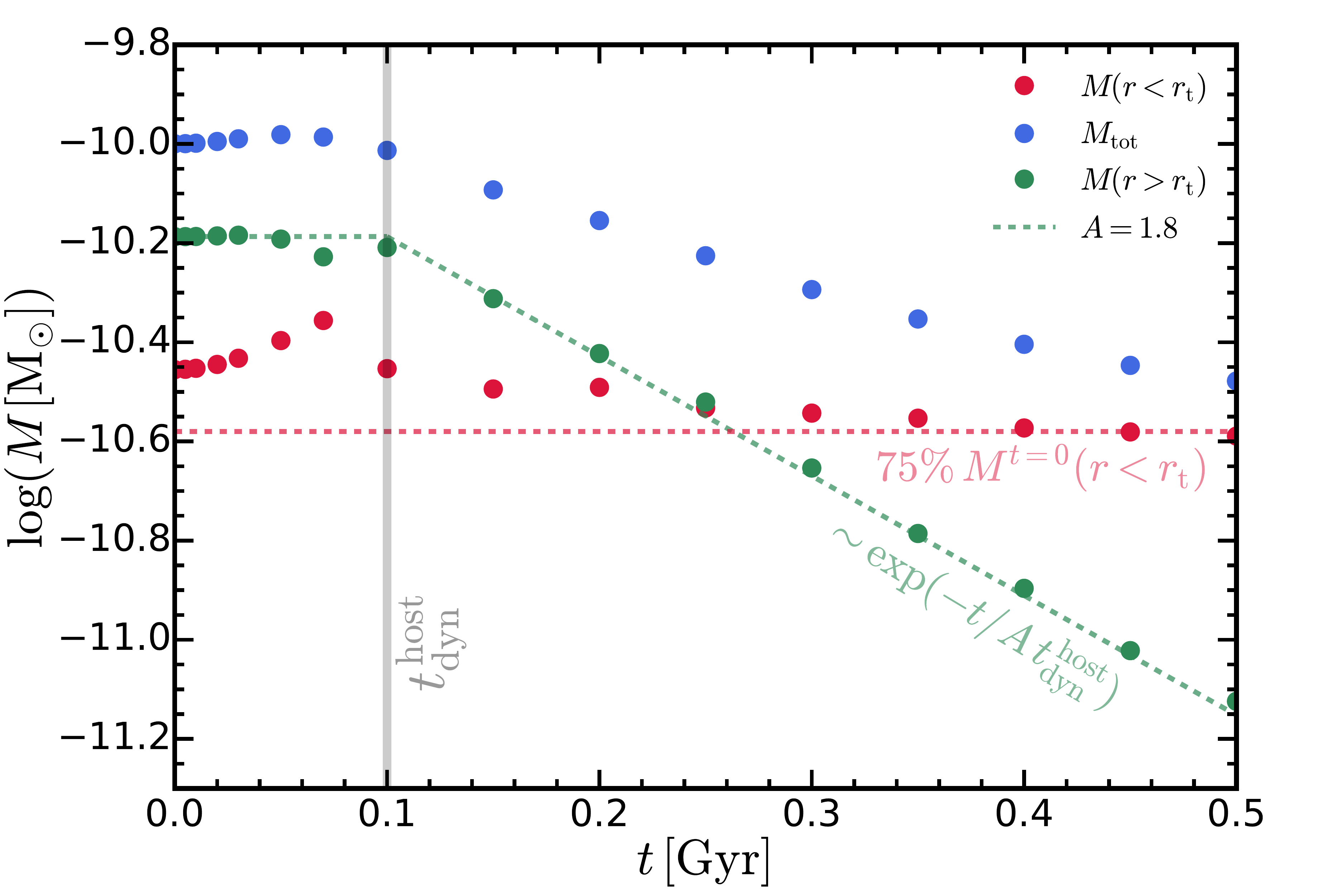}
    \caption{Mass evolution history of the minihalo in tidal disruption (the same one as in Figure~\ref{fig:tidal}). The total mass here is the sum of dark matter mass within the initial virial radius of the minihalo. In addition, we show the mass inside and outside the tidal radius. After approximately a dynamical time of the host halo, the mass outside the tidal radius starts an exponential decay as described by Equation \ref{eq:mass_tidal_loss} with the fudge factor $A\simeq 1.8$. The mass inside the tidal radius is marginally affected.}
    \label{fig:tidal-2}
\end{figure}

\subsection{Numerical test of tidal stripping}
\label{sec:tidal-sim}

To validate the semi-analytic description of tidal disruptions of minihalos in Section~\ref{sec:tidal-disruption}, we perform an idealized simulation of a minihalo traveling through the analytic gravitational potential of a Milky Way-mass host halo ($M_{\rm vir} = 10^{12}\,{\rm M}_{\odot}$). The host halo profile is modeled as an NFW profile with concentration $c=12$~\citep[{\em e.g.},][]{Klypin2002,McMillan2011,Deason2012,Bland-Hawthorn2016}. The minihalo is assumed on a circular orbit with $R = R_{\odot} \simeq 8\,{\rm kpc}$. In Figure~\ref{fig:tidal}, we show the evolution of the enclosed mass profile and density profile of the minihalo. The tidal radius calculated using Equation~\ref{eq:tidal_radius} is shown as the red vertical dashed line. After evolving for about $0.5\,{\rm Gyr}$ (for reference, the dynamical time scale of the host halo at $R=8\,{\rm kpc}$ is $t^{\rm host}_{\rm dyn} \sim 0.1\,{\rm Gyr}$, see Equation~\ref{eq:tidal_time}), the enclosed mass profile starts flattening and eventually plateaus outside the analytically evaluated tidal radius. In Figure~\ref{fig:tidal-2}, we show the mass evolutionary history of this minihalo, and specifically the mass inside and outside the tidal radius. The mass within the tidal radius is almost immune to tidal stripping, with $75\%$ of the mass remaining after $0.5\,{\rm Gyr}$. On the other hand, the mass outside the tidal radius exhibits an exponential decay after about $1\,t^{\rm host}_{\rm dyn}$. The mass loss of the minihalo can be well represented by Equation~\ref{eq:mass_tidal_loss}, implying $M(r>r_{\rm t})\sim e^{-t/A\,t^{\rm host}_{\rm dyn}}$ with a fudge factor $A$ close to $1.8$ for the minihalo tested, in broad agreement with previous numerical studies of more massive CDM subhalos~\citep[{\em e.g.},][]{Zentner2003, Pullen2014, vandenBosch2018}

The simulation presented in this section demonstrates that the semi-analytic treatment of tidal disruption works reasonably well in predicting the mass loss and the tidal stripping time scale. Given the order unity fudge factor $A$ found here, the tidal stripping time scale $t_{\rm ts}$ in Equation~\ref{eq:tidal_time} is at least an order of magnitude smaller than the Hubble time. Therefore, tidal stripping can be treated as an ``instantaneous'' process in modeling the cosmological evolution of minihalos.

\section{Modeling multiple encounters}
\label{sec:multi}
In the preceding section, we introduced a model that describes the mass loss of minihalo resulting from individual stellar interactions. In this section, we will present a model that deals with the cumulative effects of multiple encounters.
\subsection{Multiple encounters within a dynamical time}

To begin, we examine the amalgamation of encounters occurring within the dynamical time of the minihalo. A relevant illustration of such an event would be a minihalo traversing the Milky Way disk. The typical crossing time (given by Equation~\ref{eq:disk_time}) is significantly shorter than the dynamical time of the minihalo (given by Equation~\ref{eq:tdyn}). Minihalo does not have enough time to fully relax and realize the mass loss before the subsequent encounter occurs.

The change in the velocity of a particle within a minihalo of size $R$ at position $r$ (relative to the center of the minihalo) due to an encounter with a perturber of mass $m_{\ast}$ moving with relative velocity $\textbf{v}_{\ast}$ at an impact parameter $\textbf{b}$ (when $b\gg R$) can be expressed as \citep{Spitzer1958,Green2007}
\begin{equation}
    \delta \textbf{v} \simeq \dfrac{2\,G\,m_{\ast}}{v_{\ast}\,b^{2}} \left[ (\textbf{r} \cdot \textbf{e}_{\rm b})\,\textbf{e}_{\rm b} - (\textbf{r} \cdot \textbf{e}_{\rm z})\,\textbf{e}_{\rm z} \right]
    \label{eq:deltav}
\end{equation}
where $\textbf{e}_{\rm b}$ is the unit vector in the direction of $\textbf{b}$, $\textbf{e}_{\rm z}$ is the unit vector perpendicular to $\textbf{b}$ and $\textbf{u}$ ($\textbf{e}_{\rm z} \equiv \textbf{e}_{\rm b} \times  \textbf{e}_{\rm u}$). For a particle in the minihalo moving with velocity $v$, the change in energy is given by
\begin{equation}
    \delta E = \textbf{v} \cdot \delta \textbf{v} + (\delta v)^{2}/2
    \label{eq:deltaE-deltav}
\end{equation}
The ensemble average of $\delta E$ over all particles in the minihalo can be obtained in two steps. First, at a fixed position $\textbf{r}$, an average over velocities $\textbf{v}$ is taken, where the first term in Equation~\ref{eq:deltaE-deltav} vanishes if the particle velocities are assumed to be isotropic. Second, an average over $\textbf{r}$ is taken, where the assumption of spherical symmetry leads to the final form of $\Delta E$ in Equation~\ref{eq:deltaE}.

When multiple encounters occur in a short enough time compared to the dynamical time of the minihalo, particles in the minihalo can be considered frozen. The aggregation of encounters is simply $\delta \textbf{v} = \Sigma \delta \textbf{v}_{\rm i}$. According to Equation~\ref{eq:deltaE-deltav}, the aggregated $\delta E$ will have two types of terms: $\delta \textbf{v}_{\rm i} \cdot \delta \textbf{v}_{\rm i}$ and $\delta \textbf{v}_{\rm i} \cdot \delta \textbf{v}_{\rm j}$, where the former is simply energy injection from each individual encounters while the latter involves two different encounters
\begin{align}
    \delta \textbf{v}_{\rm i}&  \cdot \delta \textbf{v}_{\rm j} \propto 
    (\textbf{r} \cdot \textbf{e}^{\rm i}_{\rm b})(\textbf{r} \cdot \textbf{e}^{\rm j}_{\rm b})(\textbf{e}^{\rm i}_{\rm b} \cdot \textbf{e}^{\rm j}_{\rm b}) + (\textbf{r} \cdot \textbf{e}^{\rm i}_{\rm z})(\textbf{r} \cdot \textbf{e}^{\rm j}_{\rm z})(\textbf{e}^{\rm i}_{\rm z} \cdot \textbf{e}^{\rm j}_{\rm z}) \nonumber \\ & - (\textbf{r} \cdot \textbf{e}^{\rm i}_{\rm b})(\textbf{r} \cdot \textbf{e}^{\rm j}_{\rm z})(\textbf{e}^{\rm i}_{\rm b} \cdot \textbf{e}^{\rm j}_{\rm z}) - (\textbf{r} \cdot \textbf{e}^{\rm i}_{\rm z})(\textbf{r} \cdot \textbf{e}^{\rm j}_{\rm b})(\textbf{e}^{\rm i}_{\rm z} \cdot \textbf{e}^{\rm j}_{\rm b})
    \label{eq:dvidvj_expand}
\end{align}
When the total number of encounters aggregated is large, we can take the ensemble average over encounter parameters of the i-th and j-th events. The result depends on the local velocity distribution of stars with respect to the minihalo. Two typical scenarios are considered in the following. (1) {\it An isotropic velocity field:} Both $\textbf{e}_{\rm b}$ and $\textbf{e}_{\rm z}$ are isotropic. Assuming the i-th and j-th encounters are independent, the ensemble averages (over the orientation of incident stars) of the four terms in Equation~\ref{eq:dvidvj_expand} are the same and exactly cancel each other. (2) {\it A stream of stars:} All the stars are moving in the same direction. However, the impact parameter $\textbf{b}$ is still isotropic in the cross-section of the stream. Therefore, $\textbf{e}_{\rm b}$ and $\textbf{e}_{\rm z}$ are isotropic in this two-dimensional plane. The ensemble averages of the four terms in Equation~\ref{eq:dvidvj_expand} are the same and cancel each other. For a realistic application, if we consider a minihalo moving through a cloud of stars with an isotropic velocity distribution, the relative velocity field of stars with respect to the minihalo is a linear combination of the two scenarios above. Due to the linear nature of the ensemble average, we can conclude that the $\delta \textbf{v}_{\rm i} \cdot \delta \textbf{v}_{\rm j}$ terms can be ignored. The $\delta \textbf{v}_{\rm i} \cdot \delta \textbf{v}_{\rm i}$ terms are left and $\delta E_{\rm tot}$ is simply the summation of $\delta E$ from individual encounters.

\subsection{Multiple encounters after full relaxation}

A more complex scenario arises when the minihalo has enough time to fully relax before the next encounter, such as multiple disk crossings with an orbital period comparable to or longer than the relaxation time. In such cases, the disruption model calibrated on single encounters needs to be revised because the internal structure of the minihalo can differ significantly from the initial condition. In Section~\ref{sec:stellar_sim}, we demonstrated that the density profile post-disruption is no longer NFW-like and features steeper slopes at the outskirts.

One approach often used in the literature is to integrate the effective ``optical'' depth. This relies on the approximation that minihalos are completely destroyed when the energy imparted exceeds a certain threshold. The number fraction of minihalos ``survived'' scales exponentially with the ``optical'' depth of encounters (see the definition in Equation~\ref{eq:back-of-the-envelope-1}). This approach has been used in {\em e.g.} \citet{Schneider2010} for CDM minihalos. We will discuss it more and use it as a back-of-the-envelope estimate in the following section. 

The second approach is to ``resolve'' individual encounters and process them sequentially, relying on a model to describe the internal states of minihalos after the encounter(s). \citet{Delos:2019tsl} proposed a self-similar profile for post-encounter minihalos and calibrated disruption models accordingly. This method works reasonably well in the large ${\rm d}E/E_{\rm b}$ regime, as shown in the comparison in Figure~\ref{fig:deltamM_dE}. However, it is unknown whether it is accurate in the limit we are interested in. The typical ${\rm d}E/E_{\rm b}$ for a single stellar disk crossing is of order $10^{-2}$ (see Equation~\ref{eq:energy_injection_onep}), which is outside the calibration range of this model~\footnote{It should be noted that the definition of ${\rm d}E/E_{\rm b}$ in \citet{Delos:2019tsl} differs from the one used in this paper, as they normalize ${\rm d}E$ by the binding energy of the central dark matter core.}.

\begin{figure}
    \centering
    \includegraphics[width=1\linewidth]{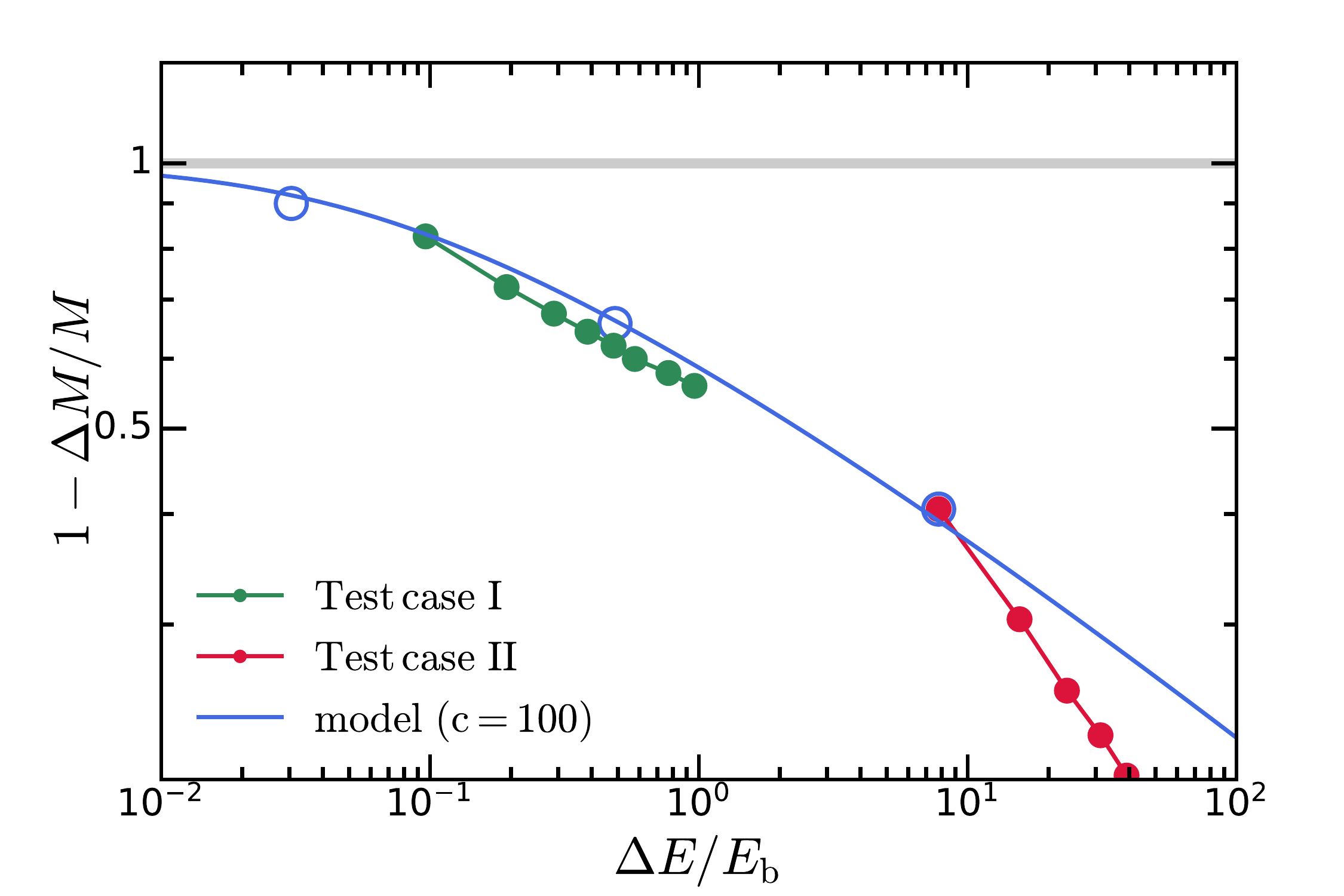}
    \caption{Mass loss of the minihalo from single/multiple encounters. The blue line and open circles show the response curve of $c=100$ minihalos we calibrated in Section~\ref{sec:stellar_sim}. The green circles show the minihalo mass loss as a function of accumulated imparted energy from successive encounters (with $\Delta E/E_{\rm b}\sim 10^{-1}$). The red circles show the same test but with typical $\Delta E/E_{\rm b}\sim 10$. Our treatment of accumulating imparted energy from independent encounters works reasonably well when individual $\Delta E/E_{\rm b} \ll 1$, but will underpredict the mass loss when $\Delta E/E_{\rm b} \gg 1$.}
    \label{fig:response-multi}
\end{figure}

Here we propose an alternative and intuitive way to treat multiple encounters which directly builds upon the knowledge about single encounters. This method requires no assumptions on parameterizing the density profiles of post-disruption minihalos and works reasonably well in a quasi-static limit. Similar to what we derived for multiple encounters within a dynamical time, we accumulate the injected energy $\Delta E$ for successive encounters and use the response function $\mathcal{F}(\Delta E_{\rm tot}/E_{\rm b})$ to obtain the total mass disruption fraction. The validity of this treatment is supported by the following observations.
\begin{itemize}
    \item The state of a minihalo after disruption could be well described by a single parameter. This parameter can be the total energy or the gravitational binding energy or some characteristic density, etc. An example of such a description is given by \citet{Delos:2019tsl}. They found a self-similar density profile for minihalos after disruption with two parameters $\rho_{\rm s}$ and $r_{\rm s}$. A scaling relation between the two parameters was also discovered. Therefore, the density profile of the minihalo is effectively determined by one single parameter. This parameter can be the total energy of the minihalo because the total energy is uniquely related to the density profile assuming virial equilibrium.
    \item For the typical ${\rm d}E/E_{\rm b}\sim 10^{-2}$ of a single stellar disk crossing, the disruption happens in a quasi-static fashion. In this limit, each encounter only causes a marginal disruption to the least bound particles. The change in the total energy of the minihalo can be expressed as $E^{\rm final}_{\rm bound} - E^{\rm init}_{\rm bound} = (1-f_{\rm ej})\, \Delta E - E^{\rm init}_{\rm unbound}$, where $f_{\rm ej}$ is the fraction of imparted energy ends up in unbound particles. In the quasi-static limit, the shell of unbound particles that eventually escape the minihalo will have negligible energy at infinity, $f_{\rm ej}\, \Delta E + E^{\rm init}_{\rm unbound} \rightarrow 0$. This is supported by both our simulations and the analytical calculations shown in \citet{kavanagh2020stellar}. For example, when $\Delta E/E_{\rm b} \sim 10^{-2}$, the corresponding $f_{\rm ej}$ is $\sim 10^{-2}$ and $E^{\rm init}_{\rm unbound}/E_{\rm i}$ is $\sim 10^{-4}$. Neglecting second order terms, we have $E^{\rm final}_{\rm bound} - E^{\rm init}_{\rm bound} \simeq \Delta E$. 
\end{itemize}
Combining the two points listed above, the final state of the minihalo is determined by the final energy of the minihalo, which is only related to the total imparted energy $\Delta E$ and has no dependence on the exact history of encounters.

Using this approach, the additional disruption from successive encounters is sensitive to the slope of the response function at large $\Delta E_{\rm tot}/E_{\rm b}$. A unique feature of the calibrated response curve is that subsequent encounters with the same energy injection will rapidly become less important ($\Delta M_{\rm mh}/M_{\rm mh}$ from each encounter rapidly decreases as the total number of encounters increases), especially for minihalos with large concentrations. The physical interpretation is that the particles which would be unbound by such an energetic encounter have already been unbound in previous encounters (the loosely bound particles at the outskirt have already been stripped and the remaining particles are more tightly bound and less vulnerable to new encounters).

To validate this treatment, We perform a series of numerical tests. We take a minihalo with $M_{\rm mh} = 10^{-10}\,{\rm M}_{\odot}$, $c=100$ and set the mass and velocity of the star as $1\,{\rm M}_{\odot}$ and $200\,{\rm km/s}$. In the first test, we pick an impact parameter such that the injected energy from a single event is $\Delta E/E_{\rm b}\sim 0.1$. We simulate eight repetitive encounters with $\Delta E/E_{\rm b} = 0.1, 0.1, 0.1, 0.1, 0.1, 0.1, 0.2, 0.2$. The minihalo is allowed to fully relax between these encounters. In Figure~\ref{fig:response-multi}, we compare the mass loss curve of successive encounters with the response curve we calibrated for single encounters. The error is at $\lesssim 0.05\,{\rm dex}$ and shows no sign of error accumulation in multiple encounters. In our real application, the maximum $\Delta E/E_{\rm b}$ from individual disk crossing will be at the level $10^{-2}$ (an estimate is given in Equation~\ref{eq:energy_injection_onep} and we have explicitly checked in our numerical sampling of minihalos). Therefore, the approach we propose can reasonably accurately approximate the disruption fraction from multiple disk crossings. In the second test, we pick an impact parameter such that the injected energy from a single event is $\Delta E/E_{\rm b}\sim 10$ and repeat the encounter five times. This is a limit where all our assumptions above should fail. In this case, the response curve calibrated underpredicts the amount of mass loss, and the prediction error clearly inflates as the number of encounters increases.

\section{Disruptions in the realistic Milky Way environment}
\label{sec:MW_disruption}

In the sections above, we described the physical process and consequence of stellar encounters and the effects of tidal fields. In the following, we will apply the model to minihalos in the realistic Milky Way environment.

\subsection{Multiple encounters in the Milky Way disk}

Disk stars are the dominant component of the stellar populations in the Milky Way, and thus will be the main contributor to the disruption of dark matter substructures. When a minihalo passes through the stellar disk, it will encounter a slab of stars within a short timescale
\begin{equation}\label{eq:disk_time}
    t_{\rm disk}^{\rm x} \simeq 2\, {\rm Myr}\, \left( \dfrac{H_{\rm d}}{400\,{\rm pc}}\right) \left( \dfrac{v^{\bot}_{\rm mh}}{200\,{\rm km/s}} \right)^{-1},
\end{equation}
where $H_{\rm d}$ is the scale height of the Milky Way thin disk and $v^{\bot}_{\rm mh}$ is the relative velocity of the minihalo perpendicular to the disk plane. As discussed in Section~\ref{sec:multi}, since this disk passage time is much shorter than the dynamical time of dark matter minihalos, the internal structure of the minihalos will not have enough time to relax between successive stellar encounters during a single passage. Therefore, a series of encounters with disk stars can be effectively considered as one encounter with the injected energy accumulated over the passage. 

\subsubsection{Back-of-the-envelope model}

First, we will consider a simplified model assuming that a minihalo will be completely destroyed after a single encounter with an impact parameter smaller than $b_{\rm min}$ given by Equation~\ref{eq:bmin}, and it will remain unperturbed if the impact parameter is larger than $b_{\rm min}$. Note that we aim for a back-of-the-envelope estimate with this simplified model \footnote{In real cases, the survival mass of the minihalo is far from negligible when $\Delta E/E_{\rm b}\sim 1$. The back-of-the-envelope model certainly breaks. This fact is shown in our calibrated response curve in Figure~\ref{fig:deltam_M} and has been highlighted in studies on CDM substructures \citep[{\em e.g.}\,][]{vandenBosch2018}.} and this model will not be used for our final analysis. Consider a minihalo with mass $M_{\rm mh}$ and an average density $\bar{\rho}_{\rm mh}$, moving through a field of stars with the differential number per unit mass $n_{m_{\ast}}  = {\rm d}N_{\ast}/({\rm d}m_{\ast} {\rm d}^3\bf{x})$ characterized by the stellar present-day mass function (PDMF), one can choose a normalization such that the total stellar mass density is $\int m_{\ast}\,n_{m_{\ast}}\,{\rm d}m_{\ast} = \rho_{\ast}$. We assume that an encounter with impact parameter $b$ has a probability $p = p(b,\,M_{\rm mh},\,c,\,m_{\ast},\,{\bf v},\,...)$ to destroy the minihalo, where ${\bf v} = {\bf v}_{\rm mh} - {\bf v}_{\ast}$, and that the stars have a locally Maxwell-Boltzmann velocity distribution function with velocity dispersion $\sigma_{\ast}$ which is independent of stellar mass. The destruction rate of the cluster is 
\begin{align}
R = \int {\rm d}^{3}{\bf v}_{\ast} & \int 2\pi b\,{\rm d}b\,\int {\rm d}m_{\ast} \,|{\bf v}| \nonumber \\
& ({\rm d}N_{\ast}/{\rm d}m_{\ast}\,{\rm d}^{3}{\bf x}\,{\rm d}^{3}{\bf v}_{\ast})\,p(b,\,\bar{\rho},\,m_{\ast},\,{\bf v}). 
\end{align}
Taking the destruction probability $p$ to be a step function between $p=0$ for $b\gg b_{\rm min}$ and $p=p_{0}\sim1$ for $b\ll b_{\rm min}$ as assumed by this simplified model, after the integration, we obtain
\begin{equation}
    R = \left(\frac{\pi \alpha^2(c) G}{\gamma(c) \bar{\rho}_{\rm mh}}\right)^{1/2}\,\rho_{\ast}. 
\end{equation}
Integrating this over time for a given minihalo gives a survival probability $f_{\rm survive} = \exp{(-\tau)}$ where $\tau \equiv \int R\,{\rm d}t = \int R {\rm d} \ell/ v_{\rm mh}$ along the minihalo trajectory. The time-integrated destruction ``optical depth'' $\tau$ is dominated by the minihalo time in the disk. If a minihalo stayed in the disk over the entire Hubble time, then 
\begin{align}
    \tau & = \left(\frac{\pi \alpha^2(c) G}{\gamma(c) \bar{\rho}_{\rm mh}}\right)^{1/2}\,\rho_{\ast}\,\dfrac{1}{H_{\rm 0}} \nonumber \\
     & \sim 8\times10^{3}\,\left( \dfrac{\langle \rho_{\ast,\,{\rm disk}} \rangle}{1\,{\rm M}_{\odot}/{\rm pc^{3}}}\right)\,\left(\dfrac{c}{100}\right)^{-1/2}\,\left( \dfrac{1+z_{\rm i}}{1+5}\right)^{-3/2},
     \label{eq:back-of-the-envelope-1}
\end{align}
where we use the asymptotic approximation $\alpha^{2}(c)/\gamma(c) \sim 1/c$ at large c. This implies the complete destruction of the minihalo if it is always in the disk. On the other hand, if a minihalo only has a single passage through the disk, then
\begin{align}
\label{eq:tau_single}
& \tau_{\rm single}  = \int R\,{\rm d}t \simeq \left(\frac{\pi \alpha^2 G}{\gamma \bar{\rho}_{\rm mh}}\right)^{1/2}\,\int \rho_{\ast}\,{\rm d}h / v^{\bot}_{\rm mh}  \\
& \sim 0.2\,\left(\dfrac{c}{100}\right)^{-1/2} \, \left(\dfrac{1+z_{\rm i}}{1+5}\right)^{-3/2}\,\left(\dfrac{\Sigma_{\ast}}{70\,{\rm M}_{\odot}\,{\rm pc^{-2}}}\right)\,\left(\dfrac{v^{\bot}_{\rm mh}}{200\,{\rm km\,s^{-1}}}\right)^{-1}, \nonumber
\end{align}
where $\Sigma_{\ast}$ is the stellar surface density of the Milky Way disk, which is reasonably well-fit by an exponential profile $\Sigma_{\ast} \approx \Sigma_{0}\,\exp{(-R / R_{\rm d})}$ with $\Sigma_{\rm 0} = 816.6\,{\rm M}_{\odot}\,{\rm pc^{-2}}$ and $R_{\rm d}= 2.9\,{\rm kpc}$ ($\Sigma_{\rm 0} = 209.5\,{\rm M}_{\odot}\,{\rm pc^{-2}}$ and $R_{\rm d}= 3.31\,{\rm kpc}$) for the thin (thick) disk of the Milky Way \citep[{\em e.g.},][]{McMillan2011,McMillan2017}. If we are only interested in the Solar neighborhood ($R_{\odot} \approx 8\,$kpc, see \citealt{Bland-Hawthorn2016} for a review of measurements on $R_{\odot}$), the total surface density of both the thin and thick disks is $\Sigma_{\ast}^{\odot} \sim 70\,{\rm M}_{\odot}\,{\rm pc^{-2}}$. Therefore, the impact of a single passage through the disk is insignificant for reasonably high halo concentrations.

Besides the two extreme cases above, the more realistic scenario is successive passages through the disk, where the number of passages of a minihalo through the disk is a critical factor. Assuming circular orbits, the number of disk passages for minihalos in the Solar neighborhood is approximately $T_{\rm Hubble}/T_{\rm circ} \sim \mathcal{O}(100)$, leading to an $\mathcal{O}(1)$ effect after accumulation. Therefore, those minihalos in the Solar neighborhood should experience significant disruption. To make a quantitative prediction for the remaining mass and structure of minihalos, we need a more complete model, which will be described in the following.

\subsubsection{General response model}

In general, the mass loss of minihalos in single (or multiple successive) encounters is characterized by the response curve $ 1 - \Delta M_{\rm mh}/M_{\rm mh} = \mathcal{F}(\Delta E_{\rm tot}/E_{\rm b}, c)$ was found in Section~\ref{sec:stellar_sim}. Assuming the surface density of stars in the disk is large enough so that we can neglect stellar shot noise (discussed further below), the total energy injection $\Delta E_{\rm tot}/E_{\rm b}$ when crossing a localized slab of stars can be calculated by integrating over all possible impact parameters, masses, and velocities of stars 
\begin{align}
\dfrac{\Delta E_{\rm tot}}{E_{\rm b}} & = \int {\rm d}^{3}{\bf v}_{\ast} \int {\rm d}\ell \, \int 2\pi b\,{\rm d}b\,\int {\rm d}m_{\ast} \, \nonumber \\
& \hspace{3cm} ({\rm d}N_{\ast}/{\rm d}m_{\ast}\,{\rm d}^{3}{\bf x}\,{\rm d}^{3}{\bf v}_{\ast})\,\dfrac{\Delta E}{E_{\rm b}}\nonumber \\
& = G \int {\rm d}\ell \, \int {\rm d}^{3}{\bf v}_{\ast} f({\bf v}_{\ast}) \dfrac{1}{|{\bf v}_{\ast}-{\bf v}_{\rm mh}|^{2}} \, \int {\rm d}m_{\ast}\, n_{m_{\ast}} \, m_{\ast}^2  \nonumber \\
&  \times\left( \int_{b_{\rm s}}^{\infty} \dfrac{\alpha^2(c)}{\pi \gamma(c) \bar{\rho}_{\rm mh}} \dfrac{2\pi b\, {\rm d}b}{b^4}  + \int_{0}^{b_{\rm s}} \dfrac{\alpha^2(c)}{\pi \gamma(c) \bar{\rho}_{\rm mh}} \dfrac{2\pi b\, {\rm d}b}{b^{4}_{\rm s}} \right) \nonumber \\
& \simeq \dfrac{G m_{\kappa} \Sigma_{\ast}}{\sigma^2_{\ast} + v^2_{\rm mh}} \, \dfrac{2 \,\alpha^2(c)}{\gamma(c) \bar{\rho}_{\rm mh} b^{2}_{\rm s}}, \nonumber \\ 
\label{eq:integrated_response}
\end{align}
where $\Delta E/E_{\rm b}$ is evaluated using Equation~\ref{eq:single_encounter}, $\sigma_{\ast}$ is the {\it one-dimensional} velocity dispersion of stars and $f(\mathbf{v_{\ast}})$ is the Maxwell-Boltzmann velocity distribution function~\footnote{The integral in the velocity space is evaluated in the $v_{\rm mh}\gg \sigma_{\ast}$ or $v_{\rm mh} \ll \sigma_{\ast}$ regime first. The leading order term in the asymptotic limits can be connected using the expression, $1/(\sigma^2_{\ast} + v^2_{\rm mh})$. The relative error between this expression and the true result is suppressed by the factor $(\sigma_{\ast}/v_{\rm mh})^{2}$ when $v_{\rm mh}\gg \sigma_{\ast}$ and vice versa.}. $m_{\kappa}$ is a characteristic mass that depends on the PDMF of stars in the Milky Way,
\begin{equation}
    m_{\kappa} \rho_{\ast} = \int {\rm d}m_{\ast}\, n_{m_{\ast}} \, m_{\ast}^2.
\end{equation}
A simple approximation that gives a good fit to the Milky Way data is to take the Kroupa IMF \citep{kroupa:2001.imf.var} as the PDMF at $m_{\ast} \leq 1\,{\rm M}_{\odot}$ (since the evolutionary effects are small at low masses) with a power-law cutoff $n_{m_{\ast}} \sim m_{\ast}^{-4.5}$ at $m_{\ast} > 1\,{\rm M}_{\odot}$~\citep[{\em e.g.},][]{scalo:imf,kroupa:1993.pdmf,sollima:2019.gaia.imf}. After the integration, we obtain $m_{\kappa} \simeq 0.6\,{\rm M}_{\odot}$, which is insensitive to the minimum or maximum star mass assumed. $m_{\kappa}$ can be viewed as the characteristic mass of the most effective disruptor, which can vary in different environments. For example, a clumpy medium may have significantly higher $m_{\kappa}$ and thus stronger disruption effects. The integration in Equation~\ref{eq:integrated_response} is carried out to infinite distances, where in principle the localized quantities we define in a disk patch no longer apply. This will not affect the results significantly since the contribution from distant stars is suppressed by the $1/b^{4}$ dependence of energy imparted. 

It is, however, still important to note that Equation~\ref{eq:integrated_response} will no longer be valid if the stellar surface density is small enough that shot noise becomes important, {\em i.e.} when $\pi b_{\rm c}^2 \Sigma_{\ast}/m_{\kappa} \sim 1$. If we take $m_{\kappa} = 0.6\,{\rm M}_{\odot}$ as estimated above, we obtain the cut-off impact parameter as
\begin{equation}
    b_{\rm c} =  0.044\,{\rm pc}\,\left(\dfrac{m_{\kappa}}{0.6\,{\rm M}_{\odot}} \right)^{1/2}\,\left( \dfrac{\Sigma_{\ast}}{100\,{\rm M}_{\odot}/{\rm pc}^2}\right)^{-1/2}.
\end{equation}
If shot noise becomes dominant, $b_{\rm c}\gg b_{\rm s}$, the total energy injection becomes~\footnote{More accurately, we can calculate $b_{\rm c}$ {\em at each value of the stellar mass} $m_{\ast}$, using the same PDMF and $\pi\,b_{\rm c}^{\prime}(m_{\ast})\,{\rm d}N_{\ast}(>m_{\ast})/{\rm d}{\rm Area}=1$, insert this into Equation~\ref{eq:integrated_response} and then integrate over all masses numerically to define an appropriately-weighted $b_{c}$. Doing so, we find that this gives an ``effective'' $b_{\rm c}$ which is only $\sim 8\%$ larger than what we obtain using $\pi\,b_{\rm c}\,\Sigma_{\ast}/m_{\kappa}=1$.}
\begin{equation}
    \dfrac{\Delta E_{\rm tot}}{E_{\rm b}} =  \dfrac{G m_{\kappa} \Sigma_{\ast}}{\sigma^2_{\ast} + v^2_{\rm mh}} \, \dfrac{\alpha^2(c)}{\gamma(c) \bar{\rho}_{\rm mh} b^{2}_{\rm c}}. 
\end{equation}
Connecting the behavior at $b_{\rm c}\ll b_{\rm s}$, the general solution of the accumulated energy injection during one disk passage can be written as
\begin{align}
    \label{eq:energy_injection_onep}
    & \dfrac{\Delta E_{\rm tot}}{E_{\rm b}} = \dfrac{G m_{\kappa} \Sigma_{\ast}}{\sigma^2_{\ast} + v^2_{\rm mh}} \, \dfrac{\alpha^2(c)}{\gamma(c) \bar{\rho}_{\rm mh}} \dfrac{2}{b^{2}_{\rm s}+2b^{2}_{\rm c}}  \nonumber \\
    &\simeq 2 \times 10^{-2}\, \left( \dfrac{c}{100} \right)^{-1} \left(\dfrac{m_{\kappa}}{0.6\,{\rm M}_{\odot}} \right) \left( \dfrac{\sqrt{\sigma^2_{\ast} + v^2_{\rm mh}}}{250\,{\rm km/s}} \right)^{-2}  \nonumber \\ 
    & \left(\dfrac{\Sigma_{\ast}}{100\,{\rm M}_{\odot}/{\rm pc}^2} \right) \left(\dfrac{\sqrt{b^{2}_{\rm s}+2b^{2}_{\rm c}}}{0.1\,{\rm pc}} \right)^{-2}\,\left( \dfrac{1+z_{\rm i}}{1+5} \right)^{-3},
\end{align}
where we use the same asymptotic approximation of $\alpha^2(c)/\gamma(c)$ as in Equation~\ref{eq:back-of-the-envelope-1}. Typically, one has $b_{\rm s} > b_{\rm c}$ for massive minihalos with low concentrations and $b_{\rm s} < b_{\rm c}$ for low-mass minihalos with high concentrations. It is also worth noting that $m_{\kappa}$ will cancel out when $b_{c} \gg b_{s}$, implying that clumpiness of the medium only matters for minihalos with physical sizes comparable to or larger than the typical spacing of the disruptors.

When the Galactic disk is dominated by stars, our treatment by integrating the cumulative perturbations from disk stars to large distances (rather than considering only close encounters like in the back-of-the-envelope model) should be physically equivalent to the disk shocking effect~\citep[{\em e.g.},][]{Ostriker1972,BT1987,Gnedin1999,Stref2017} studied in the depletion of substructures in the Milky Way \citep[{\em e.g.},][]{DOnghia2010,Stref2017,Facchinetti2022}. The disk shocking calculation implies \citep[{\em e.g.},][]{BT1987,Stref2017}
\begin{equation}
    \dfrac{\Delta E_{\rm tot}}{E_{\rm b}} \propto \dfrac{R^{2}_{\rm mh} g^{2}_{\rm z,disk}}{\sigma^{2}_{\rm mh}\,v^{2}_{\rm mh}} \propto
    \dfrac{R^{2}_{\rm mh}\,\Sigma_{\rm disk}^{2}}{M_{\rm mh}/R_{\rm mh}\,v^{2}_{\rm mh}}
    \propto
    \dfrac{\Sigma_{\ast}^{2}}{v^{2}_{\rm mh}\,\bar{\rho}_{\rm mh}},
\end{equation}
where $g_{\rm z,disk}\propto \Sigma_{\rm disk} \simeq \Sigma_{\ast}$ is the gravitational acceleration at disk vicinity, $\sigma_{\rm mh}$ is the internal velocity dispersion of the minihalo which should scale as $\sqrt{G\,M_{\rm mh}/R_{\rm mh}}$. However, taking our Equation~\ref{eq:energy_injection_onep} to a ``smooth'' disk limit ($m_{\kappa}$ being infinitesimal), we do not get the disk shocking limit naturally. The key difference is that we assume that all individual stellar encounter events are independent. We evaluate the energy injection in each independent event first before summing up, while the disk shocking evaluates the tidal field of the entire baryonic disk simultaneously before estimating the momentum and energy injection. This independence assumption we made is supported by the fact that the disk scale height, $H\sim \mathcal{O}(100)\,{\rm pc}$, is at least two orders of magnitude larger than the characteristic impact parameter $b_{\rm s}\sim R_{\rm mh}$ of the minihalo, so individual encounters should operate locally in an independent fashion. If we abandon the independence assumption, the break in energy injection at $b_{\rm s}$ in Equation~\ref{eq:deltaE_general} (which comes from resolving individual encounters) will disappear. We can integrate the $1/b^{4}$ law until reaching $b_{\rm c}$ and would obtain
\begin{equation}
    \dfrac{\Delta E_{\rm tot}}{E_{\rm b}} \propto \dfrac{m_{\kappa}\,\Sigma_{\ast}}{v^{2}_{\rm mh}\,\bar{\rho}_{\rm mh}\,b^{2}_{\rm c}} \propto \dfrac{\Sigma_{\ast}^{2}}{v^{2}_{\rm mh}\,\bar{\rho}_{\rm mh}},
\end{equation}
which is totally consistent with the disk shocking calculation.

Depending on the orbit of the minihalo, it can pass through the stellar disk multiple times after falling into the host. When the orbital time is much smaller than the relaxation time of the minihalo, the combined effect can again be modeled as one single passage with the accumulated injected energy. The total number of passages and stellar surface densities at the encounter point should take an ensemble average of all possible stellar orbits passing the location of the detector. In Appendix~\ref{app:orbit}, we calculate the relevant correction factors from the ensemble average of all possible orbits with a simplified model. The accumulated injected energy will ultimately be fed to the response function $\mathcal{F}(\Sigma \Delta E_{\rm tot}/E_{\rm b})$ to calculate the mass loss. Even if the time between passages is comparable to or larger than the minihalo relaxation time, as discussed in Section~\ref{sec:multi}, it is still reasonably accurate to use the accumulated injected energy and the response curve to approximate the disruption fraction.

\subsection{Semi-analytic model to combine stellar and tidal disruptions}

As the minihalo moves closer to the Galactic center, both $r_{\rm t}$ (Equation~\ref{eq:tidal_radius}) and $t_{\rm ts}(R)$ (Equation~\ref{eq:tidal_time}) will decrease sharply. Therefore, the total mass loss of an infalling minihalo is dominated by its pericenter passages. For minihalos of interest for detection ({\em e.g.} in the Solar neighborhood $R_{\odot} \simeq 8\rm kpc$), the tidal-stripping time scale during a pericenter passage at $R_{\odot}$ is $O(100)\,{\rm Myr}$, which is of the same order as the minihalo orbital time and much shorter than the lifetime of the minihalo in the host ($\sim T_{\rm Hubble}$). We assume that the mass of an infalling minihalo outside the tidal radius will be quickly stripped away during the first few pericenter passages before the impact of stellar disruptions start to accumulate. 

For simplicity, we evaluate the tidal radius at the target radius $r_{\rm obs}$ of observation, assuming a circular orbit
\begin{equation} \label{eq:rtidal_reduced}
    r_{ t}=r_{\rm obs}\left[\frac{M_{\rm mh}(r<r_{ t})/M_{\rm MW}(r<r_{\rm obs})}{3-\frac{{\rm d}{\rm ln}M_{\rm MW}}{{\rm d}{\rm ln}R}\big\vert_{r_{\rm obs}} }\right]^{1/3}.
\end{equation}
where the Milky Way halo mass distribution is modeled as an NFW profile for dark matter plus a Hernquist profile \citep{Hernquist1990} for the stellar content
\begin{equation}
    M_{\rm MW}(<r) = M_{\rm dm} \dfrac{f^{\rm nfw}(r/r_{\rm s})}{f^{\rm nfw}(c)} + M_{\rm b}\dfrac{r^{2}}{r^2 + a^2},
\end{equation}
where $f^{\rm nfw}(x) \equiv \ln{(1+x)}-x/(1+x)$, the host halo parameters are $c=12$, $M_{\rm dm} = 10^{12}\,{\rm M}_{\odot}$ \citep[{\em e.g.},][]{Klypin2002,McMillan2011,Deason2012,Bland-Hawthorn2016}. For baryon properties, abundance-matching studies have shown that a Milky-Way mass system typically has a stellar-to-total-mass ratio of $M_{\rm b}\simeq 0.01\,M_{\rm dm}$ \citep[{\em e.g.},][]{Moster2013} and the half mass radius $r^{\ast}_{1/2} \simeq 0.02\, R_{\rm vir}$ \citep[{\em e.g.},][]{Somerville2018}. The scale radius of the Hernquist profile is related to the stellar-half-mass-radius as $a=0.414 r^{\ast}_{1/2}$ \citep{Hernquist1990}.

The post-stripping density profile of the minihalo is assumed to be a truncated NFW profile at $r_{\rm t}$, which is equivalent to a normal NFW profile with effective virial radius, concentration, and overdensity as
\begin{align} \label{eq:tidalcorr}
    & R^{\rm eff}_{\rm vir} = r_{\rm t}, \nonumber \\
    & c_{\rm eff} = c \dfrac{r_{\rm t}}{R_{\rm vir}}, \nonumber \\
    & \Delta_{\rm eff} = \dfrac{\log{(1+c_{\rm eff})} - c_{\rm eff}/(1+c_{\rm eff})}{\log{(1+c)} - c/(1+c)} \dfrac{c^{3}}{c^{3}_{\rm eff}} \Delta_{\rm c},
\end{align}
The mass loss due to the tidal disruption is
\begin{equation} \label{eq:tidalmloss}
    1 - \Delta M_{\rm mh}/M_{\rm mh} = \dfrac{\log{(1+c_{\rm eff})} - c_{\rm eff}/(1+c_{\rm eff})}{\log{(1+c)} - c/(1+c)}.
\end{equation}
The stripped minihalo forms the initial condition for the following stellar disruptions. Therefore, for a minihalo observed at $r_{\rm obs}$, the energy accumulated from stellar encounters can be written as (following Equation~\ref{eq:energy_injection_onep} but considering multiple passages through the disk during the lifetime of the minihalo)
\begin{align}
    \label{eq:energy_injection_aftertidal}
    \dfrac{\Delta E_{\rm tot}}{E_{\rm b}} (r_{\rm obs}) = \Bigg\langle N_{\rm p} \Big\langle \dfrac{G m_{\kappa} \Sigma_{\ast}}{\sigma^2_{\ast} + v^2_{\rm mh}} & \dfrac{\alpha^2(c_{\rm eff})}{\gamma(c_{\rm eff}) \Delta_{\rm eff}\rho_{\rm crit}(z_{\rm i})} \nonumber \\
    & \dfrac{2}{b^{2}_{\rm s}(c_{\rm eff},R_{\rm vir}^{\rm eff})+2b^{2}_{\rm c}(\Sigma_{\ast})} \Big \rangle_{\rm x} \Bigg \rangle_{\rm o} \nonumber \\
    = \overline{N_{\rm p}} \, f_{\theta}\,f_{\Sigma_{\ast}}  \dfrac{G m_{\kappa} \Sigma_{\ast}(r_{\rm obs})}{\sigma^2_{\ast} + v^2_{\rm mh}} & \dfrac{\alpha^2(c_{\rm eff})}{\gamma(c_{\rm eff}) \Delta_{\rm eff}\rho_{\rm crit}(z_{\rm i})} \nonumber \\
    & \dfrac{2}{b^{2}_{\rm s}(c_{\rm eff},R_{\rm vir}^{\rm eff})+2b^{2}_{\rm c}(\Sigma_{\ast}(r_{\rm obs}))}, \nonumber \\
    \overline{N_{\rm p}} = f_{\rm N_{p}} \, N^{\rm circ}_{\rm p}, \hspace{0.4cm} N^{\rm circ}_{\rm p} = & 2\,T_{\rm Hubble}/T_{\rm circ}(r_{\rm obs})
\end{align}
where $N_{\rm p}$ is the number of passages through the stellar disk, $\langle \rangle_{\rm o}$ denotes averaging over an ensemble of minihalos observed at $r_{\rm obs}$ with all possible orbits. Therefore, $\overline{N_{\rm p}}$ represents the averaged number of passages over all possible orbits. $N^{\rm circ}_{\rm p}$ is the number of passages assuming the minihalo is on a circular orbit with radius $r_{\rm obs}$ calculated based on the Hubble time $T_{\rm Hubble}$ and the circular orbit period $T_{\rm circ}(r_{\rm obs})$. $f_{\rm N_{p}}$ characterizes the deviation of $\overline{N_{\rm p}}$ from this circular orbit estimation. In Equation~\ref{eq:energy_injection_aftertidal}, $\Sigma_{\ast}$ is the stellar surface density where the minihalo crossed the disk (the surface density profile $\Sigma_{\ast}(r)$ is given below Equation~\ref{eq:tau_single}). $\langle \rangle_{\rm x}$ denotes averaging over all past disk crossings given the orbit of the minihalo. The correction factor $f_{\Sigma_{\ast}}$ characterizes the deviation of the averaged $\Sigma_{\ast}$ at all past encounter locations for all possible orbits from $\Sigma_{\ast}(r_{\rm obs})$. $f_{\theta}$ accounts for the increased effective stellar surface density when the minihalo is not passing perpendicular to the disk, see Appendix~\ref{app:orbit} for details). In Appendix~\ref{app:orbit}, $f_{\rm N_{p}}$ and $f_{\Sigma_{\ast}}$ are estimated based on the orbital model of an isothermal halo. The combined effect of $f_{\theta}$, $f_{\rm N_{p}}$ and $f_{\Sigma_{\ast}}$ on $\Delta E_{\rm tot}/E_{\rm b}$ is $\mathcal{O}(10)$ at the Solar neighborhood. The velocity term $\sigma^2_{\ast} + v^2_{\rm mh}$ has a weak dependence on $r_{\rm obs}$ and minihalo orbits, so it is assumed to be the constant value $(250\,{\rm km/s})^{2}$ for simplicity. 

We note that $b_{\rm c}$ has an implicit dependence on the surface density at the encounter. When $b_{\rm c}\gg b_{\rm s}$ (when $\Sigma_{\ast}$ is large or $M_{\rm mh}$ is small), $\Delta E_{\rm tot}/E_{\rm b}$ will be proportional to $\Sigma_{\ast}^{2}$ and the correction factor $f_{\Sigma_{\ast}}$ should be replaced with $f_{\Sigma^{2}_{\ast}}$ (see the calculation in Appendix~\ref{app:orbit}). To properly account for this, we model the transition from  $f_{\Sigma_{\ast}}$ to  $f_{\Sigma^{2}_{\ast}}$ empirically as
\begin{equation}
    f = f_{\Sigma_{\ast}} + (f_{\Sigma^{2}_{\ast}} - f_{\Sigma_{\ast}}) \dfrac{1}{1+e^{-k \log{(\sqrt{2}b_{\rm c}/b_{\rm s})}}}
\end{equation}
where $k=3$ is assumed. We note that the value of $k$ or the detailed form of the transition does not affect the post-disruption mass function in any significant way.

\begin{figure*}
    \centering
    \includegraphics[width=1\textwidth]{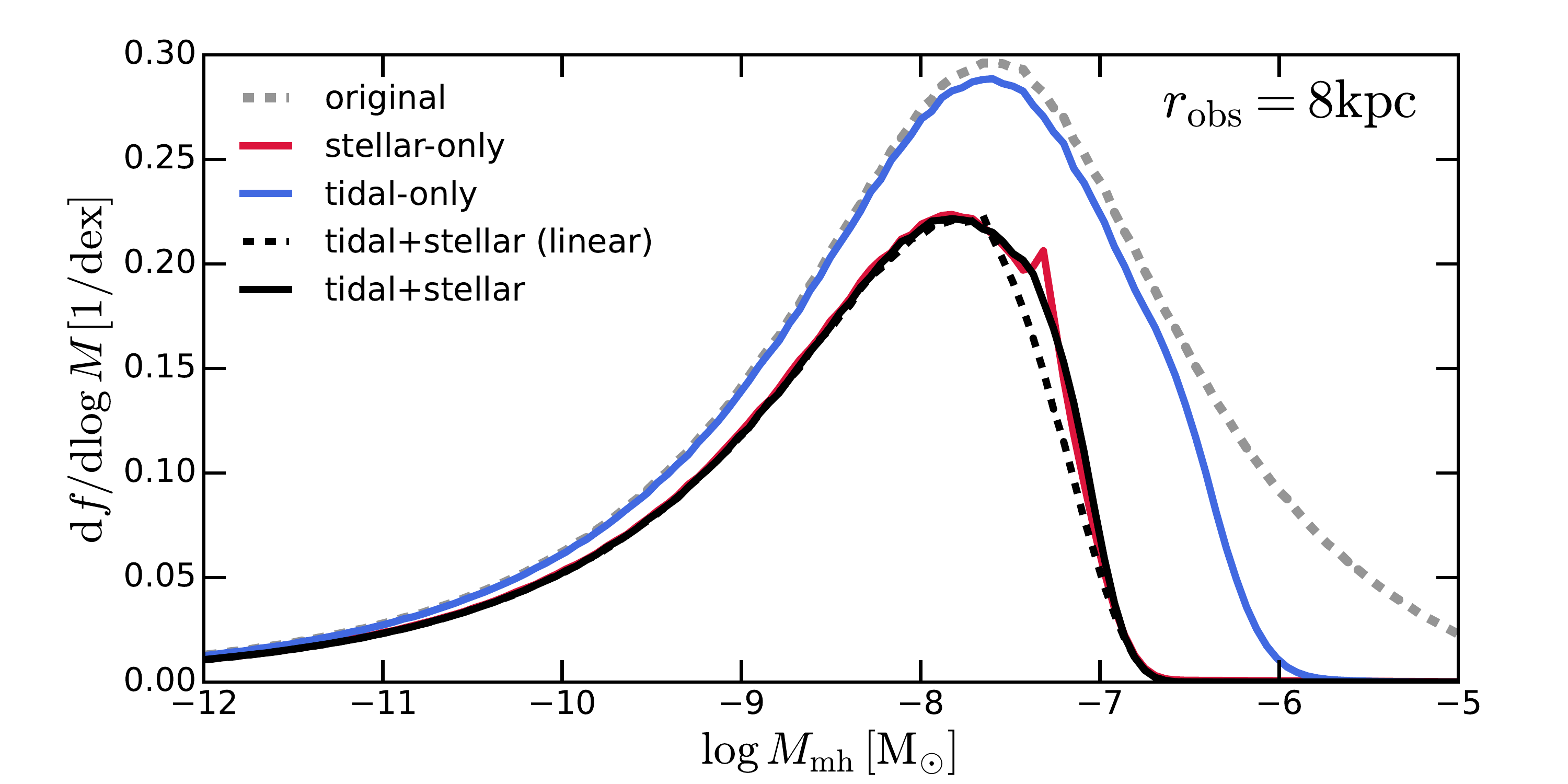}
    \includegraphics[width=0.49\textwidth]{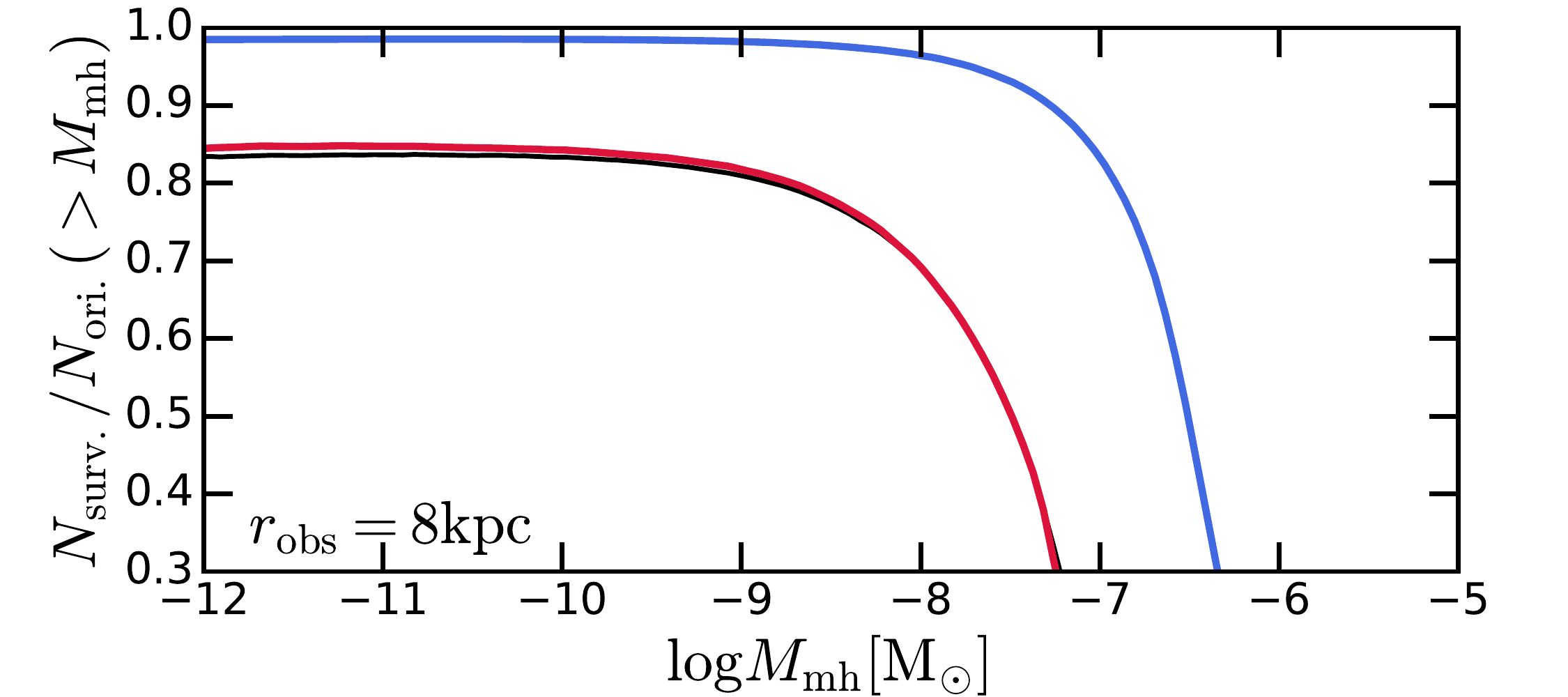}
    \includegraphics[width=0.49\textwidth]{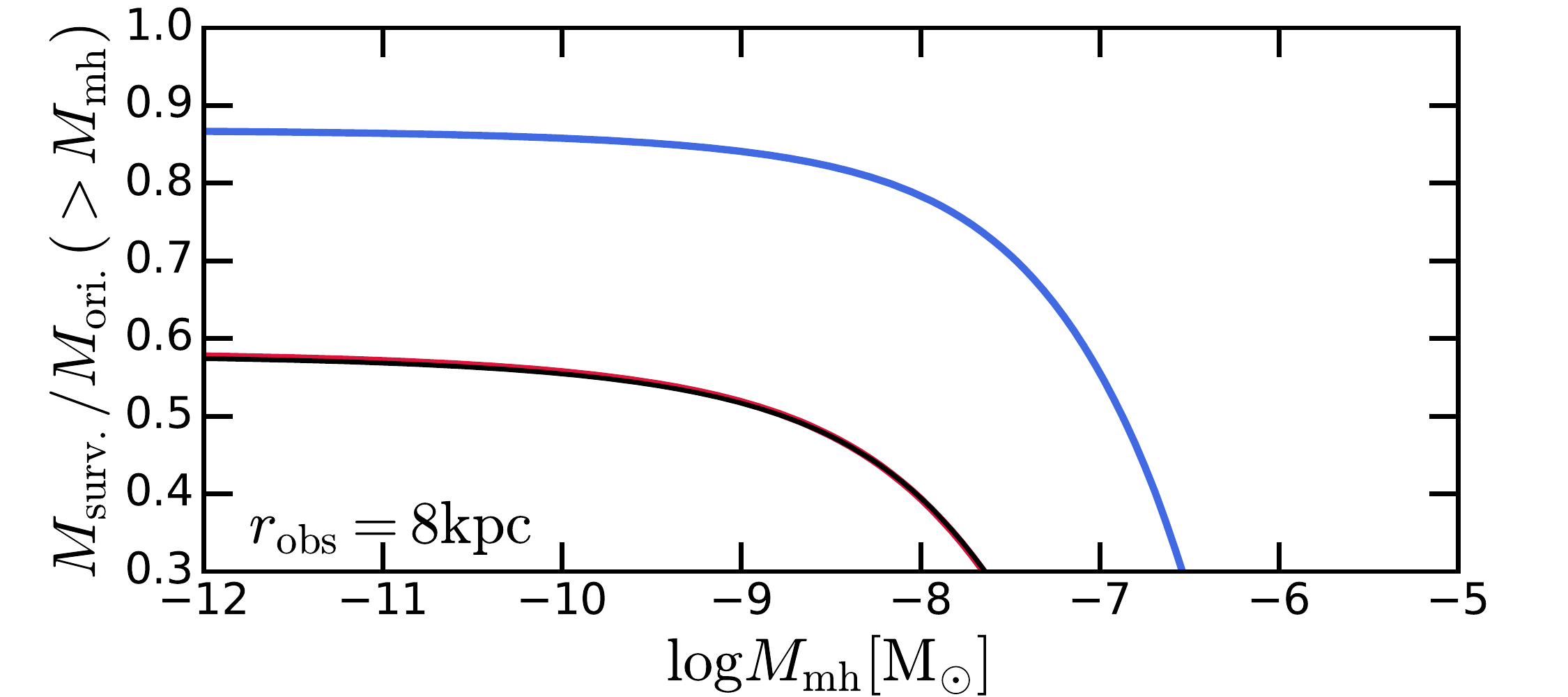}
    \caption{{\it Top}: Mass function of minihalos (from axion miniclusters) at the Solar neighborhood ($r_{\rm obs}\simeq 8\,{\rm kpc}$). We assume the AMC model with $m_{\rm a}=25\,\mu{\rm eV}$. The mass function before disruption is shown as the gray dashed line. The mass function after processing only tidal (or stellar) disruption is shown as the blue (red) solid line. The mass function post-disruption, combining both tidal and stellar disruption, is shown as the solid black line. The mass function post-disruption, combining tidal and stellar disruption linearly, is shown as the dashed black line. In general, the disruptions taken together induce approximately a $30\%$ decrease in the peak value of the mass function and shift the mass of the peak by roughly half an order of magnitude. The massive end is more strongly affected by disruption than the low-mass end. {\it Bottom}: We show the integrated number (left) and mass (right) of minihalos before and after the disruption. The typical survival fraction of minihalos with $M_{\rm mh} \geq 10^{-12}\,{\rm M}_{\odot}$ is $83\%$ in terms of number and about $58\%$ in terms of mass. Stellar disruption is the dominant disruption mechanism through the entire mass range of interest.}
    \label{fig:mass-func-8kpc}
\end{figure*}

\subsection{Monte Carlo sampling of the minihalos}

We are ready to implement all the physics of disruption discussed above to a sample of minihalos and track their mass loss. We model the evolution of the minihalo population in the Milky Way halo following the steps below:

\begin{itemize}

\item We initialized a Monte Carlo sample of minihalos. First, we construct a grid of infall redshifts from $z^{\rm min}_{\rm i}=0$ to $z^{\rm max}_{\rm i}=150$ (uniform in $\log{(1+z_{\rm i})}$), and compute the infall probability at each redshift point as $\Delta \, f^{\rm CDM}_{\rm col}(z_{\rm i})$ (using the matter power spectrum from adiabatic CDM fluctuations on small scales, see the discussion in Section~\ref{sec:models}). Then, at each redshift point, the minihalo masses are sampled uniformly over the dynamical range $10^{-14}$ to $10^{-3}\,{\rm M}_{\odot}$. The number densities of these minihalos are calculated following the redshift-dependent pre-infall mass function given in Section~\ref{sec:models} and Appendix~\ref{app:mass_function}. The weight of each individually sampled minihalo is proportional to the product of the number density and the infall probability at $z_{\rm i}$. The minihalo concentrations are calculated following the mass-concentration relation given in Section~\ref{sec:models}. These sampled physical properties represent the initial status of the minihalos upon falling into the Milky Way host.

\item Tidal stripping and structural corrections are applied to the sampled minihalos as described in Equation~\ref{eq:tidalcorr} and \ref{eq:tidalmloss}. Since the $r_{\rm t}$ solved from Equation~\ref{eq:rtidal_reduced} after normalizing over $r_{\rm s}$ is independent of minihalo mass, we calculate $r_{\rm t}/r_{\rm s}$ on a grid of $c$ and $z_{\rm i}$ for several different choices of $r_{\rm obs}$ and prepare them as lookup tables for efficient interpolation of the tidal radius. 

\item After the tidal stripping and the implementation of relevant structural corrections, we apply the stellar disruption with the mass loss given by $\mathcal{F}(\Delta E_{\rm tot}/E_{\rm b}, c)$. The cumulative energy injection from stellar encounters, $\Delta E_{\rm tot}/E_{\rm b}$, is evaluated with Equation~\ref{eq:energy_injection_aftertidal} using the structural parameters corrected after tidal disruptions. We include the orbit corrections derived in Appendix~\ref{app:orbit}. 

\end{itemize}

All the results demonstrated below have passed the convergence tests over hyperparameters in the sampling approach above, including the maximum sample redshift $z^{\rm max}_{\rm i}$, the mass range of sampling, the number of redshift grid points, the number of samples at each redshift and the resolution of tidal correction grid.

\section{Results}
\label{sec:dm_mf_change}

\subsection{Post-disruption mass functions}
\label{sec:mass_function}

In Figure~\ref{fig:mass-func-8kpc}, we show the mass function of minihalos (from AMC with the fiducial $m_{\rm a}=25\,\mu{\rm eV}$ as an example) at the Solar neighborhood ($r_{\rm obs}=8\,{\rm kpc}$). The mass functions are presented as the matter mass fraction (with respect to the total dark matter mass) in minihalos per unit logarithm interval (dex) of minihalo mass. We have assumed the axion gives the correct dark matter relic abundance. We present the initial (pre-disruption) mass function, the post-tidal/stellar disruption mass function, the mass function considering the combined effect of tidal and stellar disruptions, and the mass function aggregating tidal and stellar disruptions linearly. With both tidal and stellar disruption, the peak of the mass function is reduced by about $30\%$ accompanied by roughly half an order of magnitude mass shift of the peak. The low-mass end of minihalos is less disrupted than the massive end because lighter minihalos generically form earlier in these models and are more concentrated. Comparing the two disruption mechanisms, the stellar disruption dominates over the entire mass range, which agrees with the conclusion of previous studies of AMC \citep[{\em e.g.},][]{kavanagh2020stellar, Lee:2020wfn}. It is worth noting that tidal disruption can alter the structural parameters and enhance the average density of minihalos, which could lead to non-linear effects on the following stellar disruption. However, the effect is relatively weak on mass functions in our experiments. We note that a spike shows up in the post-disruption mass function at $M_{\rm mh} \sim 10^{-7}\,{\rm M}_{\odot}$ when considering only the stellar disruption. The spike is related to a turnover in the distribution of the minihalo initial concentrations and will be discussed in the following section. In the bottom panels of Figure~\ref{fig:mass-func-8kpc}, we show the integrated fraction of disrupted minihalos in number and total mass, respectively. The ``survival fraction'' quoted hereafter is defined as the fraction of the integrated mass/number of minihalos above a certain mass threshold retained after disruptions. Both the mass and number survival fractions show a plateau as the mass threshold becomes sufficiently low compared to the peak of the initial mass function. Thus we pick $M^{\rm lim}_{\rm mh} = 10^{-12}\,{\rm M}_{\odot}$ as the limit to measure the ``overall'' survival fraction, which is insensitive to $M^{\rm lim}_{\rm mh}$ whatsoever. The overall survival fraction of minihalos with $M_{\rm mh} \geq 10^{-12}\,{\rm M}_{\odot}$ is about $83\%$ in terms of number and about $58\%$ in terms of mass. The survival fraction will quickly diminish to $\lesssim 30\%$ at $M_{\rm mh} \gtrsim 10^{-7}\,{\rm M}_{\odot}$. Again the dominance of the stellar disruption is manifest. 

\begin{figure}
    \centering
\includegraphics[width=0.49\textwidth]{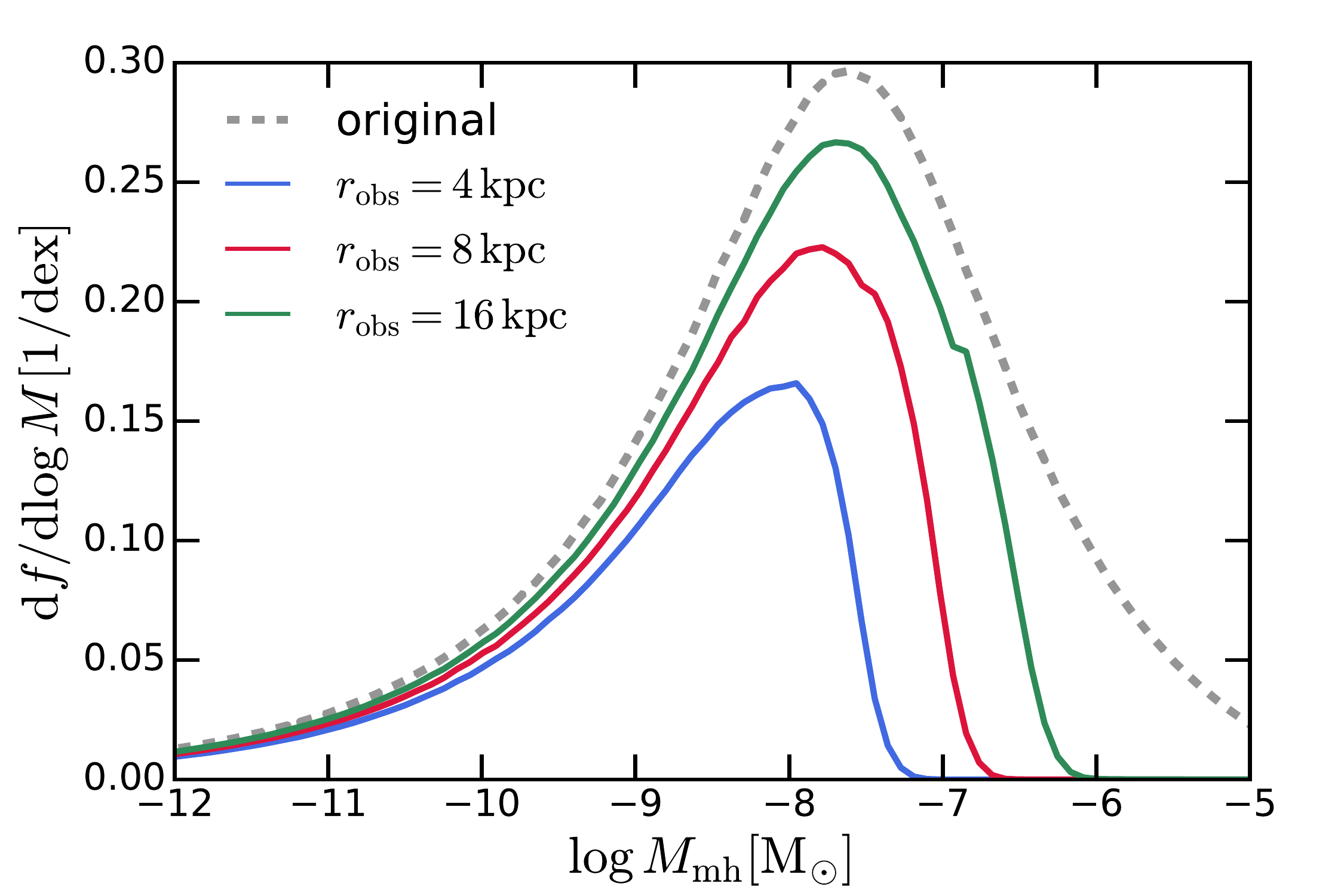}
\caption{Mass function of minihalos (from AMC with $m_{\rm a}=25\,\mu{\rm eV}$) observed at three galactocentric distances ($r_{\rm obs} = 4,8,16\,{\rm kpc}$). A similar reduction pattern in the mass density of minihalos is found at each distance, with stronger disruption at smaller radii and the massive end of the mass function. The peak of the post-disruption mass function is slightly shifted towards lower masses at smaller radii. Stellar disruption is the dominant disruption mechanism at all radii.}
\label{fig:mass-func-rvary}
\end{figure}

\begin{figure}
    \centering
    \includegraphics[width=0.49\textwidth]{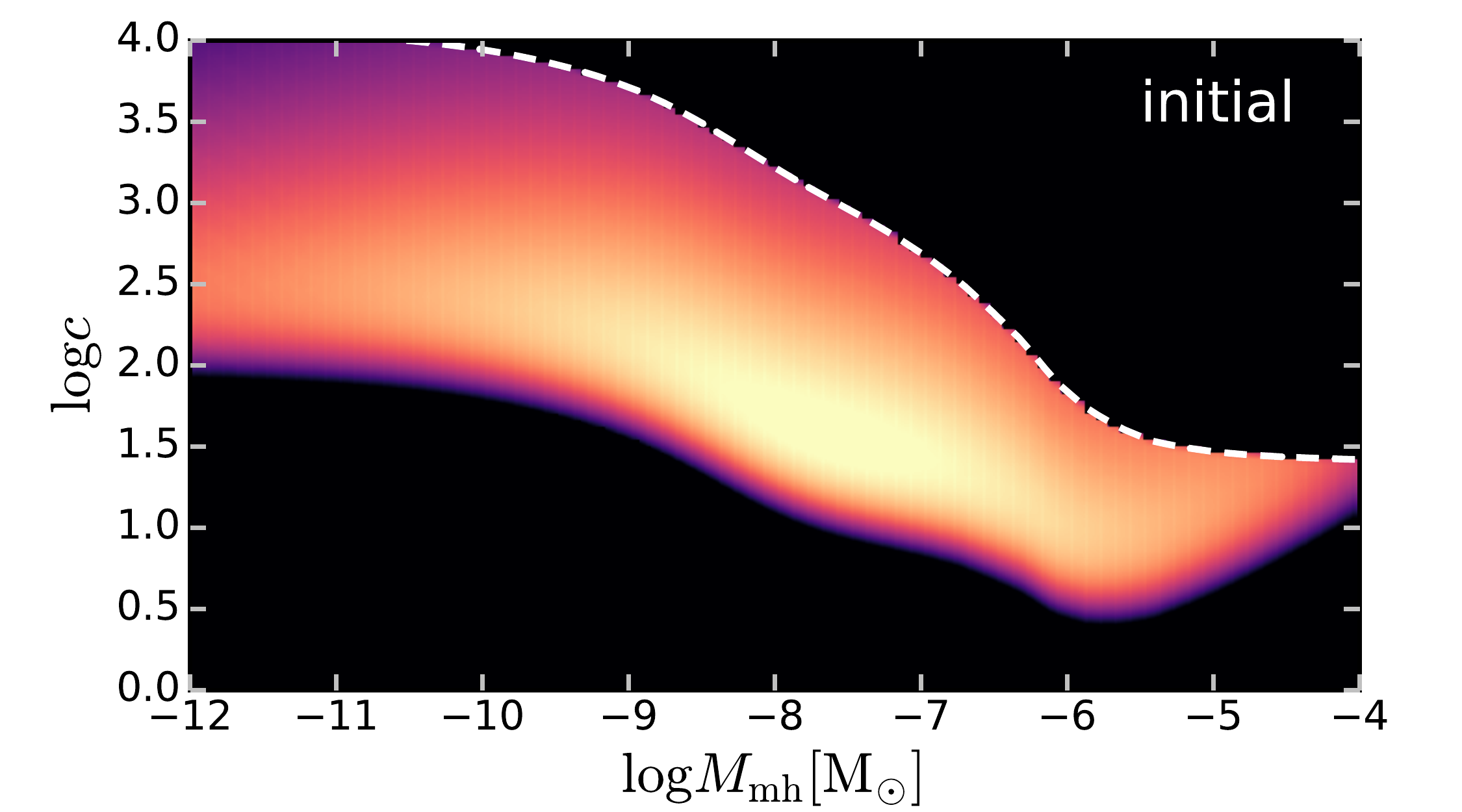}
    \includegraphics[width=0.49\textwidth]{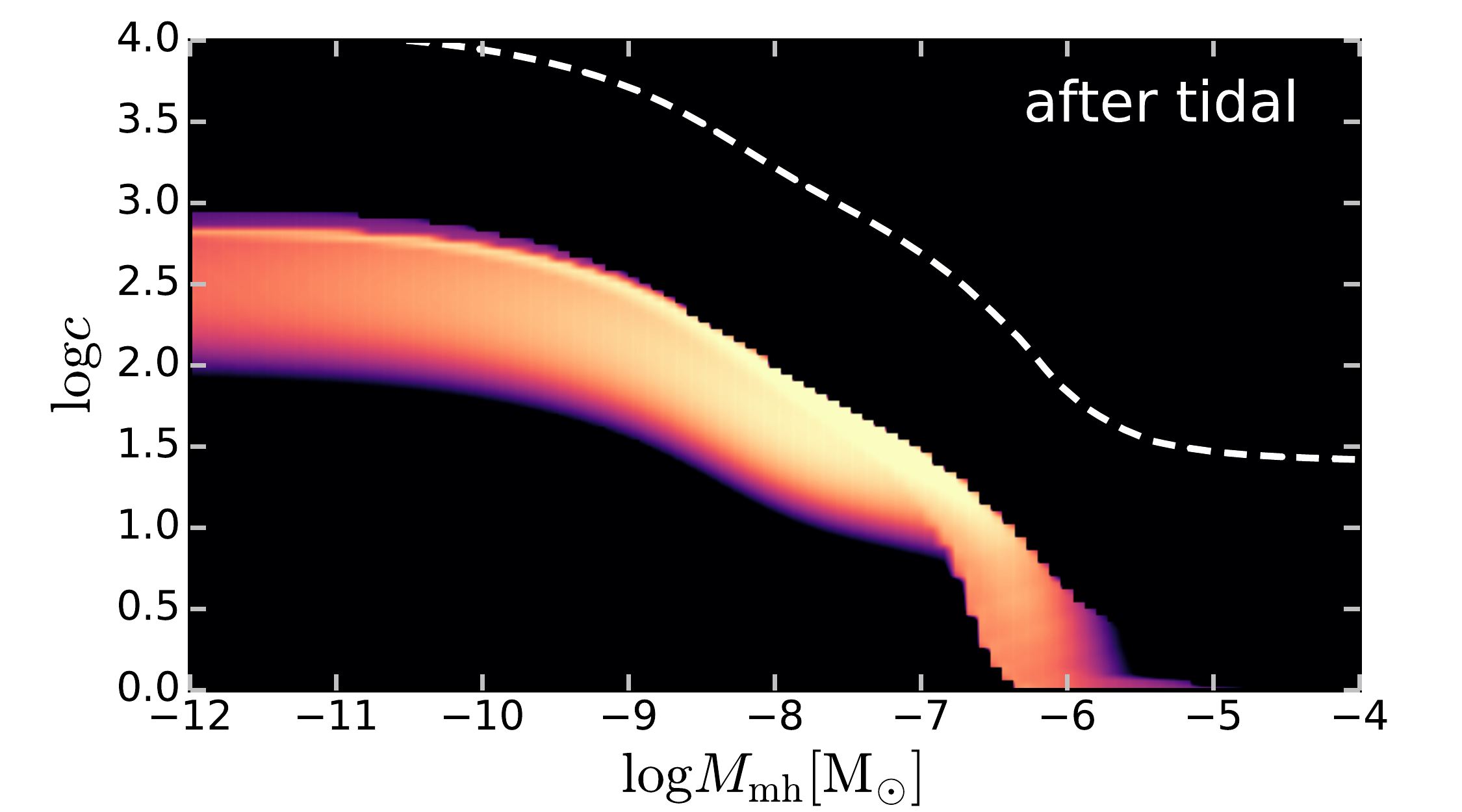}
    \caption{{\it Top:} Mass distribution of sampled minihalos in the phase space ($c$ versus $M_{\rm mh}$) before disruption. The mass concentration relation from \citet{Lee:2020wfn} is shown with the white dashed line (assuming $z_{\rm i}=0$). The down scatter of concentration on this plane is driven by the spread of the minihalo infall redshift, $z_{\rm i}$. The scatter shrinks to zero towards the massive end, since most of the massive minihalos have $z_{\rm i} \sim 0$. {\it Bottom:} Mass distribution of sampled minihalos in the phase space after tidal disruption. Effectively, tidal disruption lowers the concentration of minihalos to $r_{\rm t}/r_{\rm s}$ while increasing the averaged densities of minihalos (since the low-density outskirt has been shredded). The effect is significant in massive, low-concentration minihalos.}
    \label{fig:phase_space}
\end{figure}

\begin{figure}
    \centering
    \includegraphics[width=0.49\textwidth]{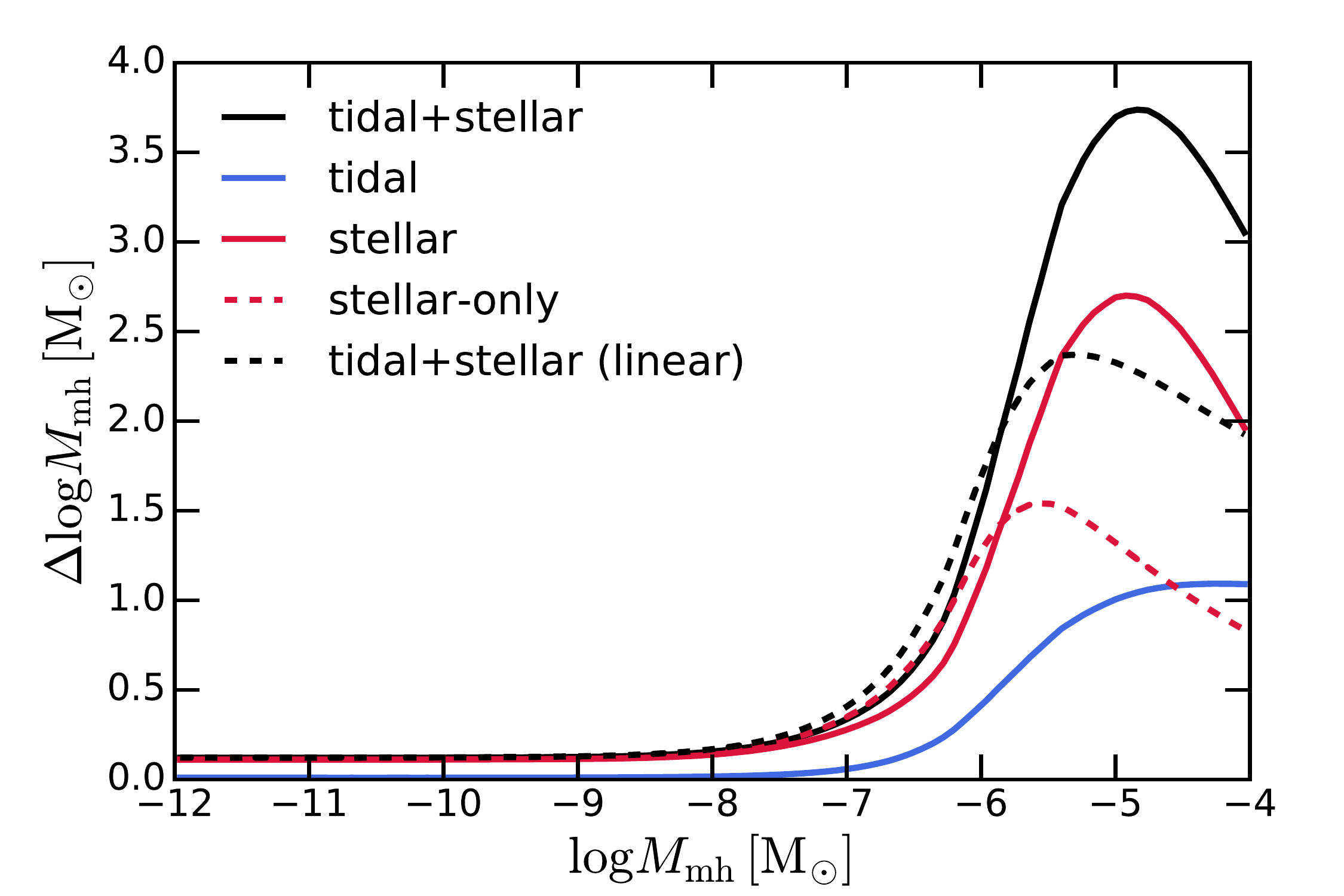}
    \caption{Median mass loss induced by disruption versus the initial mass of minihalos. We have decomposed the mass loss to tidal and stellar disruptions. The red dashed line shows the mass loss due to stellar disruption when there are no prior tidal corrections. The black dashed line shows the total mass loss if adding stellar and tidal terms linearly. In all cases, stellar disruption is the dominant mechanism. The non-linear effects of tidal disruption suppress stellar disruption at $M_{\rm mh}<10^{-6}\,{\rm M}_{\odot}$, but promote that at $M_{\rm mh}>10^{-6}\,{\rm M}_{\odot}$. Nevertheless, this does not affect the mass functions shown in Figure~\ref{fig:mass-func-8kpc}, since the differences show up at the mass where the density of minihalos is dropping rapidly.}
\label{fig:sample_dm_vs_mass}
\end{figure}

\begin{figure}
    \centering
    \includegraphics[width=0.49\textwidth]{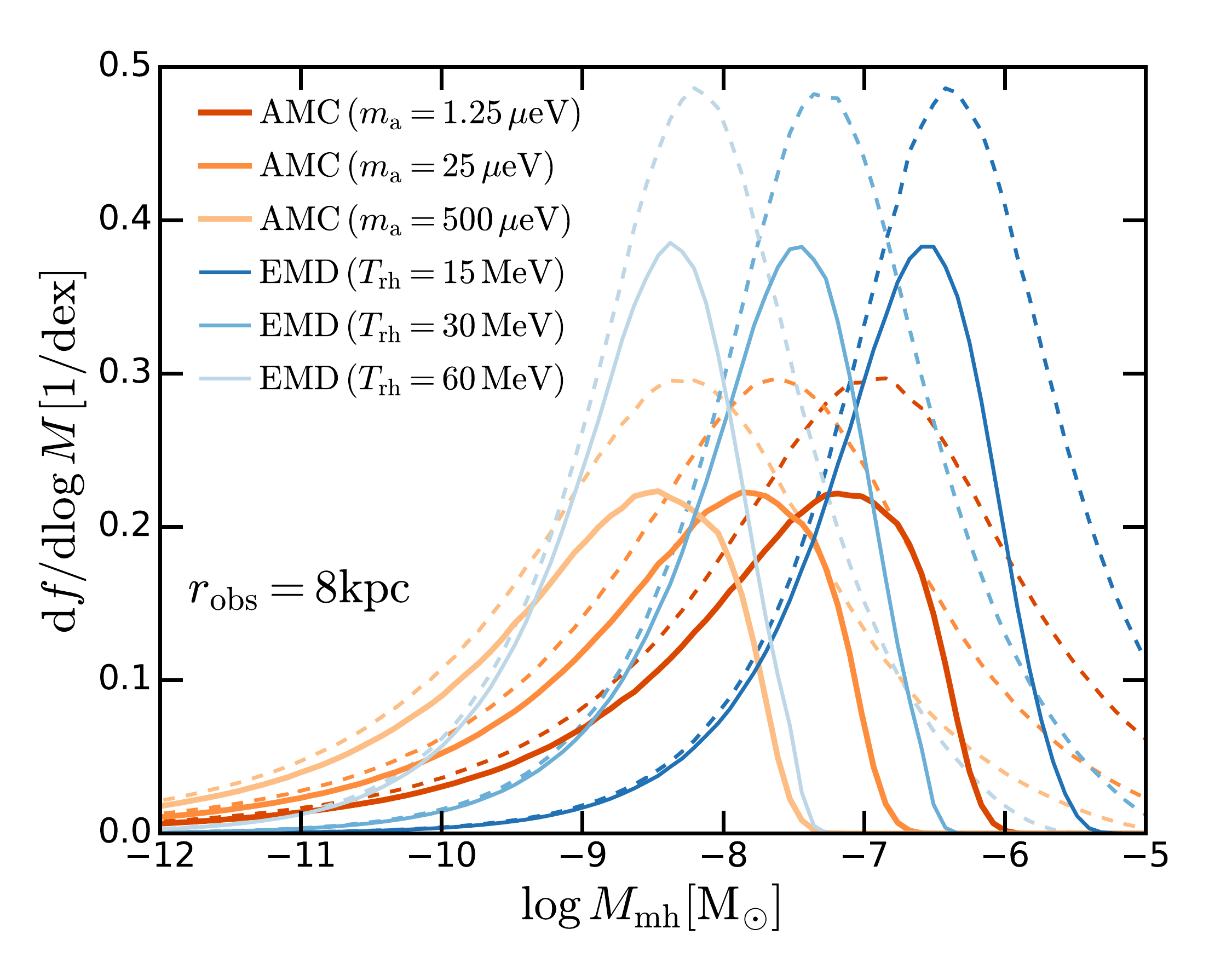}
    \includegraphics[width=0.49\textwidth]{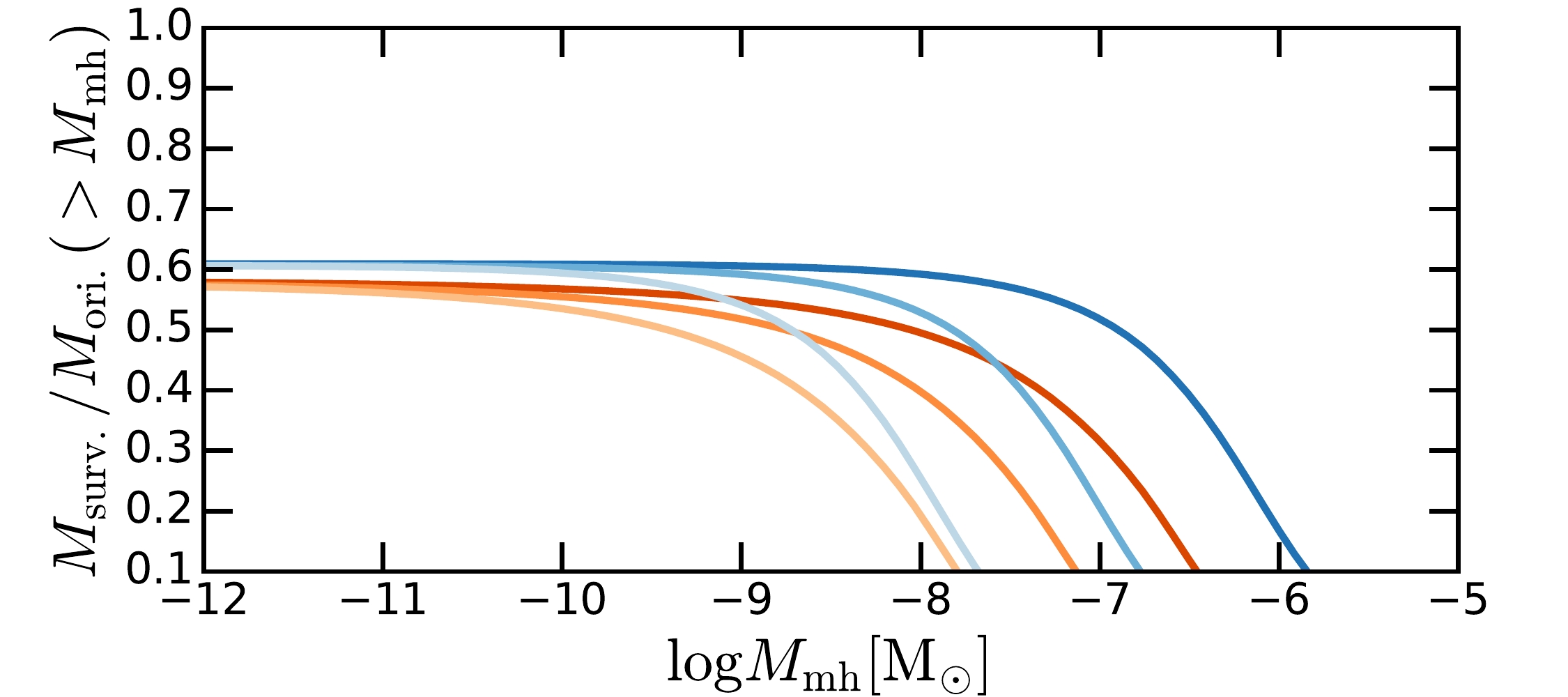}
    \caption{{\it Top:} Mass function of minihalos in different models before (dashed) and after (solid) disruption. Two different models are studied: post-inflationary AMCs and EMD. The reheating temperature for the EMD era is $T_{\rm rh}=15,30,60$ GeV while the axion mass in the post-inflationary AMC scenario is chosen as $m_a=1.25,25,500\, \mu{\rm eV}$. These parameters are purely chosen for illustrative purposes and one can use other parameters which will shift the mass range, but the shape of the mass function will remain the same. {\it Bottom}: The integrated mass survival fraction of minihalos for the models shown in the top panel. Regardless of the model choices, the overall survival fraction of minihalos with $M_{\rm mh}>10^{-12}\,{\rm M}_{\odot}$ is stably around $60\%$.}
    \label{fig:mass-func2}
\end{figure}

In Figure~\ref{fig:mass-func-rvary}, we show the post-disruption mass function of minihalos at different galactocentric distances, $r_{\rm obs}=4,8,16\,{\rm kpc}$, for the $m_{\rm a}=25\,\mu{\rm eV}$ AMC model. We find similar behavior of the post-disruption mass functions at different target radii. In all cases, the massive end is more severely disrupted, and stellar disruption dominates. The disruption at smaller galactocentric distances is stronger primarily due to enhanced stellar surface densities, resulting in the peak of the mass function shifting towards lower masses. The geometric assumptions of our model will break at the central bulge ($r_{\rm obs}\lesssim 1\,{\rm kpc}$) of the galaxy. The effective stellar density can be much higher than implied by the disk model, and minihalos will spend most of their lifetime in dense stellar environments. According to the back-of-the-envelope estimation given in Equation~\ref{eq:back-of-the-envelope-1}, the survival probability in this scenario should diminish to zero with moderately high stellar densities.

\subsection{Impacts of different disruption mechanisms}

To better illustrate the impact of tidal disruption, in the top panels of Figure~\ref{fig:phase_space}, we show the mass distribution of minihalos on the plane of $c$ versus $M_{\rm mh}$ before and after the tidal disruption. The initial mass concentration relation has a big scatter driven by the distribution of $z_{\rm i}$. At the massive end, the scatter approaches zero since all of the minihalos there have $z_{\rm i} \sim z_{\rm c} \sim 0$. Tidal disruptions (along with corrections on minihalo structural parameters, see Equation~\ref{eq:tidalcorr}) effectively lower the concentration of minihalos, especially at the massive end. We need to note that the average densities of minihalos are enhanced in the meantime, so the net effects over following stellar disruptions will depend on the trade-off between lowered concentrations and enhanced densities (see Equation~\ref{eq:energy_injection_aftertidal}).

In Figure~\ref{fig:sample_dm_vs_mass}, we show the median mass loss versus the initial mass of minihalo decomposed by the disruption mechanism. The stellar disruption is the dominant mechanism over the entire mass range of interest. However, the tidal disruptions do have non-linear effects in modifying the structural parameters of minihalos prior to stellar encounters. The net effect over stellar disruptions is a trade-off between lowered effective concentrations and enhanced average densities of minihalos. It suppresses (promotes) the stellar disruption at $M_{\rm mh}<10^{-6}\,{\rm M}_{\odot}$ ($M_{\rm mh}>10^{-6}\,{\rm M}_{\odot}$). Nevertheless, the non-linear effects do not affect much the mass functions or the overall survival probability of minihalos shown in Figure~\ref{fig:mass-func-8kpc} since most of the differences show up at the decreasing edge of the mass function. Another interesting feature to note is the peak of the mass loss curve due to stellar disruptions at around $M_{\rm mh}\sim 10^{-5}\,{\rm M}_{\odot}$ ($10^{-6}$ if tidal disruption is not introduced). The peak is related to the turnover feature (at the same mass scale) of the mass-concentration distribution shown in the upper panel of Figure~\ref{fig:phase_space}. The peak in the mass loss curve corresponds to the spike in the minihalos mass function (stellar-only case) shown in Figure~\ref{fig:mass-func-8kpc}. This is an example that convolution of the mass-concentration relation and mass function gives rise to non-linear features.

\subsection{Disruption for different physics models}

Here we explore the disruption of minihalos in different physics models summarized in Section~\ref{sec:models}. These models feature different initial mass functions and mass-concentration relations. 

In the top panel of Figure~\ref{fig:mass-func2}, we present the pre- and post-disruption mass functions of minihalos in the AMC model with axion mass $m_{\rm a}=1.25,\,25,\,500\,\mu{\rm eV}$. For the EMD case, we show the model with reheating temperature $T_{\rm rh}=15,\,30,\,60\,{\rm MeV}$. The numerical sampling experiments are all conducted at $r_{\rm obs}=8\,{\rm kpc}$. Since both the initial mass functions and mass-concentration relations are almost self-similar with a horizontal mass shift (and the relative mass change only depends on $c$ and $\rho_{\rm mh}$), the post-disruption mass functions are also similar albeit with a horizontal shift. The minihalo abundance reduction from disruptions has similar patterns between the AMC and EMD models. In the bottom panel of Figure~\ref{fig:mass-func2}, we show the mass survival fraction of minihalos. Regardless of the model choice, the overall survival fraction is about $60\%$ for minihalos with $M_{\rm mh}>10^{-12}\,{\rm M}_{\odot}$. It is visible that the AMC (EMD) model with higher $m_{\rm a}$ ($T_{\rm rh}$) suffers from stronger disruption at the massive end. This is due to the lower minihalo concentration at the same mass in these models (as shown in Figure~\ref{fig:mass_function_init}). The EMD models exhibit sharper decreases in survival fraction at the massive end. The mass function of EMD models features a steeper decline at the massive end, and therefore, at a fixed mass, the same level of a horizontal shift of mass function will give rise to a larger decrease in minihalo abundance.

\begin{figure}
    \centering
    \includegraphics[width=0.49\textwidth]{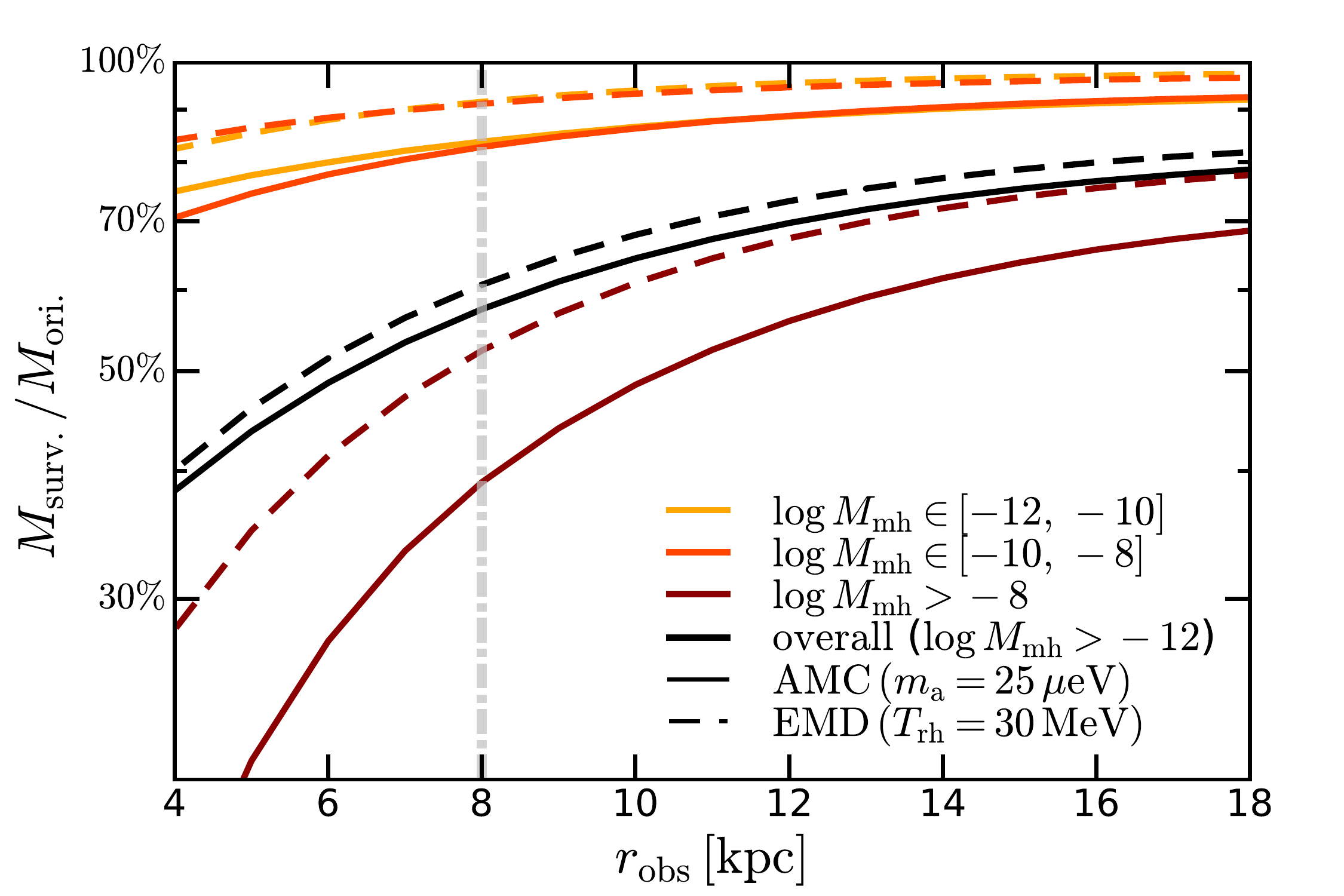}
    \caption{Mass survival fraction of minihalos as a function of galactocentric distance. In addition to the overall survival fraction after stellar and tidal disruptions, we also show the survival (mass) fraction of minihalos in three mass bins: $\log{(M_{\rm mh}/{\rm M}_{\odot})} \in [-12,-10], (-10, -8], (-8,\infty)$. The survival fraction has a strong dependence on minihalo mass. Massive minihalos are more severely disrupted, especially at small galactocentric radii. Less than a half of the minihalo in the $(-8,\infty)$ bin can survive at $r_{\rm obs}\lesssim 8\,{\rm kpc}$. On the other hand, low-mass minihalos in the $[-12,-10]$ bin have $\gtrsim 70\%$ survival fraction at any radius.}
    \label{fig:fsurvive}
\end{figure}

From the comparisons shown in this section, we can conclude that the model variations have little impact on the mass function and survival fraction of minihalos up to the mass shift noted previously. The estimated mass survival fraction of minihalos in the Solar neighborhood is about $60\%$.

\subsection{Galactic survival fraction}

In Figure~\ref{fig:fsurvive}, we show the survival (mass) fraction of minihalos after stellar and tidal disruption as a function of the radius of observation. We choose the fiducial AMC and EMD models (with $m_{\rm a}=25\,\mu{\rm eV}$ and $T_{\rm rh}=30\,{\rm MeV}$) for comparison here. The number of surviving minihalos is evaluated by integrating the mass function in three mass bins: $\log{(M_{\rm mh}/{\rm M}_{\odot})} \in [-12,-10], (-10, -8], (-8,+\infty)$. For both models, the survival fraction of minihalos significantly drops with increasing minihalo mass and decreasing galactocentric distance. Quantitatively, more than $70\%$ of minihalos in the mass range $\log{(M_{\rm mh}/{\rm M}_{\odot})} \in [-12,-10]$ survive at any radius of observation, as opposed to $\lesssim 50\%$ ($\lesssim 30\%$) survival fraction of minihalos with $\log{(M_{\rm mh}/{\rm M}_{\odot})} > -8$ at $r_{\rm obs}\lesssim 8\,{\rm kpc}$ ($4\,{\rm kpc}$). Low-mass minihalos are less vulnerable to disruption with more concentrated structures, although the survival fraction still decreases sharply at small galactocentric radii due to enhanced stellar surface density. We do not extend this to $r<4\,{\rm kpc}$. The bulge component will start to have an impact, lowering the survivability of minihalos even further. 

Even though the survival fraction of the most massive minihaloes in the Solar neighborhood can drop significantly because of their diluted structures from the hierarchical assembly, we expect a high overall survival fraction, $\gtrsim 60\%$, of QCD axion miniclusters or EMD minihalos, which is dominated by the concentrated minihalos with low masses. Local measurements like PTAs will be sensitive to minihalos after disruption since they can potentially probe a mass fraction (${\rm d}f/{\rm d}\log{M_{\rm mh}}$) well below $\sim 10\%$~\citep[{\em e.g.},][]{Dror:2019twh,Ramani:2020hdo,Lee:2020wfn}. Most of the surviving minihalos are cores with very high densities. We have shown in Figure~\ref{fig:density_profile} in Section~\ref{sec:stellar_sim} that the stellar disruption has a limited impact on the central core but primarily destroys the outer shells of the minihalo. The density profile at the outskirts of the minihalo remains steeper than the NFW profile even after the minihalo has fully relaxed from the previous encounter. Direct detection signals can be promoted further for these concentrated remnants \citep[{\em e.g.}\,][]{Lee:2020wfn}. In addition, specifically for the axion scenario, we expect a non-negligible fraction of free axions in the Solar neighborhood. These free axions could impact direct detection signals in axion haloscope experiments~\citep[{\em e.g.},][]{Asztalos2010,Graham2015,Du2018}. In summary, our findings are encouraging for the prospects of axion direct detection in the post-inflationary scenario as well as minihalos formed from EMD. We leave the calculations of astrophysical signals of dark matter minihalos in the Milky Way for a follow-up study.

\section{Conclusions}

In this paper, we systematically studied the environmental effects on dark matter minihalos (as light as $10^{-12}\,{\rm M}_{\odot}$) after infall to the Milky Way. Due to the large dynamic range, it is impossible to simultaneously track the evolution of minihalos in a standard cosmological simulation of the Milky Way-mass galaxy. Therefore, we developed a framework to combine small-scale, idealized N-body simulations with an analytic model of the Milky Way galaxy to make these sorts of predictions tractable. 

Stellar disruption and tidal disruption are the most critical environmental effects. For stellar disruption, the analytic expressions in the literature are inadequate to accurately capture the minihalo mass loss after stellar encounters. To address this, we developed a semi-analytic model calibrated by a suite of N-body simulations of idealized encounters between a star and a minihalo, varying impact parameters, minihalo masses, and concentrations. On the other hand, for tidal disruption, our N-body simulations show that the disruption effects can be predicted accurately by relatively simple analytic models. To connect to galactic scales, we apply a simplified orbital model of minihalos and derive an orbital-averaged treatment of stellar and tidal disruptions. We perform Monte Carlo simulations to model the mass evolution of minihalos with various masses, concentrations, and infall histories. The framework allows us to make predictions for the survival fraction of minihalos in the Milky Way as a function of galactocentric radius and halo parameters. This paper focuses on the miniclusters of post-inflationary axions and minihalos in the Early Matter Domination. However, the same framework can be easily applied to other types of dark matter substructure models. 

Our major findings can be summarized below:
\begin{itemize}
    \item When the energy imparted ($\Delta E$) during a stellar encounter is much smaller than the binding energy of the minihalo ($E_{\rm b}$), the mass loss is relatively insensitive to minihalo concentration. However, when $\Delta E/E_{\rm b}$ exceeds unity and continues to increase, minihalos with lower concentration will experience significantly larger mass loss in stellar encounters. This aspect has not been properly considered in many previous studies of dark matter substructures. We propose a simple and intuitive way to model successive encounters by aggregating the imparted energy from individual encounters and applying the same response curve calibrated for single encounters. This method works reasonably well when the imparted energy from individual encounters $\Delta E/E_{\rm b} \lesssim 0.1$ (quasi-static limit). 

    \item For tidal disruption, we confirm the existence of a tidal radius outside which minihalos are largely stripped. The analytical formula works remarkably well even for extremely light minihalos. On the other hand, the stellar disruption results in a shell of marginally-bound particles propagating outwards, leaving a characteristic density profile similar to the Hernquist profile at large radii. The stellar disruption can also slightly reduce the minihalo central densities. 

    \item  We show that there could be non-trivial non-linear interactions/combined effects between these disruption processes. The first tidal or stellar disruption will leave dense cores of the original minihalos, which are typically less vulnerable to future disruptions. Therefore, the assumptions on the structures of minihalos need to be revised. Our analytic method includes analytical corrections on minihalo structural parameters after tidal disruptions. 

    \item Applying our methods to well-motivated models like post-inflationary AMC and EMD minihalos, we find a relatively stable mass (number) survival fraction of $\sim 60\%$ ($\sim 80\%$) at the Solar neighborhood. The numbers are insensitive to the limiting minihalo mass we define as survival. The number is also insensitive to model choices. The survival fraction shows a strong dependence on the galactocentric radii, especially for massive minihalos. For example, the mass survival fraction of minihalos above $10^{-8}\,{\rm M}_{\odot}$ can drop to $\lesssim 50\%$ ($\lesssim 30\%$) at $r_{\rm obs}<8\,{\rm kpc}$ ($4\,{\rm kpc}$).  Over the entire mass range of interest, stellar disruption is the dominant disruption mechanism. The remaining minihalos are abundant and dense enough to give direct detection signals in {\em e.g.} upcoming PTA observations.
    
\end{itemize}

The follow-up paper will focus on the direct detection signals of minihalos. In the future, the framework built in this paper can be applied to a spectrum of interesting topics in particle astrophysics, such as the dark matter annihilation rate in minihalos and axion minicluster-neutron star encounters.

\begin{acknowledgments}
We thank Andrea Mitridate for useful discussions on the mass-concentration relation and Gabriel Aguiar for the collaboration in the early stages of this work. HX is supported in part by the United States Department of Energy (DOE) under grant number DE-SC0011637. KZ is supported by a Simons Investigator award and the U.S. DOE, Office of Science, Office of High Energy Physics, under Award No. DE-SC0011632. Support for XS and PFH was provided by the National Science Foundation (NSF) Research Grants 1911233, 20009234, 2108318, NSF CAREER grant 1455342, the National Aeronautics and Space Administration (NASA) grants 80NSSC18K0562, HST-AR-15800. Numerical calculations were run on the Caltech computing cluster ``Wheeler'', allocations AST21010 and AST20016 supported by the NSF and the Texas Advanced Computing Center (TACC), and NASA HEC SMD-16-7592. The simulation data of this work was generated and stored on the computing system ``Wheeler'' at California Institute of Technology. The code for the semi-analytic model and the summary of simulation results are available at \href{https://bitbucket.org/ShenXuejian/minicluster-disruption/src/master/}{the project repository}. The raw data of the idealized simulations will be shared on reasonable request to the corresponding author.
\end{acknowledgments}

%

\vspace{5mm}





\appendix

\section{The minihalo mass function and concentration in different models}\label{app:mass_function}

The minihalo mass function arising from an enhanced matter power spectrum at small scales can be considered separately from the adiabatic power spectrum. 
We use the Press-Schechter model to compute the minihalo mass function \citep{1974ApJ...187..425P}
\begin{equation}
    \frac{M^2{\rm d}n/{\rm d}M}{\bar{\rho}}\frac{{\rm d}M}{M}=\nu f(\nu)\frac{{\rm d}\nu}{\nu},
\end{equation}
where $\bar{\rho}$ is the comoving density of dark matter and $\nu$ is a dimensionless parameter that defines the rareness of the halo. $f(\nu)$ and $\nu$ are defined as
\begin{equation}
\begin{split}
    &\nu f(\nu)=\sqrt{\frac{\nu}{2\pi}}{\rm exp}(-\nu/2),\\
    &\nu \equiv \frac{\delta_{\rm c}^2(z)}{\sigma^2(M)},\\
\end{split}
\end{equation}
where $\delta_{\rm c} = 1.686$ is the critical density required for the formation of collapsed halos in spherical collapse models. $\sigma^2(M)$ is the variance of the initial perturbations smoothed with a top-hat filter of scale $R=(3M/4\pi\bar{\rho})^{1/3}$, which can be determined as
\begin{equation}
\sigma^2(M)\equiv \int \frac{{\rm d}k}{k}\frac{k^3P(k)}{2\pi^2}D^{2}_1(z)\vert W(kR)\vert^2,
\end{equation}
where $W(x) = (3/x^3)[{\rm sin}(x)-x{\rm cos}(x)]$ 
is the spherical top-hat window function, $D_1(z)$ is the growth function normalized in the radiation era, and $P(k)$ is the primordial matter power spectrum introduced by new physics, such as axions.
The variance of the white-noise power spectrum from the axion in the post-inflationary scenario can be expressed as
\begin{equation}
   \sigma(M)=D_1(z)\sqrt{\frac{3A_{\rm osc}}{2\pi^2}\frac{M_0}{M}},
\end{equation}
where $A_{\rm osc}$ is the amplitude of the white-noise power spectrum, which is taken to be 0.1 in this work. $M_0$ can be determined from the QCD axion mass \citep{Dai_2020,Xiao:2021nkb}
\begin{equation}\label{eq:axion_mh_mass}
      M_0 =2.3\times 10^{-10}\left(\frac{50\mu\rm eV}{m_a}\right)^{0.51}\,{\rm M}_{\odot}.
\end{equation}
Given the value of $M_0$, one can obtain the minihalo mass function using the Press-Schechter model if the adiabatic fluctuations are neglected. It is worth noting that the variance is a function of $M/M_0$, such that the shape of the mass function will remain the same when we change the model parameters but the characteristic mass will shift accordingly.

In the early matter domination scenario, the minihalo mass function can be calculated with a similar method. In this scenario, the primordial power spectrum remains the same as the adiabatic fluctuations but their growth is modified. Effectively, this enhances the primordial power spectrum at small scales. The reheating temperature $T_{\rm rh}$ is the only relevant physical parameter that determines the characteristic scale of the matter power spectrum, which corresponds to a characteristic mass scale $M_{\rm rh}$. The variance is enhanced at small scales due to early matter domination, in a manner which scales as 
\begin{equation}
\sigma(M\lesssim M_{\rm rh})\propto D_1(z)(M/M_{\rm rh})^{-(n+3)/6},
\end{equation}
where $n=0.963$ is the scalar spectral index \citep{Erickcek:2011us} and $M_{\rm rh}\approx 9.06\times 10^{-5}\,{\rm M}_{\odot}(10{\rm MeV}/T_{\rm rh})^3$. Similar to the axion minicluster scenario, the variance is only a function of $M/M_{\rm rh}$, and the shape does not change with the reheating temperature. Thus we can compute the minihalo mass function based on the variance using Press-Schechter. We can further compute the mass function including minihalos that have fallen into the massive CDM halos by including the effect of adiabatic fluctuations, as shown in Figure~\ref{fig:mass-func2}. 

\section{Convergence testing on the simulation of stellar disruptions}
\label{app:conv}

Briefly here we discuss numerical tests of the simulations of stellar disruptions. The fiducial simulations presented in the main text employed the constant gravitation softening length $10^{-9}\,{\rm kpc}$ and dark matter particle mass resolution $10^{-16}\,{\rm M}_{\odot}$ for minihalos with a mass of $10^{-10}\,{\rm M}_{\odot}$. We cap the timestep at $10^{-8}$Gyr during the stellar encounter to resolve the star trajectory since the stellar disruption is most relevant during the crossing. We justify those choices in more detail here.

\subsection{Gravitational softening}
The gravitational softening length must be chosen appropriately in the simulation so that we can resolve the relevant physical scales. In our idealized simulations of stellar disruption, the minimum halo mass is $10^{-10}\,{\rm M}_{\odot}$ which corresponds to a scale radius of $r_{\rm s}=9.6\times 10^{-8}\,{\rm kpc}$ at $z=0$. With the fiducial particle mass resolution, the convergence radius of dark matter calculated using the \citet{Power2003} criterion is $\sim 10^{-8}\,{\rm kpc}$. Therefore, the fiducial gravitational softening length, $10^{-9}\,{\rm kpc}$, and particle mass resolution should be sufficient to resolve the core profile of the halo. An additional run was performed with a gravitational softening length $2\times 10^{-9}\rm kpc$ and the results of the disruption fraction and mass profile are robust. We have also verified that the results are robust to particle number at our fiducial resolution.

\subsection{Time-stepping}
Since the crossing time of a stellar encounter is orders of magnitude shorter than the internal dynamical time of the minihalo, we need sufficiently small timesteps to resolve the trajectory of the star in the vicinity of the minihalo. The time-stepping parameter in our fiducial simulations is capped at $10^{-8}\,{\rm Gyr}$, which is roughly $1/5\,R_{\rm mh}/v_{\ast}$, where $R_{\rm mh}$ is the minihalo radius and the corresponding minihalo mass is $10^{-10}\,{\rm M}_{\odot}$. In an additional run, the time-stepping parameter is changed to $2\times 10^{-8}\,{\rm Gyr}$ while other parameters including the minihalo parameters are exactly the same. We obtain the same disruption fraction, mass profile, and energy change after the halo is fully relaxed. 

\section{Orbital model of minihalos}
\label{app:orbit}

The analytic calculation of accumulated energy in Equation~\ref{eq:energy_injection_onep} comes from only one encounter with the minihalo velocities perpendicular to the plane of the Galactic disk. In reality, minihalos after infall to the Milky Way halo will typically cross the stellar disks multiple times at various locations. To measure the accumulated energy input from a series of disk crossings, we need to evaluate the total number of passages through the disk, the stellar surface density where the encounter occurs, and the angle of incidence, with an ensemble average over all possible orbits for the minihalos eventually found at the observed radius.

For simplicity, we adopt a singular isothermal sphere model following the method in \citet{VDB1999} to estimate the uncertainties related to the orbits of minihalos. The density and potential of the system are given by
\begin{equation}
    \rho(r) = \dfrac{V^2_{\rm c}}{4\pi G r^{2}}, \hspace{0.3cm}
    \Phi(r) = V^2_{\rm c} \ln{(r/r_{\rm 0})},
\end{equation}
where $V_{\rm c}$ is the constant circular velocity assumed to be $200\,{\rm km/s}$ and $r_{0}$ is the zero-potential reference point chosen to be $10\,{\rm kpc}$. For a bound test particle in this potential, its halocentric distance $r$ will oscillate between the peri and apocenter with the period
\begin{equation}
    T = 2 \int_{r_1}^{r_2} \dfrac{{\rm d}r}{\sqrt{ 2\left[E-\Phi(r)\right] - L^2/r^2 }},
\end{equation}
where $E$ and $L$ are the energy and angular momentum of the test particle, $r_1$ and $r_2$ are the peri and apocenter of its orbit, which can be determined by solving the equation
\begin{equation}
    \dfrac{1}{r^2} + \dfrac{2\left[\Phi(r)-E\right]}{L^2} = 0.
    \label{eq:orbit}
\end{equation}
Due to the scale-free nature of the singular isothermal profile, the equations can be simplified by defining the maximum angular momentum $L_{\rm c}(E)=r_{\rm c}(E) V_{\rm c}$, where $r_{\rm c}(E)$ is the radius of the circular orbit with energy $E$ given by
\begin{equation}
    r_{\rm c}(E) = r_0 \exp{\left(\dfrac{E}{V^2_{\rm c}}-\dfrac{1}{2}\right)}
\end{equation}
Based on this, the circularity parameter is defined as $\eta = L/L_{\rm c}(E)$ and Equation~\ref{eq:orbit} is reduced to 
\begin{equation}
    \dfrac{1}{x^2} + \dfrac{2 \ln{(x)}}{\eta^2} - \dfrac{1}{\eta^2} = 0,
\end{equation}
where $x\equiv r/r_{\rm c}(E)$ and the two solutions correspond to the peri and apocenter distances, which are independent of the energy of the test particle if normalized by $r_{\rm c}(E)$. Similarly, if normalized by $r_{\rm c}(E)/V_{\rm c}$, $T$ also becomes independent of $E$. The look-up tables of $T$, $r_1$, and $r_2$ with respect to $\eta$ are computed numerically. Here we do not consider the scattering and mergers of minihalos within the parent halo and treat their orbits as unperturbed from various relaxation mechanisms.

\begin{figure*}
    \centering
    \includegraphics[width=0.33\textwidth]{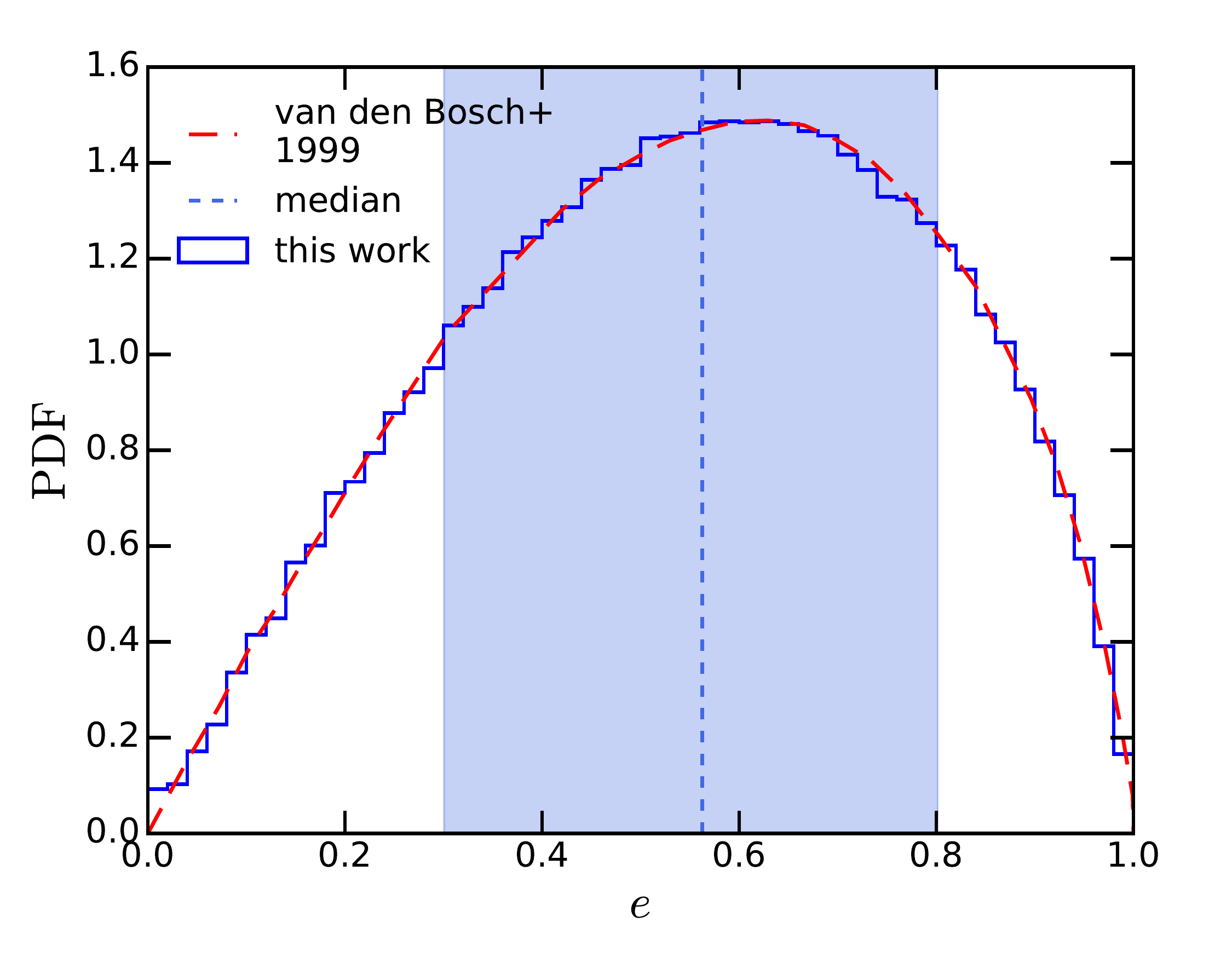}
    \includegraphics[width=0.33\textwidth]{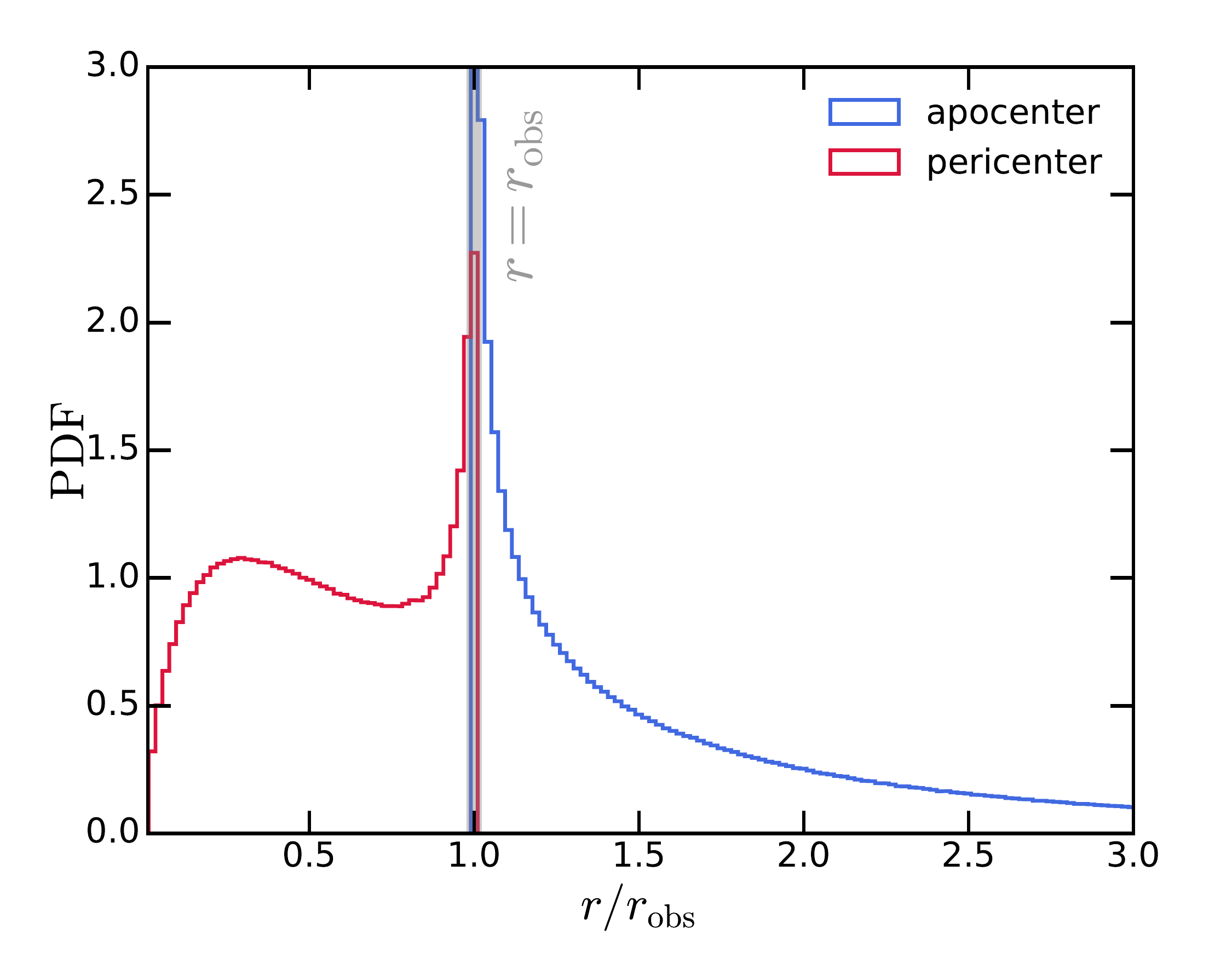}
    \includegraphics[width=0.33\textwidth]{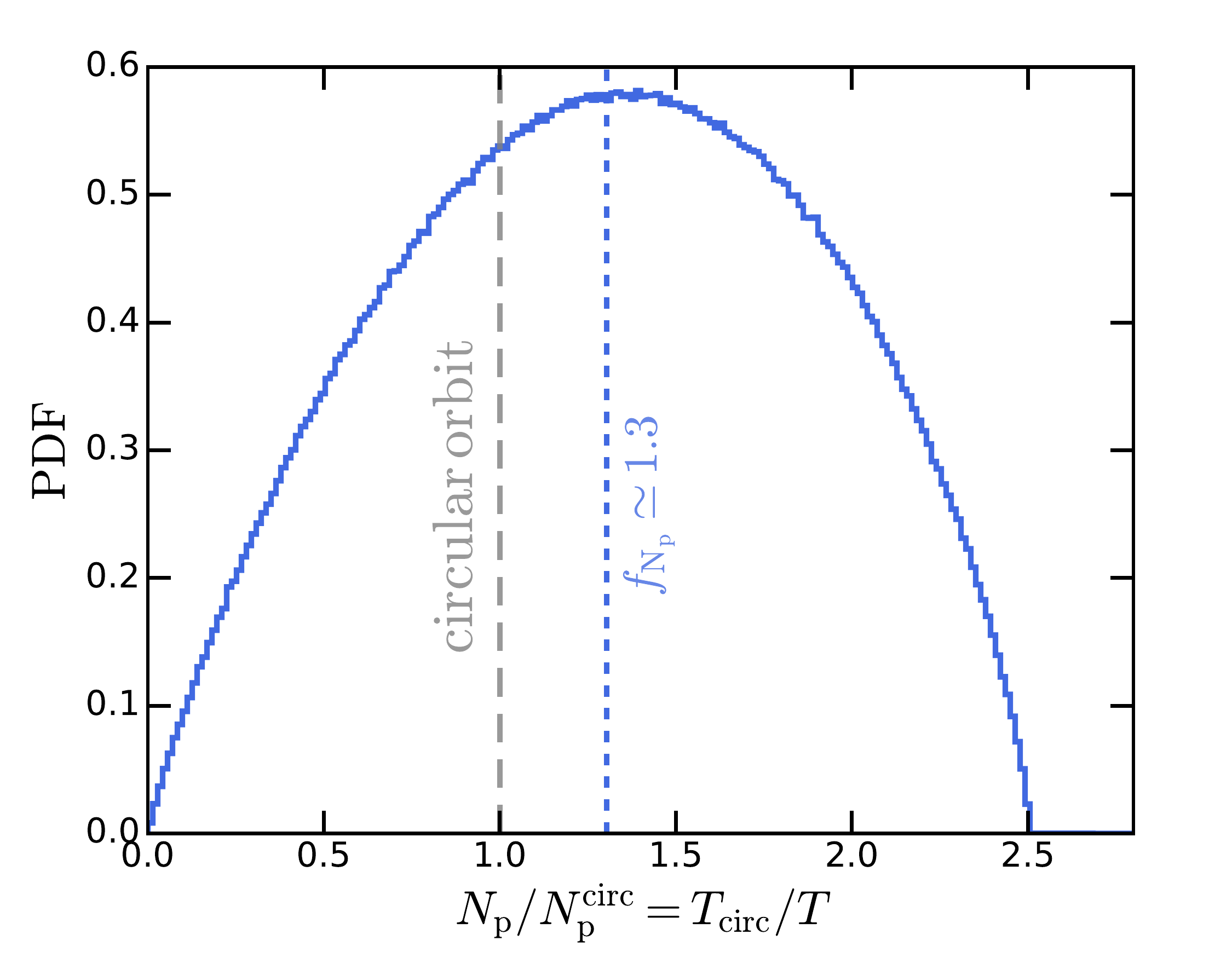}
    \caption{Distribution of eccentricities (left), pericenter and apocenter distances (middle), and orbital periods (or equivalently $N_{\rm p}$, right). The vertical blue dashed lines show the median values.}
    \label{fig:orbit}
\end{figure*}

\begin{figure}
    \centering
    \includegraphics[width=0.49\textwidth]{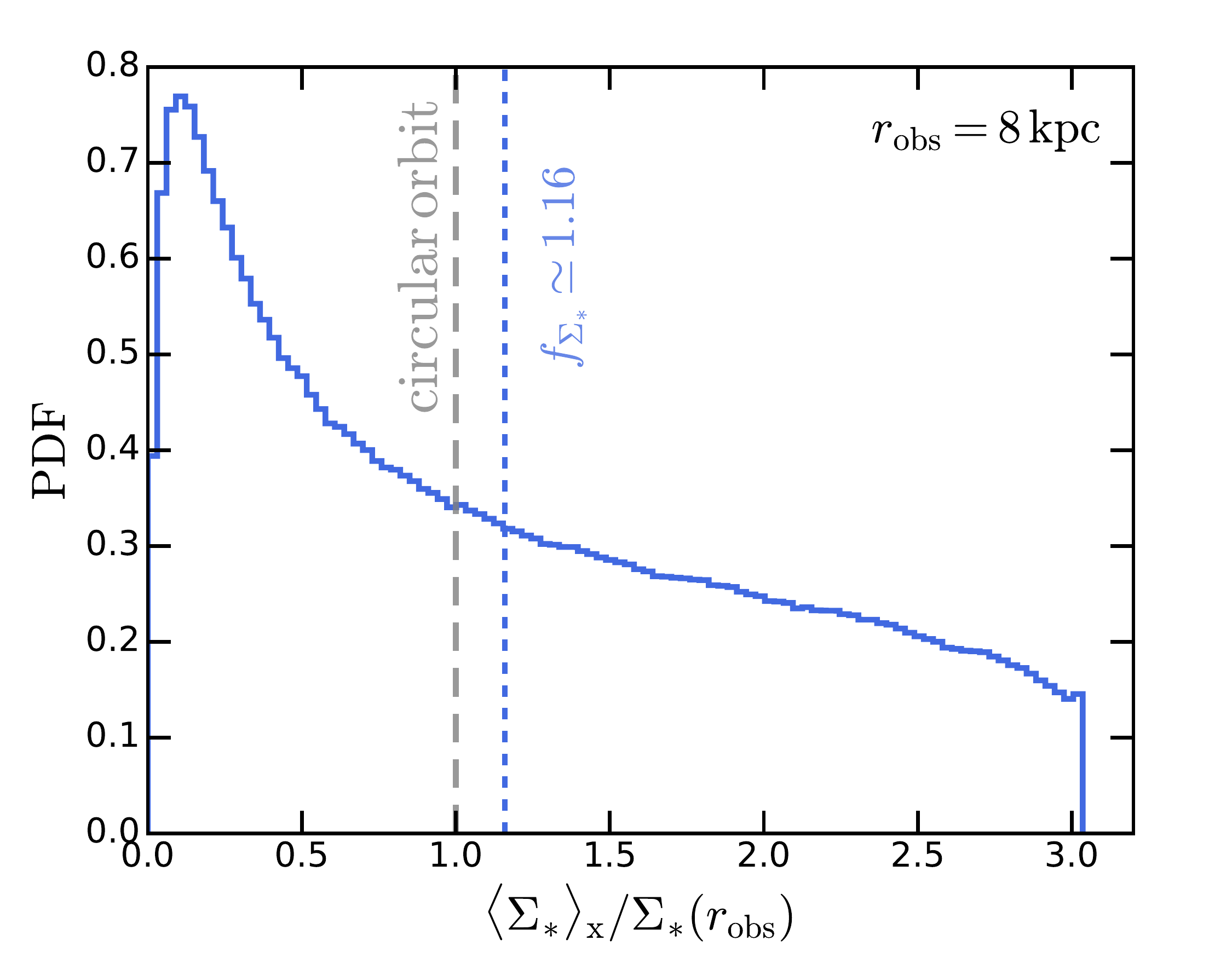}
    \includegraphics[width=0.49\textwidth]{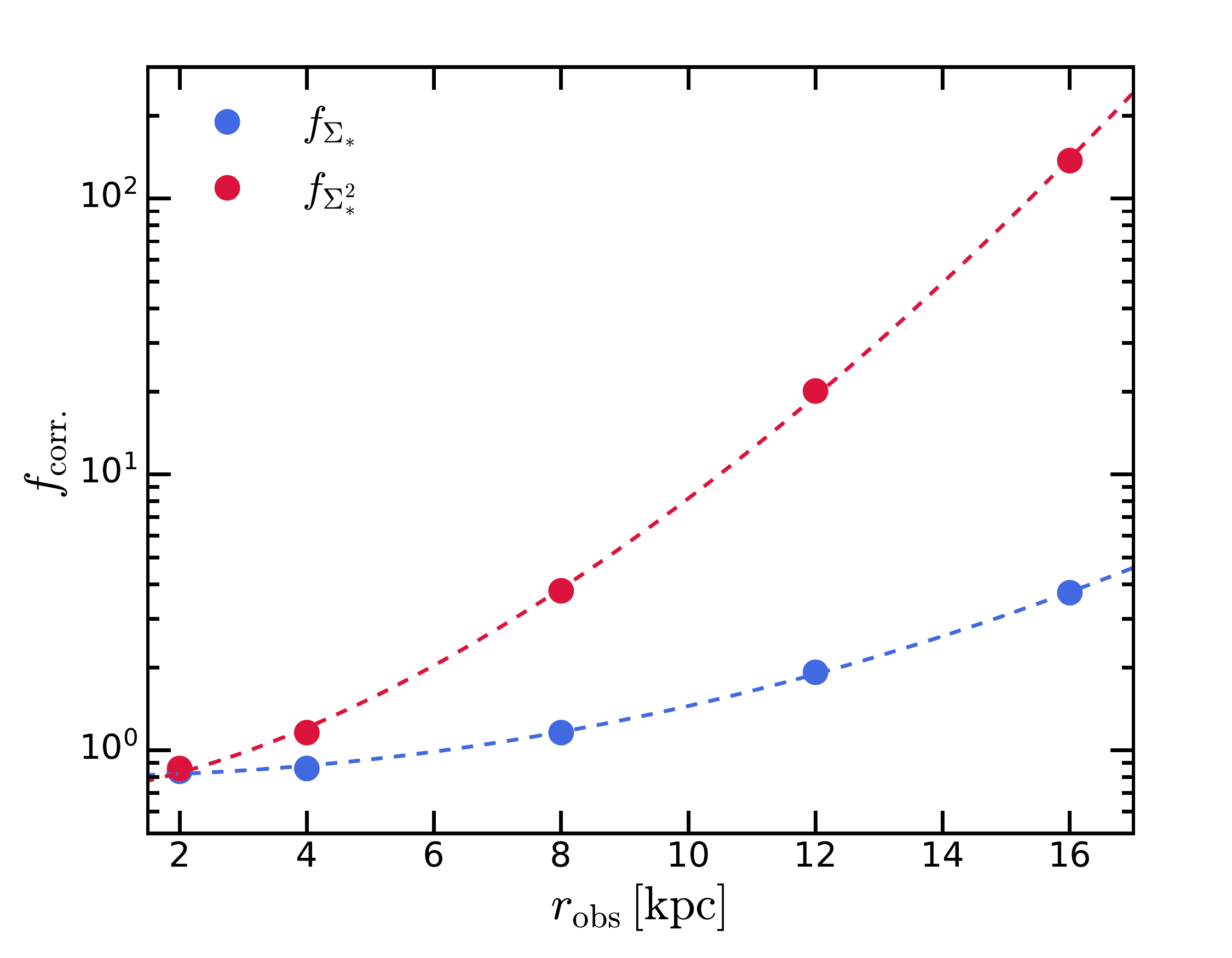}
    \caption{{\it Top}: Distribution of the surface density averaged over all past disk crossings. {\it Bottom}: The correction factor for $\Sigma_{\ast}$ and $\Sigma^{2}_{\ast}$ as a function of $r_{\rm obs}$. The dashed lines show the best-fit relations.}
    \label{fig:orbit2}
\end{figure}

Assuming spherical symmetry, the ergodic phase-space distribution function $f(\epsilon)$ can be derived through the Eddington inversion method
\begin{equation}
    f(\epsilon) = \dfrac{1}{\sqrt{8}\pi^2} \dfrac{{\rm d}}{{\rm d}\epsilon} \int_{0}^{\epsilon} \dfrac{ {\rm d}\psi }{ \sqrt{\epsilon - \psi}} \dfrac{{\rm d} \rho}{{\rm d} \psi},
    \label{eq:edd_inv}
\end{equation}
where $\psi \equiv -\Phi$ and $\epsilon \equiv -E$. For the singular isothermal profile, the solution is simply the Boltzmann distribution
\begin{equation}
    g(E) = K \exp{(-2E/V^2_{\rm c})},
\end{equation}
where $K$ is a constant normalization factor. 
Given a target radius for observation $r_{\rm obs}$, the normalized phase-space probability density function of dark matter particles localized around $r_{\rm obs}$ is
\begin{equation}
    f(E,L)|_{r_{\rm obs}} = \dfrac{4\pi}{r_{\rm obs}^2 \rho(r_{\rm obs})} \dfrac{g(E)\, L}{\sqrt{2(E-\Phi(r_{\rm obs}))-L^{2}/r_{\rm obs}}},
\end{equation}
where $E \geq \Phi(r_{\rm obs})$ and $L\leq 2r_{\rm obs}\sqrt{E-\Phi(r_{\rm obs})}$  are required. Replacing $L$ as $\eta L_{\rm c}(E)$, we obtain
\begin{equation}
    f(E, \eta)|_{r_{\rm obs}} = \dfrac{4\pi}{r_{\rm obs} \rho(r_{\rm obs})} \, g(E)\, L_{\rm c}(E)\, \dfrac{\eta}{\sqrt{\eta^2_{\rm max}-\eta^2}},
\end{equation}
where $\eta \leq \eta_{\rm max} = r_{\rm obs}\sqrt{2(E-\Phi(r_{\rm obs}))}/L_{\rm c}(E)$. Following \citet{VDB1999}, we perform a Monte Carlo sampling of particles in the phase space based on this distribution function. For each sample particle, we compute its orbital period $T$, pericenter and apocenter distances $r_1$ and $r_2$ based on the look-up table created earlier. We note that because of the self-similar nature of the isothermal sphere, the value of $r_1$ and $r_2$ with respect to $r_{\rm obs}$ is independent of $r_{\rm obs}$, similarly for $T/T_{\rm circ}(r_{\rm obs})$. In Figure~\ref{fig:orbit}, we show the probability distribution function of eccentricities, pericenter, and apocenter distances, and orbital periods of sampled minihalo orbits. The distribution should be self-similar for any target radius of observation. In the left panel of Figure~\ref{fig:orbit}, we compare the distribution of eccentricity derived here with that from \citet{VDB1999} which match perfectly. In the right panel of Figure~\ref{fig:orbit}, we compute the mean value of $N_{\rm p}/N^{\rm circ}_{\rm p} = T_{\rm circ}/T$ which gives us the correction factor for the number of passages through the stellar disk (compared to the circular orbit case) as $f_{N_{\rm p}} \equiv \langle N_{\rm p}\rangle_{\rm o}/N^{\rm circ}_{\rm p} \simeq 1.3$. $\langle \rangle_{\rm o}$ denotes averaging over all possible minihalo orbits. This correction factor is independent of the target radius of observation.

Assuming spherical symmetry, the orbit of a test particle will be confined in a plane and the precession of the orbit will eventually lead to a rosette-like pattern. The phase of the precession when the orbit crosses the disk plane is random. Considering a large ensemble of dark matter particles, the distribution of the radial location where the encounter with the stellar disk takes place will be the same as the probability of the presence of the particle at that distance, and thus equivalent to the time-averaged radial distance of the test particle. Therefore, we have the averaged surface density given the orbit parameter $E,L$ as
\begin{equation}
    \langle \Sigma_{\ast} \rangle_{\rm x}\,(E,L)  = \dfrac{2}{T} \int_{r_1}^{r_2} \dfrac{\Sigma_{\ast}(r)\, {\rm d}r}{\sqrt{ 2\left[E-\Phi(r)\right] - L^2/r^2 }},
\end{equation}
where $\langle \rangle_{\rm x}$ denotes averaging over all past disk crossings and the stellar surface density profile $\Sigma_{\ast}(r)$ is given below Equation~\ref{eq:tau_single} following the measurements in \citet{McMillan2011,McMillan2017}. The average of the second-order term $\langle \Sigma_{\ast}^2 \rangle_{\rm x}$ can be obtained similarly. In the top panel of Figure~\ref{fig:orbit2}, we show the distribution of $\langle \Sigma_{\ast} \rangle_{\rm x}/\Sigma_{\ast}(r_{\rm obs})$ at the target radius $8\,{\rm kpc}$ and the correction factor $f_{\Sigma_{\ast}}(r_{\rm obs}) \equiv \big\langle \langle \Sigma_{\ast} \rangle_{\rm x}/\Sigma_{\ast}(r_{\rm obs})\big\rangle_{\rm o}$ (averaging over all possible orbits) is about $1.16$. We compute the value of $f_{\Sigma_{\ast}}$ and $f_{\Sigma^{2}_{\ast}}$ at several different $r_{\rm obs}$ from $2$ to $16\,{\rm kpc}$ and find that both $f_{\Sigma_{\ast}}(r_{\rm obs})$ and $f_{\Sigma^{2}_{\ast}}(r_{\rm obs})$ can be fitted by the functional form $ A\, e^{B+(r_{\rm obs}/r_{\rm c})^{\alpha}}$. The best-fit parameters are $A = 0.106, B = 2.03, r_{\rm c}=12.961, \alpha=2.048$ for $f_{\Sigma_{\ast}}$ and $A = 0.318, B = 0.781, r_{\rm c}=5.740, \alpha = 1.628$ for $f_{\Sigma^{2}_{\ast}}$.

An additional correction comes from the enhanced surface density when the minihalo trajectory is not perpendicular to the disk plane. To the leading order, the effective surface density along the trajectory of the incident minihalo (as well as the time duration the minihalo stays in the disk) should scale as $1/{\rm cos}\theta$. Assuming the velocities of dark matter particles are isotropic, the correction factor is
\begin{equation}
    f_{\theta} = \Big\langle \dfrac{1}{{\rm cos}\theta} \Big\rangle
    = \int_{H/r_{\rm d}}^{1} \dfrac{{\rm d}{\rm cos}\theta}{{\rm cos}\theta} = \ln{(r_{\rm d}/H_{\rm d})} \simeq 2,
\end{equation}
where we have imposed a cut-off at ${\rm cos}\theta = H_{\rm d}/r_{\rm d}$ with $H_{\rm d}$ and $r_{\rm d}$ the scale height and length of the disk, assuming to be $400\,{\rm pc}$ and $3\,{\rm kpc}$, respectively. Particles with even smaller incidence angles stay in the disk and will become completely disrupted (see Equation~\ref{eq:tau_single}) which will have no impact on the averaged energy imparted. Combining the two effects above, we obtain the correction factor for the effective stellar surface density.





\end{document}